\documentclass[twocolumn,aps,superscriptaddress,showpacs,floatfix,prd,noshowpacs]{revtex4-2}
\usepackage[body={19.cm,24.cm}]{geometry}     
\usepackage{mathrsfs}
\usepackage{amssymb}
\usepackage{amsmath}
\usepackage{array}
\makeatletter
\@addtoreset{equation}{section}
\renewcommand{\theequation}{\arabic{section}.\arabic{equation}}
\makeatother
\usepackage{graphicx}
\usepackage[normalem]{ulem}
\usepackage{xcolor}
\usepackage{bm}
\usepackage{longtable}
\usepackage{slashed}
\usepackage{enumitem}
\usepackage{empheq}
\usepackage{booktabs}
\usepackage{multirow}

\usepackage{titlesec}
\renewcommand{\thesection}{\Roman{section}}
\titleformat{\section}{\small\bfseries\centering}{\thesection.}{0.5em}{}

\usepackage[T1]{fontenc}
\usepackage[utopia]{mathdesign}

\usepackage[breaklinks=true]{hyperref}
\hypersetup{
  colorlinks=true,
  citecolor=magenta,
  linkcolor=black,
  urlcolor=teal,
}

\renewcommand\sout{\bgroup \color{red} \ULdepth=-.5ex \ULset}

\renewcommand{\v}[1]{\textbf{#1}}
\renewcommand{\rm}[1]{\textrm{#1}}
\renewcommand{\d}{\mathrm{d}}

\usepackage{tikz}

\definecolor{lime}{HTML}{A6CE39}
\DeclareRobustCommand{\orcidicon}{
	\begin{tikzpicture}
	\draw[lime, fill=lime] (0,0) 
	circle [radius=0.16] 
	node[white] {{\fontfamily{qag}\selectfont \tiny ID}};
	\draw[white, fill=white] (-0.0625,0.095) 
	circle [radius=0.007];
	\end{tikzpicture}
	\hspace{-2mm}
}
\foreach \x in {A, ..., Z}{%
	\expandafter\xdef\csname orcid\x\endcsname{\noexpand\href{https://orcid.org/\csname orcidauthor\x\endcsname}{\noexpand\orcidicon}}
}

\DeclareSymbolFont{cmsymbols}{OMS}{cmsy}{m}{n}
\DeclareSymbolFontAlphabet{\mathcal}{cmsymbols}

\begin{document}

\title{Revisiting the Isospin Quartic Term of the Nucleonic Matter Equation of State in the Walecka Model}

\author{Jian-Hao Shi\orcidD{}}
\affiliation{Department of Physics and Center for Field Theory and Particle Physics, Fudan University, Shanghai 200438, China} 
\affiliation{Shanghai Research Center for Theoretical Nuclear Physics, NSFC and Fudan University, Shanghai 200438, China}
\author{Bao-Jun Cai\orcidA{}}\email{bjcai@fudan.edu.cn}
\affiliation{Shanghai Research Center for Theoretical Nuclear Physics, NSFC and Fudan University, Shanghai 200438, China}
\affiliation{Key Laboratory of Nuclear Physics and Ion-beam Application (MOE), Institute of Modern Physics, Fudan University, Shanghai 200433, China} 
\author{Yu-Gang Ma\orcidC{}}\email{mayugang$@$fudan.edu.cn}
\affiliation{Shanghai Research Center for Theoretical Nuclear Physics, NSFC and Fudan University, Shanghai 200438, China}
\affiliation{Key Laboratory of Nuclear Physics and Ion-beam Application (MOE), Institute of Modern Physics, Fudan University, Shanghai 200433, China} 
\affiliation{College of Physics, East China Normal University, Shanghai 200241, China}

\date{\today}

\newcommand{\x}{\mathrm{X}}
\newcommand{\y}{\mathrm{Y}}
\newcommand{\hr}{\widehat{r}}
\newcommand{\hP}{\widehat{P}}
\newcommand{\heps}{\widehat{\varepsilon}}
\newcommand{\hrho}{\widehat{\rho}}
\fontdimen2\font=1.5pt

\begin{abstract}

The fourth-order nuclear symmetry energy $E_{\rm{sym,4}}(\rho)$ is generally regarded as a small correction to the conventional (quadratic) symmetry energy. However, its magnitude, density dependence, and implications for neutron star (NS) matter still remain largely unconstrained. Within the nonlinear Walecka model, we derive a comprehensive analytical expression for $E_{\rm{sym,4}}(\rho)$ that incorporates the relevant meson interactions and cross couplings. By constructing an ensemble of 3000 equations of state (EOSs) constrained by terrestrial nuclear data and astrophysical observations, we uncover several notable features. (1) At nuclear saturation density, $E_{\rm{sym,4}}(\rho_0)$ can reach approximately $6$\,MeV or even higher, substantially exceeding the commonly expected upper limit of about $2$\,MeV. Values substantially above this conventional expectation can also remain compatible with existing microscopic calculations. (2) Despite this enhancement, $|E_{\rm{sym,4}}(\rho)|$ generally remains below about $20$\,MeV over NS densities. Nevertheless, its magnitude can reach approximately $25\%$ of the symmetry energy, indicating that the fourth-order term need not be negligible in highly neutron-rich matter. (3) Interestingly, EOSs with $E_{\rm{sym,4}}(\rho_0)\gtrsim2\,\rm{MeV}$ can exhibit positive curvature parameters $K_{\rm{sym}}$ of the symmetry energy while remaining consistent with the tidal deformability constraint from GW170817. This suggests that higher-order isospin contributions may relax the previously inferred preference for a negative $K_{\rm{sym}}$, with the fourth-order symmetry energy receiving relatively little attention in earlier analyses. (4) More strikingly, the density derivatives of $E_{\rm{sym,4}}(\rho)$ can be sizable, particularly its curvature coefficient $K_{\rm{sym,4}}$, leading to substantial contributions to the fourth-order isospin incompressibility coefficient. (5) Variations in $E_{\rm{sym,4}}(\rho_0)$ across the selected EOSs can change the low density squared speed of sound by $40\%$ or more. Furthermore, higher-order symmetry energies can appreciably alter the proton fraction and the core-crust transition properties of NS matter. These findings reveal an intriguing distinction between the magnitude of the fourth-order symmetry energy and its physical consequences: Even when its contribution to the EOS is relatively modest, its density dependence can produce pronounced effects on the response and composition of highly asymmetric nucleonic matter (ANM). Our results highlight the importance of going beyond the conventional quadratic approximation for a deeper understanding of the EOS of ANM and its implications for NS properties.

\end{abstract}

\pacs{21.65.-f, 21.30.Fe, 24.10.Jv}
\maketitle

\section{Introduction and Conclusions}

A small contribution to the energy does not necessarily imply a negligible effect on its density derivatives or the resulting physical response. This distinction is particularly relevant for the isospin dependence of nucleonic matter, where the equation of state (EOS) is commonly approximated by a quadratic function of the neutron-proton asymmetry\,\cite{LCK08}. The success of this parabolic approximation is largely supported by the relatively small fourth-order symmetry energy found in conventional nuclear models\,\cite{Bombaci1991,Lee1998,Zuo1999,Steiner2006,Cai2012,GonzalezBoquera2017,GonzalezBoquera2018,Vinas2021Gogny,SeifHashem2025,Bednarek2023,Pu2017,Zabari2019,CaiWangZhangChen2022}. However, neutron star (NS) matter is highly neutron-rich, and its composition, pressure, and stability depend not only on the energy itself but also on its derivatives with respect to density and isospin asymmetry\,\cite{Kubis2007,Xu2009,Lattimer2021}. Consequently, even a modest departure from the parabolic approximation may have appreciable physical consequences.

For cold asymmetric nucleonic matter (ANM), the EOS defined as the energy per nucleon can be expanded as $
E(\rho,\delta)\approx E_0(\rho)+E_{\rm{sym}}(\rho)\delta^2+E_{\rm{sym,4}}(\rho)\delta^4+\cdots$,
where $\rho$ is the total baryon density and $\delta$ is the isospin asymmetry between neutrons and protons. The quadratic symmetry energy $E_{\rm{sym}}(\rho)$ characterizes the leading energy cost of neutron-proton imbalance, while $E_{\rm{sym,4}}(\rho)$ describes the first correction to the parabolic approximation. At nuclear saturation density $\rho_0$, the latter is conventionally expected to be no larger than about $2$\,MeV\,\cite{Bombaci1991,Lee1998,Zuo1999,Steiner2006,Cai2012,GonzalezBoquera2017,GonzalezBoquera2018,Vinas2021Gogny,SeifHashem2025,Bednarek2023,Pu2017,Zabari2019,CaiWangZhangChen2022}. 

Yet this expectation is empirical and does not constitute a general upper bound, as no fundamental physical principle or symmetry requires the fourth-order symmetry energy to be small. In particular, the behavior of $E_{\rm{sym,4}}(\rho)$ at higher densities remains uncertain, as is the extent to which existing nuclear and astrophysical observations restrict its magnitude.

The physical origin of the fourth-order symmetry energy is also worth examining. In the dilute limit, its leading contribution arises from the nucleon kinetic energy, while interaction contributions enter at higher orders in density\,\cite{Cai2012}. At finite densities, however, the response of the nuclear mean fields to isospin asymmetry becomes increasingly important. Within the nonlinear Walecka model framework\,\cite{Walecka1974,SW1986,MullerSerot96,Li2022, Cai2017,Chin1977,ToddRutel2005,Dutra2014RMF,Long2004,Meng2006,MengZhou2015,Qin25,Chen2007}, the scalar $\sigma$ and $\delta$ mesons, the vector $\omega$ and $\rho$ mesons, and their nonlinear interactions can contribute differently to the fourth-order term\,\cite{Cai2012}. In particular, the $\sigma$-$\rho$ and $\omega$-$\rho$ cross couplings need not affect $E_{\rm{sym,4}}(\rho)$ in the same manner, even though both modify the effective isovector part. Better understanding these competing contributions requires an analytical formulation that retains their coupled responses to isospin asymmetry.

An equally important question is how the fourth-order symmetry energy enters different physical observables. Its contribution to the energy per nucleon is proportional to $\delta^4$, but differentiation with respect to isospin asymmetry introduces additional numerical factors. Thus, the same fourth-order term can have different relative weights in the EOS, the neutron-proton chemical potential difference, and the thermodynamic stability condition\,\cite{Kubis2007,Xu2009,Bombaci1991,Lee1998,Zuo1999,Steiner2006,Cai2012,GonzalezBoquera2017,GonzalezBoquera2018,Vinas2021Gogny,SeifHashem2025,Bednarek2023,Pu2017,Zabari2019,CaiWangZhangChen2022}. Its density derivatives introduce further effects on the incompressibility and sound speed. These observations suggest that the physical importance of $E_{\rm{sym,4}}(\rho)$ cannot be judged from its magnitude alone. They also motivate distinguishing its effects on the internal composition and stability of NS matter from those on global stellar properties, such as masses, radii, and tidal deformabilities\,\cite{Lindblom1992,LP01,LattimerPrakash2016,Baym2018,Oertel2017,Alford2008,Ann23,LiCaiXieZhang2021,Tews2018,Essick2021,Annala2020,Greif2019,XieLi2021-a,ZhangLi2025Twin,Han2021,Patra2022,ZhangLi2018NST,XieLi2019,BedaqueSteiner2015,Altiparmak2022,EckerRezzolla2022,Fujimoto2022,Marczenko2023,CaiLi2025Review,Shi2026}.

In this work, we derive a comprehensive analytical expression for $E_{\rm{sym,4}}(\rho)$ in the nonlinear Walecka model incorporating the $\sigma$, $\omega$, $\rho$, and $\delta$ mesons, together with their nonlinear self-interactions and cross couplings\,\cite{Li2022}. We examine its low- and high-density behavior, identify the contributions of individual meson fields, and investigate the conditions under which the fourth-order symmetry energy can become negative or appreciably enhanced. We then construct an ensemble of 3000 EOSs satisfying terrestrial nucleonic matter constraints\,\cite{Danielewicz2002,Tsang2009,Tsang2012,Danielewicz2014,LiHan2013} and astrophysical observations\,\cite{Demorest2010,Antoniadis2013,Cromartie2020,Fonseca2021,Romani2022,Abbott2017,Abbott2018,Abbott2020,Miller19,Riley19,Miller21,Riley21,Salmi24,Ditt24,Choud24,Mauviard26,Linares2018,Miller26,Salmi22,Mauviard25,Miller26-a} to assess the allowed magnitude and density dependence of $E_{\rm{sym,4}}(\rho)$. Finally, we explore its implications for the proton fraction, core-crust transition, fourth-order incompressibility, NS mass-radius relations, tidal deformabilities, and sound speed, with particular attention to the distinction between the magnitude of the fourth-order symmetry energy and the effects of its derivatives.

Our main findings in this work can be summarized as follows. (1) The fourth-order symmetry energy at saturation, $E_{\rm{sym,4}}(\rho_0)$, can reach approximately 6\,MeV or even higher, considerably exceeding the conventional expectation of about 2\,MeV. This indicates that the fourth-order contribution at saturation density can be substantially larger than conventionally assumed. (2) Despite this enhancement, $|E_{\rm{sym,4}}(\rho)|$ generally remains below about 20\,MeV over NS densities, with $|E_{\rm{sym,4}}(\rho)/E_{\rm{sym}}(\rho)|$ reaching approximately $25\%$, consistent with nuclear and astrophysical constraints. Thus, the fourth-order term need not be negligible in highly neutron-rich matter. The analytical expression further reveals competing contributions from different meson fields and their nonlinear interactions, which can substantially alter the magnitude, sign, and density dependence of $E_{\rm{sym,4}}(\rho)$, giving rise to nonmonotonic behavior and considerable model dependence. (3) Interestingly, EOSs with sizable $E_{\rm{sym,4}}(\rho_0)$ can exhibit positive curvature parameters $K_{\rm{sym}}$ of the symmetry energy while remaining consistent with the tidal deformability constraint from GW170817. This suggests that accounting for higher-order isospin contributions may relax the preference for a negative $K_{\rm{sym}}$ found in previous analyses of tidal deformability constraints, where the role of the fourth-order symmetry energy has received relatively little attention. (4) More strikingly, although $E_{\rm{sym,4}}(\rho)$ itself is relatively modest, its density derivatives, particularly the curvature parameter $K_{\rm{sym,4}}$, can be sizable and significantly influence the fourth-order isospin incompressibility coefficient $K_{\rm{sat,4}}$. This illustrates how a relatively small energy contribution can produce pronounced effects on the density response of highly ANM. (5) Variations associated with $E_{\rm{sym,4}}(\rho_0)$ can produce relative changes of about $40\%$ or more in the low-density squared speed of sound. Furthermore, higher-order symmetry energies can appreciably alter the proton fraction and core-crust transition properties of NS matter through their contributions to beta equilibrium and thermodynamic stability, while their effects on global NS mass-radius relations and tidal deformabilities remain comparatively modest. This contrast highlights the different sensitivities of local thermodynamic properties, matter composition, and global stellar observables to higher-order symmetry energies.

Taken together, these findings demonstrate that the physical relevance of higher-order symmetry energies depends not only on their magnitude but also on their density dependence and the observables considered. They further underscore the importance of going beyond the conventional parabolic approximation for a deeper understanding of the EOS of dense matter\,\cite{Dri21,Soren23,Lovato22,Li2026PRL,MaYG_arxiv, Li2018EffectiveMass,Akmal1998,Fang2026NST,Marino2024,Alp2025,Zou2026PRC,Xu2025CPL,Ding2024NST,Drischler2026,Li2026CPL,XuWang2024SCPMA,Yang2025CPL} and its implications for NS physics\,\cite{Li2019,Chat25,Alar25,Wat16,LiA25,Sun2025,Ofengeim2024,Ferreira2025,Tsang24,Huth2022,CaiLiMa2026phi,CaiLiMa2026X,CaiLi2024SSS,Cui2025NST,Xu2026SCPMA,Xie2026SCPMA,LiGrundler2024,RichterLi2023,GrundlerLi2026,Shi2026SSS}.

The remainder of this paper is organized as follows. In Section \ref{SEC2}, we introduce the theoretical framework and derive the analytical expression for $E_{\rm{sym,4}}(\rho)$ in the nonlinear Walecka model. In Section \ref{SEC3}, we investigate its low- and high-density properties and the roles of different meson interactions. In Section \ref{SEC4}, we constrain its magnitude and density dependence using terrestrial nuclear data and astrophysical observations, and examine its implications for NS properties. In Section \ref{SEC5}, we summarize our findings and discuss future directions. Appendix \ref{app:E4} provides the detailed derivation of $E_{\rm{sym,4}}(\rho)$, while Appendix \ref{app:2} examines the fine-tuned cancellations among extremely large higher-order symmetry energies and their potential inconsistencies at intermediate isospin asymmetries.

\section{Theoretical Framework and Analytical Formulation of the Fourth-Order Nuclear Symmetry Energy}\label{SEC2}

In this section, we give the theoretical framework for investigating the fourth-order nuclear symmetry energy $E_{\rm{sym,4}}(\rho)$ within an extended an extended nonlinear relativistic mean-field (RMF) model. In Subsection \ref{subs_EOSANM}, we introduce the basic definitions of the EOS of ANM and its expansions in isospin asymmetry and baryon density. In Subsection \ref{subs_E4-expression}, we formulate the nonlinear Walecka model incorporating the $\sigma$, $\omega$, $\rho$, and $\delta$ mesons, together with their self-interactions and cross couplings. We then derive the corresponding mean field equations, energy density, and pressure, followed by an analytical expression for $E_{\rm{sym,4}}(\rho)$ that separates the nucleon kinetic contribution from the meson-field responses. These results provide the basis for examining the magnitude and density dependence of the fourth-order symmetry energy and the roles of different meson interactions.

\subsection{Basic Definitions of the EOS of ANM}\label{subs_EOSANM}

For cold ANM, the neutron and proton
number densities are denoted by $\rho_{\rm n}$ and $\rho_{\rm p}$,
respectively. The total baryon density and isospin asymmetry are
defined as $\rho=\rho_{\rm n}+\rho_{\rm p}$ and
$\delta=(\rho_{\rm n}-\rho_{\rm p})/(\rho_{\rm n}+\rho_{\rm p})$.
The limits $\delta=0$ and $\delta=1$ correspond to symmetric nuclear
matter (SNM) and pure neutron matter (PNM), respectively.
The binding energy per nucleon is related to the nuclear energy
density $\varepsilon(\rho,\delta)$ through\,\cite{ChenCai2012}
\begin{equation}
    E(\rho,\delta)
    =\frac{\varepsilon(\rho,\delta)}{\rho}-M_{\rm N},
\end{equation}
where $M_{\rm N}$ is the nucleon mass.

Owing to the approximate charge symmetry of the strong interaction,
the EOS of ANM can be expanded in even powers of the isospin
asymmetry $\delta$ as\,\cite{LCK08}
\begin{equation}
    E(\rho,\delta)
    \approx E_0(\rho)+E_{\mathrm{sym}}(\rho)\delta^2
       +E_{\mathrm{sym},4}(\rho)\delta^4
       +\mathcal{O}(\delta^6),
\end{equation}
where $E_0(\rho)\equiv E(\rho,0)$ denotes the SNM EOS.
The (quadratic) symmetry energy and the fourth-order symmetry energies are defined by\,\cite{LCK08}
\begin{align}
    E_{\mathrm{sym}}(\rho)
    &=
    \left.\frac{1}{2}
    \frac{\partial^2 E(\rho,\delta)}{\partial\delta^2}
    \right|_{\delta=0},\\
    E_{\mathrm{sym},4}(\rho)
    &=
    \left.\frac{1}{24}
    \frac{\partial^4 E(\rho,\delta)}{\partial\delta^4}
    \right|_{\delta=0}.
\end{align}
The former characterizes the leading energy cost of isospin
asymmetry, while the latter quantifies the leading correction
to the widely used parabolic approximation.

The density dependence of the EOS near the SNM saturation density
$\rho_0$ can be characterized by introducing
$\chi\equiv(\rho-\rho_0)/3\rho_0$.
The SNM EOS is then expanded as\,\cite{LCK08}
\begin{equation}
    E_0(\rho)
    \approx E_0(\rho_0)+\frac{1}{2}K_0\chi^2
    +\frac{1}{6}J_0\chi^3+\mathcal{O}(\chi^4),
    \label{eq:anm_snm_density_expansion}
\end{equation}
where $E_0(\rho_0)$ is the saturation binding energy, and\,\cite{LCK08}
\begin{align}
    K_0
    &\equiv
    \left.9\rho_0^2
    \frac{\d^2 E_0(\rho)}{\d\rho^2}
    \right|_{\rho=\rho_0},~~
    J_0
    \equiv
    \left.27\rho_0^3
    \frac{\d^3 E_0(\rho)}{\d\rho^3}
    \right|_{\rho=\rho_0},
\end{align}
denote the incompressibility and skewness coefficients, respectively.
The former characterizes the stiffness of SNM against compression near saturation, while the latter describes the leading departure from a quadratic density dependence. The absence of a linear term follows from the saturation condition,
under which the pressure of SNM vanishes at $\rho_0$.
Similarly, the symmetry energy can be expanded as\,\cite{LCK08}
\begin{equation}
    E_{\mathrm{sym}}(\rho)
    \approx E_{\mathrm{sym}}(\rho_0)+L\chi+\mathcal{O}(\chi^2),
    \label{eq:anm_symmetry_density_expansion}
\end{equation}
where the slope parameter is defined by\,\cite{LCK08}
\begin{equation}
    L\equiv
    \left.3\rho_0
    \frac{\mathrm{d}E_{\mathrm{sym}}(\rho)}{\mathrm{d}\rho}
    \right|_{\rho=\rho_0}.
    \label{eq:anm_symmetry_slope}
\end{equation}
The parameters $E_{\mathrm{sym}}(\rho_0)$ and $L$ characterize the magnitude and leading density dependence of the symmetry energy near saturation, respectively, and play important roles in determining the properties of neutron-rich matter.

Together, these characteristic parameters describe the density dependence of SNM and the leading isospin dependence of ANM near saturation. However, they do not fully determine the EOS at large isospin asymmetries, where the convergence of the expansion in $\delta$ becomes particularly important. In particular, $E_{\mathrm{sym},4}(\rho)$ characterizes the leading departure from the parabolic approximation and may become appreciable in strongly neutron-rich matter, where the suppression by $\delta^4$ is considerably weakened. Its density dependence may also introduce additional contributions to the pressure and other thermodynamic properties that cannot be inferred from the quadratic symmetry energy alone. Moreover, retaining the fourth-order term does not necessarily guarantee an accurate description of PNM, since sixth- and higher-order contributions may also become important as $\delta$ approaches unity. These considerations motivate a systematic investigation of $E_{\mathrm{sym},4}(\rho)$, including its density dependence and physical implications, together with a quantitative assessment of the truncated isospin expansion against the complete EOS.
\subsection{Analytical Expression for the Fourth-order Symmetry Energy $E_{\textmd{sym},4}(\rho)$ in the Nonlinear Walecka Model with $\sigma$, $\omega$, $\rho$, and $\delta$ Mesons and Their Couplings}\label{subs_E4-expression}

In this work, we adopt an extended nonlinear Walecka model incorporating the $\sigma$, $\omega$, $\rho$, and $\delta$ mesons, together with their relevant nonlinear self-interactions and cross couplings\,\cite{MullerSerot96,Cai2012,Li2022}. By including all four meson fields and a broad set of meson interactions, this framework represents one of the most comprehensive nonlinear Walecka formulations employed in the literature, providing considerable flexibility in describing the density and isospin dependence of nucleonic matter. The corresponding Lagrangian density reads\,\cite{Li2022}
\begin{align}
    \mathcal{L}
    ={}&
    \overline{\psi}
    \left(i\partial_\mu\gamma^\mu-M_{\rm N}\right)\psi
    \notag\\
    &+g_\sigma\sigma\overline{\psi}\psi
    -g_\omega\omega_\mu\overline{\psi}\gamma^\mu\psi
    -g_\rho\vec{\rho}_\mu
    \overline{\psi}\gamma^\mu\vec{\tau}\psi
    \notag\\
    &+\frac{1}{2}
    \left(\partial_\mu\sigma\partial^\mu\sigma
    -m_\sigma^2\sigma^2\right)
    -\frac{1}{3}b_\sigma M_{\rm N}\left(g_\sigma\sigma\right)^3
    -\frac{1}{4}c_\sigma\left(g_\sigma\sigma\right)^4
    \notag\\
    &-\frac{1}{4}\omega_{\mu\nu}\omega^{\mu\nu}
    +\frac{1}{2}m_\omega^2\omega_\mu\omega^\mu
    +\frac{1}{4}c_\omega
    \left(g_\omega^2\omega_\mu\omega^\mu\right)^2
    \notag\\
    &-\frac{1}{4}\vec{\rho}_{\mu\nu}\vec{\rho}^{\mu\nu}
    +\frac{1}{2}m_\rho^2\vec{\rho}_\mu\vec{\rho}^{\,\mu}
    \notag\\
    &+\frac{1}{2}g_\rho^2
    \vec{\rho}_\mu\vec{\rho}^{\,\mu}
    \left(\Lambda_{\rm S}g_\sigma^2\sigma^2
    +\Lambda_{\rm V}g_\omega^2\omega_\nu\omega^\nu\right)
    \notag\\
    &+\frac{1}{2}   \left(\partial_\mu\vec{\delta}\partial^\mu\vec{\delta}
    -m_\delta^2\vec{\delta}^{\,2}\right)
    +g_\delta\vec{\delta}
    \overline{\psi}\vec{\tau}\psi
    +\frac{1}{2}C_{\delta\sigma}g_\sigma^2\sigma^2g_\delta^2\vec{\delta}^{\,2}.
    \label{eq:L}
\end{align}
Here, $M_{\rm N}$ and $m_\sigma$, $m_\omega$, $m_\rho$, and $m_\delta$ denote the nucleon and corresponding meson masses, respectively. The coupling constants $g_\sigma$, $g_\omega$, $g_\rho$, and $g_\delta$ characterize the respective nucleon-meson interactions. The coefficients $b_\sigma$ and $c_\sigma$ govern the cubic and quartic $\sigma$ self-interactions, while $c_\omega$ controls the quartic $\omega$ self-interaction. The parameters $\Lambda_{\rm S}$ and $\Lambda_{\rm V}$ describe the $\sigma$-$\rho$ and $\omega$-$\rho$ cross couplings, respectively, whereas $C_{\delta\sigma}$ characterizes the $\sigma$-$\delta$ mixing interaction. The inclusion of these nonlinear couplings provides additional flexibility in describing the density and isospin dependence of the nuclear EOS, particularly the higher-order symmetry energies.

Within the RMF approximation, meson-field
fluctuations are neglected, and the meson fields are replaced by their
expectation values. For uniform nucleonic matter at rest, the non-vanishing
mean fields are $\overline{\sigma}$, $\overline{\omega}_0$,
$\overline{\rho}_0^{(3)}$, and $\overline{\delta}^{(3)}$.
The isoscalar $\sigma$ and $\omega$ fields describe the scalar attraction
and vector repulsion, respectively, while the isovector $\rho$ and
$\delta$ fields govern the isospin dependence of the vector and scalar
interactions. Applying the Euler--Lagrange equations to Eq.\,\eqref{eq:L}
yields the coupled mean-field equations\,\cite{Li2022}
\begin{align}
    m_\sigma^2\overline{\sigma}
    ={}&g_\sigma\bigg[
    \rho_{\rm p}^{\rm s}+\rho_{\rm n}^{\rm s}
    -b_\sigma M_{\rm N}\left(g_\sigma\overline{\sigma}\right)^2
    -c_\sigma\left(g_\sigma\overline{\sigma}\right)^3
    \notag\\
    &\quad+\Lambda_{\rm S}g_\sigma\overline{\sigma}
    \left(g_\rho\overline{\rho}_0^{(3)}\right)^2
    +C_{\delta\sigma}g_\sigma\overline{\sigma}\left(g_\delta\overline{\delta}^{(3)}\right)^2
    \bigg],
    \label{eq:seqm}\\
    m_\omega^2\overline{\omega}_0
    ={}&g_\omega\left[
    \rho-c_\omega\left(g_\omega\overline{\omega}_0\right)^3
    -\Lambda_{\rm V}g_\omega\overline{\omega}_0
    \left(g_\rho\overline{\rho}_0^{(3)}\right)^2
    \right],
    \label{eq:oeqm}\\
    m_\rho^2\overline{\rho}_0^{(3)}
    ={}&g_\rho\left[
    \rho_{\rm p}-\rho_{\rm n}
    -g_\rho\overline{\rho}_0^{(3)}
    \left(\Lambda_{\rm S}g_\sigma^2\overline{\sigma}^2 + \Lambda_{\rm V}g_\omega^2\overline{\omega}_0^2\right)
    \right],
    \label{eq:reqm}\\
    m_\delta^2\overline{\delta}^{(3)}
    ={}&g_\delta\left[
    \rho_{\rm p}^{\rm s}-\rho_{\rm n}^{\rm s}
    +C_{\delta\sigma}g_\delta\overline{\delta}^{(3)}
    \left(g_\sigma\overline{\sigma}\right)^2
    \right].
    \label{eq:deqm}
\end{align}
Here, the overline denotes a mean-field expectation value, the subscript
``0'' specifies the temporal component of a four-vector, and the
superscript ``(3)'' denotes the third component in isospin space.
These equations determine the meson mean fields self-consistently
through the nucleon number and scalar densities. The nonlinear
self-interactions and cross couplings modify the response of these
fields to changes in density and isospin asymmetry, thereby influencing
the density dependence of the nuclear EOS.

The neutron and proton scalar densities $\rho_J^{\rm s}$,
with $J=\mathrm{n,p}$, are given by\,\cite{Cai2012,Chen2007}
\begin{align}
    \rho_J^{\rm s}
    &=
    \frac{2}{(2\pi)^3}
    \int_{|\v{k}|\le k_{\rm F}^J}
    \d\v{k}
    \frac{M_J^*}{\sqrt{\v{k}^2+M_J^{*,2}}}
    \notag\\
    &=
    \frac{M_J^*}{2\pi^2}
    \left[
    k_{\rm F}^J\varepsilon_{\rm F}^{*,J}
    -M_J^{*,2}
    \ln\left(
    \frac{k_{\rm F}^J+\varepsilon_{\rm F}^{*,J}}{M_J^*}
    \right)
    \right].
    \label{eq:rhos}
\end{align}
The nucleon Dirac effective mass is determined by the isoscalar
$\sigma$ field and the isovector $\delta$ field\,\cite{Cai2012},
\begin{equation}
    M_J^*
    =M_{\rm N}-g_\sigma\overline{\sigma}
    +g_\delta\overline{\delta}^{(3)}\tau_3^J.
    \label{eq:effective_mass}
\end{equation}
The corresponding effective Fermi energy is
\begin{equation}
    \varepsilon_{\rm F}^{*,J}
    =\sqrt{k_{\rm F}^{J,2}+M_J^{*,2}},
    \label{eq:effective_fermi_energy}
\end{equation}
where $\tau_3^J=-1$ ($+1$) for protons (neutrons), and
$k_{\rm F}^J=k_{\rm F}(1+\tau_3^J\delta)^{1/3}$ is the Fermi
momentum of species $J$, with
$k_{\rm F}=(3\pi^2\rho/2)^{1/3}$ denoting the SNM Fermi momentum
at the same baryon density. The corresponding number density is
$\rho_J=k_{\rm F}^{J,3}/3\pi^2$.
While the $\sigma$ field shifts the neutron and proton Dirac effective
masses equally, the $\delta$ field generates an isospin-dependent
mass splitting in ANM. This splitting introduces
an additional contribution to the isospin dependence of the EOS,
particularly its higher-order symmetry energies.

Once the meson mean fields are determined self consistently, the energy
density and pressure follow directly from the energy-momentum tensor\,\cite{SW1986},
\begin{equation}
    \mathcal{T}^{\mu\nu}
    =
    \sum_{\phi_a}
    \frac{\partial\mathcal{L}}
    {\partial(\partial_\mu\phi_a)}
    \partial^\nu\phi_a
    -\mathcal{L}g^{\mu\nu}.
    \label{eq:energy_momentum_tensor}
\end{equation}
For uniform matter at rest,
$\varepsilon=\langle\mathcal{T}^{00}\rangle$ and
$P=3^{-1}\sum_{j=1}^3\langle\mathcal{T}^{jj}\rangle$.
Using the mean-field equations gives\,\cite{SW1986}
\begin{align}
    \varepsilon
    ={}&
    \varepsilon_{\rm p}^{\mathrm{kin}}
    +\varepsilon_{\rm n}^{\mathrm{kin}}
    \notag\\
    &+\frac{1}{2}m_\sigma^2\overline{\sigma}^2
    +\frac{1}{2}m_\omega^2\overline{\omega}_0^2
    +\frac{1}{2}m_\rho^2\overline{\rho}_0^{(3),2}
    +\frac{1}{2}m_\delta^2\overline{\delta}^{(3),2}
    \notag\\
    &+\frac{1}{3}M_{\rm N}b_\sigma
    \left(g_\sigma\overline{\sigma}\right)^3
    +\frac{1}{4}c_\sigma\left(g_\sigma\overline{\sigma}\right)^4
    +\frac{3}{4}c_\omega\left(g_\omega\overline{\omega}_0\right)^4
    \notag\\
    &+\frac{1}{2}
    \left(g_\rho\overline{\rho}_0^{(3)}\right)^2
    \left[\Lambda_{\rm S}\left(g_\sigma\overline{\sigma}\right)^2
    +3\Lambda_{\rm V}\left(g_\omega\overline{\omega}_0\right)^2\right]
    \notag\\
    &-\frac{1}{2}C_{\delta\sigma}
    \left(g_\delta\overline{\delta}^{(3)}\right)^2
    \left(g_\sigma\overline{\sigma}\right)^2,
    \label{eq:eps}
\end{align}
and\,\cite{SW1986}
\begin{align}
    P
    ={}&
    P_{\rm p}^{\mathrm{kin}}+P_{\rm n}^{\mathrm{kin}}
    \notag\\
    &-\frac{1}{2}m_\sigma^2\overline{\sigma}^2
    +\frac{1}{2}m_\omega^2\overline{\omega}_0^2
    +\frac{1}{2}m_\rho^2\overline{\rho}_0^{(3),2}
    -\frac{1}{2}m_\delta^2\overline{\delta}^{(3),2}
    \notag\\
    &-\frac{1}{3}M_{\rm N}b_\sigma
    \left(g_\sigma\overline{\sigma}\right)^3
    -\frac{1}{4}c_\sigma\left(g_\sigma\overline{\sigma}\right)^4
    +\frac{1}{4}c_\omega\left(g_\omega\overline{\omega}_0\right)^4
    \notag\\
    &+\frac{1}{2}
    \left(g_\rho\overline{\rho}_0^{(3)}\right)^2
    \left[\Lambda_{\rm S}\left(g_\sigma\overline{\sigma}\right)^2
    +\Lambda_{\rm V}\left(g_\omega\overline{\omega}_0\right)^2\right]
    \notag\\
    &+\frac{1}{2}C_{\delta\sigma}
    \left(g_\delta\overline{\delta}^{(3)}\right)^2
    \left(g_\sigma\overline{\sigma}\right)^2.
    \label{eq:P}
\end{align}

These expressions contain the nucleonic kinetic contributions, the
meson mean-field energies, and the nonlinear interaction terms.
In particular, the scalar and vector interactions contribute
differently to the energy density and pressure, while the nonlinear
meson couplings provide additional flexibility in controlling the
stiffness and isospin dependence of the EOS.
The nucleonic kinetic energy density and pressure are given by\,\cite{Chen2007}
\begin{align}
    \varepsilon_J^{\mathrm{kin}}
    &=
    \frac{2}{(2\pi)^3}
    \int_{|\v{k}|\le k_{\rm F}^J}
    \d\v{k}
    \sqrt{\v{k}^2+M_J^{*,2}}
    \notag\\
    &=
    \frac{1}{4}
    \left(
    3\varepsilon_{\rm F}^{*,J}\rho_J
    +M_J^*\rho_J^{\rm s}
    \right),
    \label{eq:kinetic_energy_density}
\\
    P_J^{\mathrm{kin}}
    &=
    \frac{1}{3}\frac{2}{(2\pi)^3}
    \int_{|\v{k}|\le k_{\rm F}^J}
    \d\v{k}
    \frac{\v{k}^2}
    {\sqrt{\v{k}^2+M_J^{*,2}}}
    \notag\\
    &=
    \frac{1}{4}
    \left(
    \varepsilon_{\rm F}^{*,J}\rho_J
    -M_J^*\rho_J^{\rm s}
    \right),
    ~~ J=\mathrm{n,p}.
    \label{eq:kinetic_pressure}
\end{align}
The kinetic energy density and pressure satisfy the basic relation $
\varepsilon_J^{\mathrm{kin}}+P_J^{\mathrm{kin}}
=\varepsilon_{\rm F}^{\ast,J}\rho_J$\,\cite{Cai2012PLB}.
Moreover, the scalar density can be obtained by differentiating the kinetic energy density with respect to the Dirac effective mass,
\begin{equation}
\rho_J^{\rm s}
=\left.\frac{\partial\varepsilon_J^{\mathrm{kin}}}{\partial M_J^*}\right|_{k_{\rm F}^J},\end{equation} 
where the derivative is taken at fixed Fermi momentum, or equivalently, fixed nucleon density. This relation can be verified directly from the analytical expression for $\varepsilon_J^{\mathrm{kin}}$ by using
\begin{equation}\label{def:pp}
\left.\frac{\partial\rho_J^{\rm s}}{\partial M_J^*}\right|_{k_{\rm F}^J}
=3\left(\frac{\rho_J^{\rm s}}{M_J^*}
-\frac{\rho_J}{\varepsilon_{\rm F}^{*,J}}\right).
\end{equation}
This identity also provides a useful basis for evaluating the scalar density response to variations in the effective mass.

Analytical expressions for $E_0(\rho)$ and its density derivatives, including $K_0$ and $J_0$, have been derived for the corresponding nonlinear RMF models in Refs.\,\cite{Cai2012,Cai2017,Li2022,Shi2026SSS}. Extending the analysis to the fourth-order symmetry energy requires expanding the nucleon contributions and self-consistent meson fields to higher orders in isospin asymmetry, incorporating both the direct isovector-scalar response and induced isoscalar field responses. The resulting expression for $E_{\mathrm{sym},4}(\rho)$ in the current model is given by:
\begin{align}
    E_{\mathrm{sym},4}(\rho)
    ={}&
    \frac{
    k_{\rm F}^2
    (
    4M_0^{*,4}+11M_0^{*,2}k_{\rm F}^2+10k_{\rm F}^4
    )}
    {648\varepsilon_{\rm F}^{*,5}}
    \notag\\
    &+
    \frac{
    M_0^*k_{\rm F}^2
    (M_0^{*,2}+4k_{\rm F}^2)}
    {54\varepsilon_{\rm F}^{*,5}}
    S_\delta^{[1]}
    +
    \frac{
    k_{\rm F}^2(2M_0^{*,2}-k_{\rm F}^2)}
    {12\varepsilon_{\rm F}^{*,5}}
    S_\delta^{[1], 2}
    \notag\\
    &
    -
    \frac{M_0^*k_{\rm F}^2}
    {2\varepsilon_{\rm F}^{*,5}}
    S_\delta^{[1], 3}
   +
    \left.
    \frac{1}{24\rho}
    \frac{\partial^2 A}{\partial M_0^{\ast,2}}
    \right|_{k_{\rm F}}
    \cdot S_\delta^{[1], 4}\notag\\
    &
    +\frac{1}{2\rho}\left(
\frac{B_\omega^2}{H_\omega}
-\frac{B_\sigma^2}{H_\sigma}
\right).
    \label{eq:text_Esym4}
\end{align}
The first term represents the kinetic contribution, while the remaining terms arise from the response of the isovector scalar field, characterized by
\begin{equation}
S_\delta^{[1]}=
    -{\rho M_0^*}/
    {\varepsilon_{\rm F}^*H_\delta},
\end{equation}
where the $\delta$-meson response factor is 
\begin{equation}
    H_\delta
=a_\delta
-C_{\delta\sigma}S_\sigma^{[0],2}+A,
\label{eq:text_Hdelta}
\end{equation}
with $a_\phi=m_\phi^2/g_\phi^2$ for $\phi=\sigma,\omega,\delta,\rho$.
The last two contributions of Eq.\,(\ref{eq:text_Esym4}) originate from the induced $\omega$- and $\sigma$-meson field responses, respectively, through which nonlinear interactions and cross couplings can influence the result.
Here, the effective nucleon mass $M_0^*$ and effective Fermi energy $\varepsilon_\rm F^*$ in SNM are given by
\begin{equation}
    M_0^*=M_{\rm N}-S_\sigma^{[0]},
~~
    \varepsilon_{\rm F}^*=\sqrt{k_{\rm F}^2+M_0^{*,2}}.
\end{equation}
The quantities $S_\sigma^{[0]}=g_\sigma\overline{\sigma}|_{\delta=0}$
and $V_\omega^{[0]}=g_\omega\overline{\omega}_0|_{\delta=0}$
denote the scaled $\sigma$- and $\omega$-meson mean fields in SNM at the same baryon density $\rho$, respectively. These background fields determine the effective nucleon properties and serve as the reference state for the expansion in isospin asymmetry. The coefficient $S_\delta^{[1]}$ represents the first-order term in the expansion of the scaled $\delta$ field with respect to the isospin asymmetry $\delta$. The scalar density response function $A$ is defined as\,\cite{Liu2002}
\begin{align}\label{def_A}
    A\equiv&
    \left.
    \frac{\partial\rho_{\rm s}^{[0]}}{\partial M_0^*}
    \right|_{k_{\rm F}}
    =
    3\left(
    \frac{\rho_{\rm s}^{[0]}}{M_0^*}
    -\frac{\rho}{\varepsilon_{\rm F}^*}
    \right)\notag\\
    =&\frac{3}{\pi^2}
    \left[
    k_{\rm F}\varepsilon_{\rm F}^*
    -M_0^{*,2}
    \ln\left(
    \frac{k_{\rm F}+\varepsilon_{\rm F}^*}{M_0^*}
    \right)
    \right]-\frac{3\rho}{\varepsilon_{\rm{F}}^\ast}
\end{align}
where $\rho_{\rm s}^{[0]}$ is the total scalar density in SNM. This function characterizes the response of the scalar density to variations in the effective nucleon mass at fixed Fermi momentum. Its higher-order derivatives also enter the fourth-order expansion. The first and second derivatives of $A$ with respect to $M_0^*$, evaluated at fixed $k_{\rm F}$, are
\begin{align}
    \left.\frac{\partial A}{\partial M_0^*}\right|_{k_\rm F}
    &=
    -\frac{6M_0^*}{\pi^2}
    \left[
    \ln\left(
    \frac{k_{\rm F}+\varepsilon_{\rm F}^*}{M_0^*}
    \right)
    -\frac{k_{\rm F}}{\varepsilon_{\rm F}^*}
    -\frac{k_{\rm F}^3}{3\varepsilon_{\rm F}^{*,3}}
    \right],
    \label{eq:app_AM}\\
    \left.\frac{\partial^2 A}{\partial M_0^{*,2}}\right|_{k_\rm F}
    &=
    \frac{6}{\pi^2}
    \left[
    -\ln\left(
    \frac{k_{\rm F}+\varepsilon_{\rm F}^*}{M_0^*}
    \right)
    +\frac{k_{\rm F}}{\varepsilon_{\rm F}^*}
    +\frac{k_{\rm F}^3}{3\varepsilon_{\rm F}^{*,3}}
    +\frac{k_{\rm F}^5}{\varepsilon_{\rm F}^{*,5}}
    \right].
    \label{eq:AMM}
\end{align}
The effective $\sigma$- and $\omega$-meson response factors are defined by
\begin{align}
    H_\sigma
    &=
    A+a_\sigma
    +2b_\sigma M_{\rm N}S_\sigma^{[0]}
    +3c_\sigma S_\sigma^{[0],2},
    \label{eq:text_Hsigma}\\
    H_\omega
    &=
    a_\omega+3c_\omega V_\omega^{[0],2},
    \label{eq:text_Homega}
\end{align}
The corresponding response functions $B_\sigma$ and $B_\omega$ collect the sources driving the induced isoscalar field responses:
\begin{align}
    B_\sigma
    &=
    B_{\mathrm{kin}}
    -C_{\delta\sigma}S_\sigma^{[0]}
    S_\delta^{[1],2}
    -\Lambda_{\rm S}S_\sigma^{[0]}
   V_\rho^{[1],2},
    \label{eq:text_Bsigma}\\
    B_\omega
    &=
    -\Lambda_{\rm V}V_\omega^{[0]}
    V_\rho^{[1],2},
    \label{eq:text_Bomega}
\end{align}
where $B_\mathrm{kin}$ collects the contributions associated with the nucleon kinetic and scalar density responses,
\begin{equation}
    B_\mathrm{kin}
    =
    \frac{\rho M_0^*k_{\rm F}^2}
    {6\varepsilon_{\rm F}^{*,3}}
    -
    \frac{\rho k_{\rm F}^2}
    {\varepsilon_{\rm F}^{*,3}}S_\delta^{[1]}
    -
   \left. \frac{1}{2}\frac{\partial A}{\partial M_0^*}\right|_{k_\rm F}\cdot S_\delta^{[1],2}.
    \label{eq:app_Bkin}
\end{equation}
The quantity 
\begin{equation}
V_\rho^{[1]}=-\rho/H_\rho,
\end{equation}
denotes the first-order coefficient in the expansion of the $\rho$-meson mean field with respect to the isospin asymmetry, $V_\rho\approx V_\rho^{[1]}\delta+O(\delta^3)$, 
where the effective $\rho$-meson response factor is defined as
\begin{equation}
    H_\rho=a_\rho+\Lambda_{\rm{S}}S_\sigma^{[0],2}+\Lambda_{\rm V}V_\omega^{[0],2}.\label{eq:text_Hrho}
\end{equation}
Together, these response quantities describe the interplay between the isoscalar and isovector meson fields.

Substituting the meson field responses into Eq.\,(\ref{eq:text_Esym4}), we obtain the explicit analytical expression for the fourth-order symmetry energy:
\begin{widetext}
\begin{align}
E_{\mathrm{sym},4}(\rho)
={}&
\frac{k_{\rm F}^2
(4M_0^{*,4}+11M_0^{*,2}k_{\rm F}^2+10k_{\rm F}^4)}
{648\varepsilon_{\rm F}^{*,5}}
-
\frac{\rho M_0^{*,2}k_{\rm F}^2
(M_0^{*,2}+4k_{\rm F}^2)}
{54\varepsilon_{\rm F}^{*,6}}\frac{1}{a_\delta-C_{\delta\sigma}S_\sigma^{[0],2}+A}
\notag\\
&+
\frac{\rho^2M_0^{*,2}k_{\rm F}^2}
{12\varepsilon_{\rm F}^{*,7}}\frac{1}{
(a_\delta-C_{\delta\sigma}S_\sigma^{[0],2}+A)^2}
\left(
2M_0^{*,2}-k_{\rm F}^2
+\frac{6\rho M_0^{*,2}}
{\varepsilon_{\rm F}^*}\frac{1}{
a_\delta-C_{\delta\sigma}S_\sigma^{[0],2}+A}
\right)
\notag\\
&+
\left.
\frac{\rho^3M_0^{*,4}}
{24\varepsilon_{\rm F}^{*,4}}\frac{1}{
(a_\delta-C_{\delta\sigma}S_\sigma^{[0],2}+A)^4}
\frac{\partial^2 A}{\partial M_0^{\ast,2}}
\right|_{k_{\rm F}}
+
\frac{\Lambda_{\rm V}^2V_\omega^{[0],2}\rho^3}
{2(a_\omega+3c_\omega V_\omega^{[0],2})
(a_\rho+\Lambda_{\rm S}S_\sigma^{[0],2}
+\Lambda_{\rm V}V_\omega^{[0],2})^4}
\notag\\
&-
\frac{\rho}
{2}\frac{1}{A+a_\sigma+2b_\sigma M_{\rm N}S_\sigma^{[0]}
+3c_\sigma S_\sigma^{[0],2}}
\left[
\frac{M_0^*k_{\rm F}^2}{6\varepsilon_{\rm F}^{*,3}}
+\frac{\rho M_0^*k_{\rm F}^2}
{\varepsilon_{\rm F}^{*,4}}\frac{1}{
a_\delta-C_{\delta\sigma}S_\sigma^{[0],2}+A}
\right.
\notag\\
&\hspace{1.5cm}
\left.
-\frac{\rho M_0^{*,2}}
{2\varepsilon_{\rm F}^{*,2}}\frac{1}{
(a_\delta-C_{\delta\sigma}S_\sigma^{[0],2}+A)^2}
\left[
\left.\frac{\partial A}{\partial M_0^*}\right|_{k_{\rm F}}
+2C_{\delta\sigma}S_\sigma^{[0]}
\right]
-\frac{\Lambda_{\rm S}S_\sigma^{[0]}\rho}
{(a_\rho+\Lambda_{\rm S}S_\sigma^{[0],2}
+\Lambda_{\rm V}V_\omega^{[0],2})^2}
\right]^2.
\label{eq:Esym4-full}
\end{align}
\end{widetext}

The six contributions to $E_{\rm{sym,4}}(\rho)$ in Eq.\,(\ref{eq:Esym4-full}) can be identified in their order of appearance. The first represents the kinetic contribution, followed by three terms associated with the $\delta$-meson response, involving different inverse powers of the effective response factor $H_\delta=a_\delta-C_{\delta\sigma}S_\sigma^{[0],2}+A$, see Eq.\,(\ref{eq:text_Hdelta}). Specifically, these terms involve $H_\delta^{-1}$ (immediately following the kinetic term), both $H_\delta^{-2}$ and $H_\delta^{-3}$ (second line), and $H_\delta^{-4}$ (first term of the third line), respectively. The last two contributions arise from the induced $\omega$- and $\sigma$-meson field responses, characterized by prefactors proportional to $\Lambda_{\rm{V}}^2V_\omega^{[0],2}\rho^3$ (second term of the third line) and $-\rho/2H_\sigma$, respectively. Notably, the induced $\omega$-meson contribution is nonnegative for a positive response factor $H_\omega$, whereas the induced $\sigma$-meson contribution is nonpositive for $H_\sigma>0$. The competing contributions of these meson fields, together with the higher-order dependence on $H_\delta$, can generate substantial variations in both the magnitude and density dependence of $E_{\rm{sym,4}}(\rho)$. This decomposition reveals how the interplay among different meson interactions can produce a nontrivial fourth-order symmetry energy, even when the lower-order saturation properties of the EOS are constrained.

The detailed derivation of $E_{\mathrm{sym},4}(\rho)$ is presented in Appendix \ref{app:E4}. Several analytical features of $E_{\rm{sym},4}(\rho)$, including its behavior in different density regimes and the roles of the meson interactions, will be discussed in the next section.

For later reference, the quadratic symmetry energy in the same RMF model takes the compact form\,\cite{Liu2002}
\begin{equation}
E_{\rm{sym}}(\rho)
=\frac{k_{\rm{F}}^2}{6\varepsilon_{\rm{F}}^*}
+\frac{\rho}{2H_\rho}
-\frac{\rho M_0^{*,2}}{2\varepsilon_{\rm{F}}^{*,2}H_\delta},
\label{eq:Esym-compact}
\end{equation}
Equivalently, it can be written explicitly as using the definitions for $H_\rho$ and $H_\delta$:
\begin{align}
E_{\rm{sym}}(\rho)
=&\frac{k_{\rm{F}}^2}{6\varepsilon_{\rm{F}}^*}
+\frac{\rho}
{2}\frac{1}{a_\rho+\Lambda_{\rm{S}}S_\sigma^{[0],2}
+\Lambda_{\rm{V}}V_\omega^{[0],2}}\notag\\
&-\frac{\rho M_0^{*,2}}
{2\varepsilon_{\rm{F}}^{*,2}}\frac{1}{
a_\delta-C_{\delta\sigma}S_\sigma^{[0],2}+A}.
\end{align}

\section{Properties of Fourth-Order Symmetry Energy}\label{SEC3}

The analytical expression for $E_{\rm{sym,4}}(\rho)$ derived in the preceding section, Eq.\,(\ref{eq:Esym4-full}), reveals the distinct roles of the meson cross couplings. In particular, the two isoscalar-isovector couplings, $\Lambda_{\rm{S}}$ and $\Lambda_{\rm{V}}$, enter the effective $\rho$-meson response through the same combination, $a_\rho+\Lambda_{\rm{S}}S_\sigma^{[0],2}+\Lambda_{\rm{V}}V_\omega^{[0],2}$, but contribute differently to $E_{\rm{sym,4}}(\rho)$. The former modifies the $\sigma$-field response, while the latter generates an explicit contribution through the induced $\omega$-field response. Their relative importance depends on density and the interplay with other meson interactions.
In this section, we investigate the resulting properties of $E_{\rm{sym,4}}(\rho)$. Subsections \ref{subs_lowrho_E4} and \ref{subs_lowrho_ratio} examine its low-density behavior and its ratio to the quadratic symmetry energy, respectively. Subsection \ref{subs_special} considers special model limits and consistency with previous results, while Subsection \ref{subs_positive} discusses the possible negative fourth-order symmetry energy and its sensitivity to nonlinear meson couplings. Finally, Subsection \ref{subs_highrho} analyzes the high-density asymptotic behavior under different meson interactions.

\subsection{Low Density Behavior of $E_{\textmd{sym,4}}(\rho)$}\label{subs_lowrho_E4}

The low-density expansion helps explain why the fourth order symmetry energy is generally small. In the limit $\rho\to0$, the Fermi momentum $k_{\rm{F}}=(3\pi^2\rho/2)^{1/3}\sim\rho^{1/3}$ approaches zero, and the isoscalar meson fields satisfy
\begin{align}
S_\sigma^{[0]}
\approx&\frac{\rho}{a_\sigma}
+\mathcal{O}(\rho^{5/3}),~~
V_\omega^{[0]}
\approx \frac{\rho}{a_\omega}
-\frac{c_\omega}{a_\omega^4}\rho^3
+\mathcal{O}(\rho^5).
\end{align}
Consequently, the effective nucleon mass and the scalar density response behave as
\begin{align}
M_0^*
\approx&M_{\rm{N}}-\frac{\rho}{a_\sigma}
+\mathcal{O}(\rho^{5/3}),~~
A
\approx\frac{3\rho k_{\rm{F}}^2}{5M_{\rm{N}}^3}
+\mathcal{O}(\rho^{7/3}),
\end{align}
see Eq.\,(\ref{apxA}) for the approximation for $A$.
Expanding the general RMF expression for the fourth order symmetry energy in powers of density, while retaining all meson couplings, gives
\begin{align}
E_{\rm{sym,4}}(\rho)
\approx&\frac{k_{\rm{F}}^2}{162M_{\rm{N}}}
+\frac{k_{\rm{F}}^4}{648M_{\rm{N}}^3}
\notag\\
&+\frac{\rho k_{\rm{F}}^2}{M_{\rm{N}}^2}
\left(
\frac{1}{162a_\sigma}
-\frac{1}{54a_\delta}
\right)
+\mathcal{O}(\rho^2).
\end{align}
The first two terms arise from the kinetic contribution, while the leading interaction corrections appear at order $\rho^{5/3}$ and involve the $\sigma$ and $\delta$ mesons. Nonlinear meson interactions contribute at higher orders in density.

To examine the effects of the meson cross couplings, we first consider $C_{\delta\sigma}$, which enters $E_{\rm{sym,4}}(\rho)$ through both the $\delta$- and $\sigma$-meson responses. In the explicit $\delta$-meson terms of Eq.\,(\ref{eq:Esym4-full}), it modifies the response denominator $H_\delta$ as:
\begin{equation}
a_\delta-C_{\delta\sigma}S_\sigma^{[0],2}+A
\approx a_\delta+A
-{C_{\delta\sigma}\rho^2}/{a_\sigma^2}
+\cdots.
\end{equation}
Meanwhile, $C_{\delta\sigma}$ also enters the last term of Eq.\,(\ref{eq:Esym4-full}) through the induced $\sigma$-field response. The corresponding leading contributions, arising from the explicit $\delta$-meson term and the $\sigma$-field response, respectively, are
\begin{align}
E_{\rm{sym,4}}^{(C_{\delta\sigma},\delta)}(\rho)
\approx&-\frac{C_{\delta\sigma}\rho^3k_{\rm{F}}^2}
{54a_\sigma^2a_\delta^2M_{\rm{N}}^2}
+\cdots,\\
E_{\rm{sym,4}}^{(C_{\delta\sigma},\sigma)}(\rho)
\approx&\frac{C_{\delta\sigma}\rho^3k_{\rm{F}}^2}
{6a_\sigma^2a_\delta^2M_{\rm{N}}^2}
+\cdots.
\end{align}
These two contributions have opposite signs, with the $\sigma$-field contribution dominating at leading order. Their sum gives
\begin{equation}
E_{\rm{sym,4}}^{(C_{\delta\sigma})}(\rho)
\approx\frac{4}{27}\frac{C_{\delta\sigma}\rho^3k_{\rm{F}}^2}
{a_\sigma^2a_\delta^2M_{\rm{N}}^2}
+\cdots.
\end{equation}
Thus, the leading contribution proportional to $C_{\delta\sigma}$ appears at order $\rho^{11/3}$, reflecting the combined effects of the $\delta$- and $\sigma$-meson responses.

Similarly, the coupling $\Lambda_{\rm{S}}$ affects the second order response of the $\sigma$ field through
$
-\Lambda_{\rm{S}}S_\sigma^{[0]}
V_\rho^{[1],2}
\approx
-{\Lambda_{\rm{S}}\rho^3}/
{a_\sigma a_\rho^2}$.
The corresponding leading contribution to the fourth order symmetry energy is
\begin{equation}
E_{\rm{sym,4}}^{(\Lambda_{\rm{S}})}(\rho)
\approx\frac{\Lambda_{\rm{S}}\rho^3k_{\rm{F}}^2}
{6a_\sigma^2a_\rho^2M_{\rm{N}}^2}
+\cdots,
\end{equation}
which also begins at order $\rho^{11/3}$.

The coupling $\Lambda_{\rm{V}}$ contributes explicitly through the second order response of the $\omega$ field,
$
E_{\rm{sym,4}}^{(\omega)}(\rho)
=
{\Lambda_{\rm{V}}^2V_\omega^{[0],2}\rho^3}/
{2H_\omega H_\rho^4
}$.
Its leading density dependence is
\begin{equation}
E_{\rm{sym,4}}^{(\Lambda_{\rm{V}})}(\rho)
\approx\frac{\Lambda_{\rm{V}}^2\rho^5}
{2a_\omega^3a_\rho^4}
+\cdots.
\end{equation}
Therefore, the first contribution involving $\Lambda_{\rm{V}}$ alone appears at order $\rho^5$.

At order $\rho^5$, the squared $\sigma$-field response generates contributions quadratic in $C_{\delta\sigma}$ and $\Lambda_{\rm{S}}$, while the induced $\omega$-field response introduces a term proportional to $\Lambda_{\rm{V}}^2$. These contributions can be written as
\begin{align}
E_{\rm{sym,4}}^{(\rm{cross},2)}(\rho)\approx&-\frac{\rho^5}{2a_\sigma^3}
\left(
\frac{C_{\delta\sigma}}{a_\delta^2}
+\frac{\Lambda_{\rm{S}}}{a_\rho^2}
\right)^2
+\frac{\Lambda_{\rm{V}}^2\rho^5}
{2a_\omega^3a_\rho^4}
+\cdots.
\end{align}
The $\sigma$-field contribution is non-positive and contains terms proportional to $C_{\delta\sigma}^2$, $\Lambda_{\rm{S}}^2$, and $C_{\delta\sigma}\Lambda_{\rm{S}}$, whereas the $\omega$-field contribution is nonnegative. These terms represent the quadratic cross-coupling contributions from the corresponding field responses, rather than the complete expansion at order $\rho^5$.

In addition, $\Lambda_{\rm{V}}$ modifies the effective $\rho$-meson response and thereby generates mixed contributions involving $\Lambda_{\rm{S}}$. The leading such contribution is
\begin{equation}
E_{\rm{sym,4}}^{(\Lambda_{\rm{S}}\Lambda_{\rm{V}})}(\rho)
=-\frac{\Lambda_{\rm{S}}\Lambda_{\rm{V}}
\rho^5k_{\rm{F}}^2}
{3a_\sigma^2a_\omega^2a_\rho^3M_{\rm{N}}^2}
+\cdots,
\end{equation}
which first appears at order $\rho^{17/3}$.

\renewcommand*{\tablename}{\small TAB.}
\begin{table}[h!]
\centering
\renewcommand{\arraystretch}{1.1}
\setlength{\tabcolsep}{8pt}
\begin{tabular}{ccc}
\hline\hline
Contribution & Coupling & Scaling \\
\hline
leading kinetic term
& none & $\rho^{2/3}$ \\
relativistic kinetic correction
& none & $\rho^{4/3}$ \\
ordinary meson interactions
& $\sigma,\delta$ & $\rho^{5/3}$ \\
scalar cross coupling
& $C_{\delta\sigma}$ & $\rho^{11/3}$ \\
scalar-vector cross coupling
& $\Lambda_{\rm{S}}$ & $\rho^{11/3}$ \\
quadratic scalar response
& $C_{\delta\sigma}^2$ & $\rho^5$ \\
quadratic scalar response
& $\Lambda_{\rm{S}}^2$ & $\rho^5$ \\
mixed scalar response
& $C_{\delta\sigma}\Lambda_{\rm{S}}$ & $\rho^5$ \\
induced vector response
& $\Lambda_{\rm{V}}^2$ & $\rho^5$ \\
mixed cross coupling
& $\Lambda_{\rm{S}}\Lambda_{\rm{V}}$
& $\rho^{17/3}$ \\
\hline\hline
\end{tabular}
\caption{Leading low-density scaling of the kinetic
and meson interaction contributions to
$E_{\rm{sym,4}}(\rho)$. The listed orders refer
to the first contributions associated with the
indicated interactions or field responses,
rather than the complete expansion at each order.}
\label{tab:E4_lowdensity}
\end{table}

The leading low-density orders of the various kinetic and meson interaction contributions are summarized in TAB.\,\ref{tab:E4_lowdensity}. The leading kinetic term scales as $\rho^{2/3}$, followed by the relativistic correction at order $\rho^{4/3}$ and the ordinary $\sigma$- and $\delta$-meson contributions at order $\rho^{5/3}$. The contributions from $C_{\delta\sigma}$ and $\Lambda_{\rm{S}}$ first appear at order $\rho^{11/3}$, while the explicit contribution proportional to $\Lambda_{\rm{V}}^2$ from the induced $\omega$-field response begins at order $\rho^5$. This hierarchy illustrates the suppression of nonlinear meson interactions and cross couplings in the dilute limit.

The relatively small leading kinetic contribution and the higher-order suppression of interaction terms help explain why $E_{\rm{sym,4}}(\rho)$ is generally small at low densities. However, this asymptotic behavior does not impose an upper bound at finite densities, where nonlinear interactions and meson field responses may become increasingly important. In particular, strong isovector scalar interactions or enhanced scalar field responses can substantially increase the magnitude of $E_{\rm{sym,4}}(\rho)$. Thus, values of $E_{\rm{sym,4}}(\rho_0)$ reaching several MeV are compatible with the universal low-density behavior, even though they exceed the conventional expectation of $\lesssim2\,\rm{MeV}$.

\subsection{Low Density Behavior of the Ratio $E_{\textmd{sym,4}}(\rho)/E_{\textmd{sym}}(\rho)$}\label{subs_lowrho_ratio}

We next examine the ratio of the fourth order and quadratic symmetry energies at low densities, approximately $\rho\lesssim0.1\,\mathrm{fm}^{-3}$. In the dilute limit, both quantities are dominated by their kinetic contributions, which scale as $\rho^{2/3}$, yielding the universal relation\,\cite{Cai2012}
\begin{equation}
\lim_{\rho\to0}\frac{E_{\rm{sym,4}}(\rho)}{E_{\rm{sym}}(\rho)}=\frac{1}{27}.
\end{equation}
As shown in FIG.\,\ref{fig:E4_E2}, the ratio can fall below $1/27$ at low but finite densities. Thus, this limiting value is not necessarily a lower bound.
To understand this behavior analytically, we consider the leading low-density expansions,
\begin{equation}
E_{\rm{sym}}(\rho)\approx\frac{k_{\rm{F}}^2}{6M_{\rm{N}}}+\frac{\rho}{2}\left(\frac{1}{a_\rho}-\frac{1}{a_\delta}\right)+\mathcal{O}(\rho^{4/3}),
\end{equation}
and
\begin{equation}
E_{\rm{sym,4}}(\rho)\approx\frac{k_{\rm{F}}^2}{162M_{\rm{N}}}+\mathcal{O}(\rho^{4/3}).
\end{equation}
Their ratio can therefore be expressed as
\begin{equation}
\frac{E_{\rm{sym,4}}(\rho)}{E_{\rm{sym}}(\rho)}
=\frac{1}{27}\left(
1-\frac{3M_{\rm{N}}\rho}{k_{\rm{F}}^2}
\frac{a_\delta-a_\rho}{a_\rho a_\delta}
+\mathcal{O}(\rho^{2/3})
\right).
\end{equation}
Since $\rho/k_{\rm{F}}^2\sim\rho^{1/3}$, the leading correction to the universal ratio is governed by the relative strengths of the $\rho$ and $\delta$ meson interactions. In particular, for $a_\delta>a_\rho>0$, the correction is negative, implying
$
{E_{\rm{sym,4}}(\rho)}/{E_{\rm{sym}}(\rho)}<1/27$,
at sufficiently low but finite densities.

This suppression reflects the dominance of the repulsive $\rho$ meson contribution over the attractive $\delta$ meson contribution to the quadratic symmetry energy. Their net contribution enters $E_{\rm{sym}}(\rho)$ already at order $\rho$, whereas the interaction corrections to $E_{\rm{sym,4}}(\rho)$ appear only at higher orders. Consequently, the quadratic symmetry energy receives a relatively stronger enhancement, reducing the ratio below its universal dilute limit.
Although the two symmetry energies share a common kinetic origin as $\rho\to0$, their ratio is sensitive to the competition between the isovector meson interactions at finite densities. For $a_\delta>a_\rho$, the universal value $1/27$ is approached from below, consistent with the low-density behavior observed in the numerical calculations.

\subsection{Special Limits of $E_{\textmd{sym,4}}(\rho)$}\label{subs_special}

Two special limits help identify the physical contributions to $E_{\rm{sym,4}}(\rho)$ and check the general derivation.

We first consider the simplest RMF model containing only the $\sigma$ and $\omega$ mesons, with all nonlinear self interactions and meson cross couplings switched off. In this case, the fourth order symmetry energy reduces to
\begin{align}
E_{\rm{sym,4}}^{(\sigma\omega)}(\rho)
=&\frac{k_{\rm{F}}^2
(4M_0^{*,4}+11M_0^{*,2}k_{\rm{F}}^2+10k_{\rm{F}}^4)}
{648\varepsilon_{\rm{F}}^{*,5}}
-\frac{\rho M_0^{*,2}k_{\rm{F}}^4}
{72\varepsilon_{\rm{F}}^{*,6}(A+a_\sigma)}.
\end{align}
The first term is the kinetic contribution evaluated with the nucleon effective mass $M_0^*$, while the second term arises from the response of the $\sigma$ field to isospin asymmetry. The latter is negative for a positive scalar response denominator $A+a_\sigma$. The $\omega$ field does not contribute explicitly because it couples only to the total baryon density, which remains fixed when isospin asymmetry is varied.
Adding a linear $\rho$ meson interaction does not change this result. Without meson cross couplings, the $\rho$ meson contributes only to the quadratic symmetry energy and does not generate a fourth order term. Thus, the fourth order symmetry energy depends not simply on the strength of the isovector interaction, but also on how the meson fields respond to isospin asymmetry.

A second important limit is obtained by removing the $\delta$ meson while retaining the nonlinear meson self interactions and the cross couplings $\Lambda_{\rm{S}}$ and $\Lambda_{\rm{V}}$. The general expression then becomes\,\cite{Cai2012}
\begin{align}
E_{\rm{sym,4}}^{(\sigma\omega\rho)}(\rho)
=&\frac{k_{\rm{F}}^2
(4M_0^{*,4}+11M_0^{*,2}k_{\rm{F}}^2+10k_{\rm{F}}^4)}
{648\varepsilon_{\rm{F}}^{*,5}}
+\frac{\Lambda_{\rm{V}}^2V_\omega^{[0],2}\rho^3}
{2H_\omega H_\rho^4}
\notag\\
&-\frac{\rho}
{2H_\sigma}
\left(
\frac{M_0^*k_{\rm{F}}^2}
{6\varepsilon_{\rm{F}}^{*,3}}
-\frac{\Lambda_{\rm{S}}S_\sigma^{[0]}\rho}
{H_\rho^2}
\right)^2.
\end{align}
This expression is equivalent to the analytical result obtained in Ref.\,\cite{Cai2012}. Expanding the last squared term reproduces the contribution quadratic in $\Lambda_{\rm{S}}$ and its interference with the kinetic response of the $\sigma$ field. The second term represents the induced $\omega$ field response associated with $\Lambda_{\rm{V}}$.

The two cross couplings therefore play different roles. The $\Lambda_{\rm{S}}$ coupling modifies the scalar response and can either enhance or reduce the fourth order symmetry energy through interference with the kinetic contribution. In contrast, the explicit $\omega$ field response associated with $\Lambda_{\rm{V}}$ is nonnegative for a positive $\omega$ field response denominator. Both effects arise because the isovector $\rho$ field induces changes in the isoscalar fields when isospin asymmetry is varied.

\subsection{Positiveness of the Quadratic Symmetry Energy and Possible Negative Fourth Order Symmetry Energy}\label{subs_positive}

Although the quadratic symmetry energy $E_{\rm{sym}}(\rho)$ is empirically positive, the fourth-order symmetry energy $E_{\rm{sym,4}}(\rho)$ can become negative, as demonstrated by our numerical results. This behavior arises from the competition between the positive kinetic contribution and the meson interaction contributions, particularly the negative $\sigma$-field response. Such negative contributions were already identified in RMF models without the $\delta$ meson,  although they were insufficient to overcome the kinetic contribution in the parameter sets considered. Including the $\delta$ meson introduces additional scalar contributions that can further alter this balance.

The $\sigma$-$\delta$ cross coupling $C_{\delta\sigma}$ is particularly important, as it modifies both the effective $\delta$-meson response denominator $H_\delta=a_\delta-C_{\delta\sigma}S_\sigma^{[0],2}+A$ and the $\sigma$-field response through $\partial A/\partial M_0^*+2C_{\delta\sigma}S_\sigma^{[0]}$. At fixed density, the SNM background fields are independent of $C_{\delta\sigma}$, giving
\begin{equation}
{\partial H_\delta^{-n}}/{\partial C_{\delta\sigma}}
=n{S_\sigma^{[0],2}}/{H_\delta^{n+1}}.
\end{equation}
Since $E_{\rm{sym,4}}(\rho)$ contains terms up to $H_\delta^{-4}$, its sensitivity to $C_{\delta\sigma}$ can be enhanced when $H_\delta$ is small. However, competing contributions may lead to a nonmonotonic dependence on this coupling, including a possible sign change of $E_{\rm{sym,4}}(\rho)$.

\subsection{High-density Asymptotic Behavior of $E_{\textmd{sym,4}}(\rho)$}\label{subs_highrho}

The high-density limit shows how nonlinear meson interactions can change the behavior of $E_{\rm{sym,4}}(\rho)$. Unlike the low-density limit, the leading high-density behavior depends on the meson cross couplings. In particular, $\Lambda_{\rm{S}}$ can produce a strongly negative fourth order symmetry energy even when the quadratic symmetry energy remains positive. As shown below, a nonzero $\Lambda_{\rm{V}}$ can suppress this negative contribution and lead to a different asymptotic behavior.

We begin with the scalar sector. In SNM, $\rho=2k_{\rm{F}}^3/3\pi^2$. For the conventional high-density branch of the RMF model, the nucleon effective mass approaches zero, while the equilibrium $\sigma$ field approaches a finite value. The scalar density then satisfies
$
\rho_{\rm{s}}^{[0]}
\approx{M_0^*k_{\rm{F}}^2}/{\pi^2}$.
Provided that the scalar field equation has a finite and nonzero limiting source, one obtains
\begin{equation}
M_0^*\sim k_{\rm{F}}^{-2},
~~
S_\sigma^{[0]}\to M_{\rm{N}},
~~
\varepsilon_{\rm{F}}^*\approx k_{\rm{F}}.
\end{equation}
Moreover, the scalar response function behaves as
\begin{equation}
A=3\left(
    {\rho_{\rm s}^{[0]}}/{M_0^*}
    -{\rho}/{\varepsilon_{\rm F}^*}\right)
\approx{k_{\rm{F}}^2}/{\pi^2},
\end{equation}
using Eq.\,(\ref{def_A}) and $\varepsilon_{\rm{F}}^\ast\approx k_{\rm{F}}$ at asymptoticly large $\rho$.
Consequently, the effective $\sigma$ and $\delta$ meson response denominators satisfy $
H_\sigma
\approx
H_\delta\approx{k_{\rm{F}}^2}/{\pi^2}$.
We use these limits to identify the leading contributions to the  $E_{\rm{sym,4}}(\rho)$.

The kinetic contribution has the simple asymptotic form
\begin{equation}
E_{\rm{sym,4}}^{\rm{kin}}(\rho)
\approx\frac{5k_{\rm{F}}}{324}
=\frac{5}{324}
\left(\frac{3\pi^2\rho}{2}\right)^{1/3}.
\end{equation}
It is positive and grows as $\rho^{1/3}$. By contrast, the explicit $\delta$-meson contributions are strongly suppressed at high densities, owing to the vanishing nucleon Dirac effective mass and the increasing scalar response denominator $H_\delta$. Using the asymptotic relations $M_0^*\sim k_{\rm F}^{-2}$, $\varepsilon_{\rm F}^*\sim k_{\rm F}$, and $H_\delta\approx k_{\rm F}^2/\pi^2$, the leading contributions involving $H_\delta^{-1}$ and $H_\delta^{-2}$ scale as $ M_0^{*,2}/k_{\rm F}\sim k_{\rm F}^{-5}$, while the sub-leading $H_\delta^{-3}$ term decreases as $k_{\rm F}^{-11}$. Moreover, since $\partial^2 A/\partial M_0^{*,2}\sim\ln(k_{\rm F}/M_0^*)$, the contribution involving $H_\delta^{-4}$ scales as $k_{\rm F}^{-11}\ln k_{\rm F}$ and is therefore further suppressed. Thus, although the $\delta$ meson and its coupling to the $\sigma$ meson can significantly influence the fourth-order symmetry energy at finite densities, their explicit contributions become asymptotically negligible and do not affect the leading high-density behavior along the scalar field branch considered here.

The remaining competition is mainly between the positive kinetic contribution and the $\sigma$ and $\omega$ field responses. In particular, the $\sigma$ field contributes a negative squared term,
\begin{align}
E_{\rm{sym,4}}^{(\sigma)}(\rho)
=&-\frac{\rho}
{2}\frac{1}{A+a_\sigma+2b_\sigma M_{\rm{N}}S_\sigma^{[0]}
+3c_\sigma S_\sigma^{[0],2}}\notag\\
&\times
\left(-
\frac{\Lambda_{\rm{S}}S_\sigma^{[0]}\rho}
{(a_\rho+\Lambda_{\rm{S}}S_\sigma^{[0],2}
+\Lambda_{\rm{V}}V_\omega^{[0],2})^2}+\cdots
\right)^2,
\end{align}
where the omitted terms involve the effective nucleon mass and decrease as $k_{\rm{F}}^{-3}$ or faster. Importantly, the $\Lambda_{\rm{S}}$ term need not vanish as the effective nucleon mass approaches zero. Its behavior depends on the effective $\rho$ meson response denominator, leading to the different asymptotic limits.

\renewcommand*{\tablename}{\small TAB.}
\begin{table}[h!]
\centering
\renewcommand{\arraystretch}{1.2}
\setlength{\tabcolsep}{4pt}
\begin{tabular}{ccc}
\hline\hline
Contribution & $k_{\rm{F}}$ scaling & $\rho$ scaling \\
\hline
kinetic term
& $5k_{\rm{F}}/324$
& $\sim\rho^{1/3}$ \\
term proportional to $H_\delta^{-1}$
& $\mathcal{O}(k_{\rm{F}}^{-5})$
& $\mathcal{O}(\rho^{-5/3})$ \\
term involving $H_\delta^{-2}$ and $H_\delta^{-3}$
& $\mathcal{O}(k_{\rm{F}}^{-5})$
& $\mathcal{O}(\rho^{-5/3})$ \\
term involving $H_\delta^{-4}$
& $\mathcal{O}(k_{\rm{F}}^{-11}\ln k_{\rm{F}})$
& $\mathcal{O}(\rho^{-11/3}\ln\rho)$ \\
\hline
$\omega$ response, $\Lambda_{\rm{V}}=0$
& $0$ & $0$ \\
$\omega$ response, $\Lambda_{\rm{V}}>0,\ c_\omega>0$
& $\sim k_{\rm{F}}$
& $\sim\rho^{1/3}$ \\
$\omega$ response, $\Lambda_{\rm{V}}>0,\ c_\omega=0$
& $\mathcal{O}(k_{\rm{F}}^{-9})$
& $\mathcal{O}(\rho^{-3})$ \\
\hline
$\sigma$ response, $\Lambda_{\rm{S}}=0$
& $\mathcal{O}(k_{\rm{F}}^{-5})$
& $\mathcal{O}(\rho^{-5/3})$ \\
$\sigma$ response, $\Lambda_{\rm{S}}\neq0,\ \Lambda_{\rm{V}}=0$
& $\sim-k_{\rm{F}}^{7}$
& $\sim-\rho^{7/3}$ \\
$\sigma$ response, $\Lambda_{\rm{S}}\neq0,\ \Lambda_{\rm{V}}>0,\ c_\omega>0$
& $\mathcal{O}(k_{\rm{F}}^{-1})$
& $\mathcal{O}(\rho^{-1/3})$ \\
$\sigma$ response, $\Lambda_{\rm{S}}\neq0,\ \Lambda_{\rm{V}}>0,\ c_\omega=0$
& $\mathcal{O}(k_{\rm{F}}^{-5})$
& $\mathcal{O}(\rho^{-5/3})$ \\
\hline\hline
\end{tabular}
\caption{Leading high-density asymptotic behavior of the six contributions to $E_{\rm{sym,4}}(\rho)$, listed in the same order as in its full expression. Here, $H_\delta=a_\delta-C_{\delta\sigma}S_\sigma^{[0],2}+A$. The results assume the conventional high-density scalar-field branch, $M_0^*\sim k_{\rm{F}}^{-2}$, with $k_{\rm{F}}\sim\rho^{1/3}$ and nonsingular meson response denominators.}
\label{tab:E4-asymptotic-terms}
\end{table}

The high-density asymptotic behaviors of the six contributions to $E_{\rm{sym,4}}(\rho)$ are summarized in TAB.\,\ref{tab:E4-asymptotic-terms}, following the order established in Eq.\,(\ref{eq:Esym4-full}). The leading behavior of the first four contributions is independent of the cross-coupling scenarios considered here, whereas the last two depend on $\Lambda_{\rm{S}}$, $\Lambda_{\rm{V}}$, and $c_\omega$. In particular, the $\sigma$-field contribution can become strongly negative when $\Lambda_{\rm{S}}\neq0$ and $\Lambda_{\rm{V}}=0$, while a nonzero $\Lambda_{\rm{V}}$ suppresses this behavior at high densities.

We first consider the case without either cross coupling, $\Lambda_{\rm{S}}=\Lambda_{\rm{V}}=0$. The potentially enhanced $\sigma$ field contribution is absent, and the explicit $\omega$ field contribution vanishes. The remaining meson contributions decrease at high densities, leaving the kinetic term dominant:
\begin{equation}
\text{Case 1: }
E_{\rm{sym,4}}(\rho)
\approx\frac{5k_{\rm{F}}}{324}
\sim\rho^{1/3},
~~
\Lambda_{\rm{S}}=\Lambda_{\rm{V}}=0.
\end{equation}
Thus, the fourth order symmetry energy is asymptotically positive. In comparison, the symmetry energy behaves as $E_{\rm{sym}}(\rho)\approx\rho/2a_\rho$, and the ratio $E_{\rm{sym,4}}(\rho)/E_{\rm{sym}}(\rho)$ approaches zero.

A qualitatively different behavior emerges for $\Lambda_{\rm{S}}\neq0$ and $\Lambda_{\rm{V}}=0$. In this case, the effective $\rho$-meson response denominator approaches a finite constant,
\begin{equation}
a_\rho+\Lambda_{\rm{S}}S_\sigma^{[0],2}
\to a_\rho+\Lambda_{\rm{S}}M_{\rm{N}}^2.
\end{equation}
Assuming this limit is positive, the $\Lambda_{\rm{S}}$ term in the squared $\sigma$-field response grows linearly with density,
\begin{equation}
-\frac{\Lambda_{\rm{S}}S_\sigma^{[0]}\rho}
{(a_\rho+\Lambda_{\rm{S}}S_\sigma^{[0],2})^2}
\approx
-\frac{\Lambda_{\rm{S}}M_{\rm{N}}\rho}
{(a_\rho+\Lambda_{\rm{S}}M_{\rm{N}}^2)^2}.
\end{equation}
Since the $\sigma$-meson response denominator grows as $k_{\rm{F}}^2$, the resulting negative contribution scales as $\rho^3/k_{\rm{F}}^2\sim\rho^{7/3}$, overwhelming the positive kinetic contribution at sufficiently high densities. Consequently, the leading asymptotic behavior is
\begin{equation}\label{E4:case2}
\text{Case 2: }
E_{\rm{sym,4}}(\rho)
\approx
-\frac{\pi^2\Lambda_{\rm{S}}^2M_{\rm{N}}^2}{2(a_\rho+\Lambda_{\rm{S}}M_{\rm{N}}^2)^4}\frac{\rho^3}{k_{\rm{F}}^2}\sim-\rho^{7/3},
~~
\Lambda_{\rm{S}}\neq0,~\Lambda_{\rm{V}}=0.
\end{equation}
Thus, under the stated assumptions, $E_{\rm{sym,4}}(\rho)$ eventually becomes negative and diverges to $-\infty$.

The quadratic symmetry energy behaves differently in the same limit:
\begin{equation}
E_{\rm{sym}}(\rho)
\approx
\frac{\rho}
{2(a_\rho+\Lambda_{\rm{S}}M_{\rm{N}}^2)}
\sim\rho.
\end{equation}
It remains positive and grows linearly with density, whereas the fourth order symmetry energy becomes increasingly negative. Their ratio satisfies
\begin{equation}
\frac{E_{\rm{sym,4}}(\rho)}{E_{\rm{sym}}(\rho)}
\approx
-\frac{\pi^2\Lambda_{\rm{S}}^2M_{\rm{N}}^2}
{(a_\rho+\Lambda_{\rm{S}}M_{\rm{N}}^2)^3}
\frac{\rho^2}{k_{\rm{F}}^2}
\sim-\rho^{4/3}.
\end{equation}
Thus, $E_{\rm{sym,4}}(\rho)/E_{\rm{sym}}(\rho)$ also approaches negative infinity.
This behavior originates from the coupling between the $\sigma$ and $\rho$ fields. Through $\Lambda_{\rm{S}}$, a change in isospin asymmetry modifies the $\rho$ field and induces a response of the $\sigma$ field. This response contributes negatively to $E_{\rm{sym,4}}(\rho)$ and can eventually dominate the positive kinetic contribution. The negative divergence therefore comes from the nonlinear meson interaction, rather than directly from the decreasing nucleon effective mass.

The situation changes when $\Lambda_{\rm{V}}>0$. The $\omega$ and $\rho$ meson cross coupling makes the effective $\rho$ meson response denominator increase with density, suppressing the negative contribution induced by $\Lambda_{\rm{S}}$. The resulting asymptotic behavior depends on the quartic $\omega$ meson self interaction.

For $c_\omega>0$, the cubic term dominates the $\omega$ field equation at sufficiently high densities, giving
$
V_\omega^{[0]}
\approx({\rho}/{c_\omega})^{1/3}$.
Consequently,
\begin{equation}
a_\rho+\Lambda_{\rm{S}}S_\sigma^{[0],2}
+\Lambda_{\rm{V}}V_\omega^{[0],2}
\approx
\Lambda_{\rm{V}}
({\rho}/{c_\omega})^{2/3}.
\end{equation}
The $\Lambda_{\rm{S}}$ term inside the $\sigma$ field response decreases as $\rho^{-1/3}$, and the complete negative $\sigma$ field contribution decreases as $\rho^{-1/3}$. It is therefore subleading compared with the kinetic term.

The $\omega$ field response, however, remains important. Its contribution approaches
\begin{equation}
E_{\rm{sym,4}}^{(\omega)}(\rho)
\approx
\frac{c_\omega^{5/3}}
{6\Lambda_{\rm{V}}^2}\rho^{1/3},
\end{equation}
which is positive and has the same density dependence as the kinetic contribution. The total fourth order symmetry energy becomes\,\cite{CaiLi2022PRC}
\begin{equation}
\text{Case 3: }
E_{\rm{sym,4}}(\rho)
\approx
\frac{5k_{\rm{F}}}{324}
+\frac{c_\omega^{5/3}}
{6\Lambda_{\rm{V}}^2}\rho^{1/3},
~~
\Lambda_{\rm{V}}>0,~ c_\omega>0.
\end{equation}
Thus, the fourth order symmetry energy is asymptotically positive, regardless of whether $\Lambda_{\rm{S}}$ vanishes, provided the assumed high-density branch remains valid.
The quadratic symmetry energy also grows as $\rho^{1/3}$\,\cite{CaiLi2022PRC},
\begin{equation}
E_{\rm{sym}}(\rho)
\approx
\frac{k_{\rm{F}}}{6}
+\frac{c_\omega^{2/3}}
{2\Lambda_{\rm{V}}}\rho^{1/3}.
\end{equation}
Consequently, the ratio $E_{\rm{sym,4}}(\rho)/E_{\rm{sym}}(\rho)$ approaches a finite positive constant determined by the kinetic and vector meson contributions\,\cite{CaiLi2022PRC}:
\begin{equation}
    \frac{E_{\rm{sym,4}}(\rho)}{E_{\rm{sym}}(\rho)}=\frac{5\Lambda_{\rm V}^2+54c_\omega^{5/3}(2/3\pi^2)^{1/3}}{54\Lambda_{\rm V}^2+162\Lambda_{\rm V}c_\omega^{2/3}(2/3\pi^2)^{1/3}}.
\end{equation}
It should be emphasized that this asymptotic expression is derived for nonzero $\Lambda_{\rm V}$ and therefore cannot be used to obtain the corresponding result for $\Lambda_{\rm V}=0$ by direct substitution. The latter case requires a separate asymptotic analysis, since the leading density dependence of the mean fields changes when the $\omega$-$\rho$ cross coupling is absent.

\renewcommand*{\tablename}{\small TAB.}
\begin{table}[h!]
\centering
\renewcommand{\arraystretch}{1.25}
\setlength{\tabcolsep}{6pt}
\begin{tabular}{cc}
\hline\hline
Quantity & High-density asymptotic form \\
\hline
\multicolumn{2}{c}{$\Lambda_{\rm{S}}=\Lambda_{\rm{V}}=0$} \\
$E_{\rm{sym}}(\rho)$ & $\rho/2a_\rho$ \\
$E_{\rm{sym,4}}(\rho)$ & $5k_{\rm{F}}/324$ \\
$E_{\rm{sym,4}}(\rho)/E_{\rm{sym}}(\rho)$ & $\to0$ \\
\hline
\multicolumn{2}{c}{$\Lambda_{\rm{S}}\neq0,\ \Lambda_{\rm{V}}=0$} \\
$E_{\rm{sym}}(\rho)$ & $\rho/2H_{\rho,\infty}$ \\[4pt]
$E_{\rm{sym,4}}(\rho)$ & $\displaystyle-\frac{\pi^2\Lambda_{\rm{S}}^2M_{\rm{N}}^2}{2H_{\rho,\infty}^4}\frac{\rho^3}{k_{\rm{F}}^2}$ \\[6pt]
$E_{\rm{sym,4}}(\rho)/E_{\rm{sym}}(\rho)$ & $\to-\infty$ \\
\hline
\multicolumn{2}{c}{$\Lambda_{\rm{V}}>0,\ c_\omega>0$} \\
$E_{\rm{sym}}(\rho)$ & $\displaystyle\frac{k_{\rm{F}}}{6}+\frac{c_\omega^{2/3}\rho^{1/3}}{2\Lambda_{\rm{V}}}$ \\[6pt]
$E_{\rm{sym,4}}(\rho)$ & $\displaystyle\frac{5k_{\rm{F}}}{324}+\frac{c_\omega^{5/3}\rho^{1/3}}{6\Lambda_{\rm{V}}^2}$ \\[6pt]
$E_{\rm{sym,4}}(\rho)/E_{\rm{sym}}(\rho)$ & $\to\displaystyle\frac{5\Lambda_{\rm V}^2+54c_\omega^{5/3}(2/3\pi^2)^{1/3}}{54\Lambda_{\rm V}^2+162\Lambda_{\rm V}c_\omega^{2/3}(2/3\pi^2)^{1/3}}$ \\
\hline
\multicolumn{2}{c}{$\Lambda_{\rm{V}}>0,\ c_\omega=0$} \\
$E_{\rm{sym}}(\rho)$ & $k_{\rm{F}}/6$ \\
$E_{\rm{sym,4}}(\rho)$ & $5k_{\rm{F}}/324$ \\
$E_{\rm{sym,4}}(\rho)/E_{\rm{sym}}(\rho)$ & $\to5/54$ \\
\hline\hline
\end{tabular}
\caption{Leading high-density asymptotic behavior of the quadratic and fourth order symmetry energies for different meson cross couplings. The results assume the conventional high-density scalar-field branch and positive, nonsingular response denominators. Here, $H_{\rho,\infty}=a_\rho+\Lambda_{\rm{S}}M_{\rm{N}}^2$ and $k_{\rm{F}}\sim\rho^{1/3}$.}
\label{tab:E4-asymptotic-cases}
\end{table}

If $c_\omega=0$ while $\Lambda_{\rm{V}}>0$, the $\omega$ field grows linearly with density,
\begin{equation}
V_\omega^{[0]}={\rho}/{a_\omega}.
\end{equation}
The effective $\rho$ meson response denominator then increases as $\rho^2$, producing stronger suppression of the $\Lambda_{\rm{S}}$ contribution. The explicit $\omega$ field contribution also decreases as $\rho^{-3}$. Both meson response contributions are subleading, and the kinetic term again dominates:
\begin{equation}
\text{Case 4: }
E_{\rm{sym,4}}(\rho)
\approx\frac{5k_{\rm{F}}}{324},
~~
\Lambda_{\rm{V}}>0,~ c_\omega=0.
\end{equation}
In the same limit, $E_{\rm{sym}}(\rho)\approx k_{\rm{F}}/6$, yielding\,\cite{CaiLi2022PRC}
\begin{equation}
\lim_{\rho\to\infty}
\frac{E_{\rm{sym,4}}(\rho)}{E_{\rm{sym}}(\rho)}
=\frac{5}{54}\approx9.3\%.
\end{equation}
This is the ratio obtained from the ultra-relativistic free Fermi gas contributions.

The leading asymptotic behaviors of the quadratic and fourth order symmetry energies in these four cases are summarized in TAB.\,\ref{tab:E4-asymptotic-cases}.
The results in TAB.\,\ref{tab:E4-asymptotic-cases} show that $\Lambda_{\rm{S}}$ and $\Lambda_{\rm{V}}$ can lead to very different high-density behavior. Without $\Lambda_{\rm{V}}$, a nonzero $\Lambda_{\rm{S}}$ can produce a negative scalar contribution that grows as $-\rho^{7/3}$. When $\Lambda_{\rm{V}}>0$, the increasing effective $\rho$ meson response denominator suppresses this contribution, and the leading fourth order symmetry energy is positive and proportional to $\rho^{1/3}$. Thus, depending on the nonlinear meson interactions, the ratio $E_{\rm{sym,4}}(\rho)/E_{\rm{sym}}(\rho)$ can approach zero, a finite positive constant, or negative infinity.

It is important to emphasize that the negative divergence of $E_{\rm{sym,4}}(\rho)/E_{\rm{sym}}(\rho)$ characterizes the high density behavior of the fourth-order expansion coefficient around SNM, rather than that of the full EOS at finite isospin asymmetry. In particular, the increasingly large magnitude of this ratio suggests that the expansion in $\delta$ may become less accurate at high densities, with higher-order terms becoming increasingly important. Therefore, the negative divergence does not necessarily imply a corresponding negative divergence of the full EOS.
Overall, the high-density analysis shows that the fourth order symmetry energy is sensitive to the nonlinear responses of the meson fields. Its asymptotic behavior can distinguish different scalar and vector cross couplings and provide information that is not contained in the quadratic symmetry energy alone.

An additional implication concerns the possible density at which the fourth order symmetry energy changes sign. In the dilute limit, $E_{\rm{sym,4}}(\rho)\approx k_{\rm{F}}^2/162M_{\rm{N}}>0$, whereas for $\Lambda_{\rm{S}}\neq0$ and $\Lambda_{\rm{V}}=0$, its high-density behavior is given by Eq.\,(\ref{E4:case2}).
If the corresponding mean field solution remains continuous over the relevant density range, there must be at least one density $\rho_{\rm{cross}}$ satisfying $
E_{\rm{sym,4}}(\rho_{\rm{cross}})=0$.
The location of this zero, and the possible occurrence of multiple sign changes, depend on the competition among the meson interactions. Their dependence on the cross couplings may therefore provide additional information beyond the value of $E_{\rm{sym,4}}(\rho)$ at saturation density.

Another useful aspect is the density dependence of the fourth order symmetry energy, characterized by its slope and curvature parameters at saturation density,
\begin{align}\label{def-Lsym4Ksym4}
L_{\rm{sym,4}}=&\left.3\rho_0\frac{\d E_{\rm{sym,4}}(\rho)}{\d\rho}\right|_{\rho_0},~~
K_{\rm{sym,4}}=\left.9\rho_0^2\frac{\d^2E_{\rm{sym,4}}(\rho)}{\d\rho^2}\right|_{\rho_0}.
\end{align}
Although $E_{\rm{sym,4}}(\rho)$ is generally smaller than the quadratic symmetry energy, its density derivatives need not be correspondingly small. In particular, the fourth order contribution to the pressure of ANM at fixed isospin asymmetry is
\begin{equation}\label{def:P4}
P_{\rm{sym,4}}(\rho,\delta)=\rho^2\delta^4\frac{\d E_{\rm{sym,4}}(\rho)}{\d\rho},
\end{equation}
which gives $P_{\rm{sym,4}}(\rho_0,\delta)=\rho_0L_{\rm{sym,4}}\delta^4/3$. Thus, even a relatively small fourth-order symmetry energy can produce a noticeable pressure correction if it varies rapidly with density. The curvature parameter $K_{\rm{sym,4}}$ further characterizes its density dependence near saturation. Both the magnitude and density derivatives of $E_{\rm{sym,4}}(\rho)$ are therefore important for understanding the EOS of highly ANM. We discuss these effects in more detail in Subsection \ref{subs_xp}.

\section{$E_{\textmd{sym,4}}(\rho)$ in the Nonlinear Walecka Model Constrained by Astrophysical and Terrestrial Observations}\label{SEC4}

In this section, we revisit the magnitude, density dependence, and astrophysical implications of the fourth-order nuclear symmetry energy $E_{\rm{sym,4}}(\rho)$ within the nonlinear Walecka model. In Subsection \ref{subs_model}, we construct an ensemble of RMF EOSs constrained by terrestrial nuclear experiments and astrophysical observations, allowing us to explore the range of $E_{\rm{sym,4}}(\rho)$ compatible with current constraints. In Subsection \ref{subs_RMFE4}, we investigate its density dependence, its relative magnitude compared with the conventional quadratic symmetry energy, and the competing contributions from different meson interactions. Particular attention is paid to the possibility of a substantially larger $E_{\rm{sym,4}}(\rho_0)$ than conventionally expected, while its overall magnitude remains relatively constrained at densities relevant to NSs. Finally, in Subsection \ref{subs_xp}, we examine the effects of the fourth-order symmetry energy on the proton fraction $x_{\rm p}$, core-crust transition boundary, NS mass-radius (M-R) relations, tidal deformabilities and speed of sound profiles of NSs, with an emphasis on the importance and limitations of the parabolic approximation in highly neutron-rich matter.

\subsection{Model Constraints and Ensemble Construction}\label{subs_model}

We construct an ensemble of nonlinear RMF models by uniformly  sampling nucleonic matter properties near saturation within ranges compatible with terrestrial constraints. The saturation density $\rho_0$ and binding energy $E_0(\rho_0)$ per nucleon are varied within
$0.14\,\mathrm{fm}^{-3}\lesssim \rho_0\lesssim0.16\,\mathrm{fm}^{-3}$
and $-17\,\mathrm{MeV}\lesssim E_0(\rho_0)\lesssim-15\,\mathrm{MeV}$\,\cite{Dutra2014RMF}, respectively, while the Dirac effective mass at saturation is restricted to
$0.5\lesssim M_0^*/M_{\rm N}
\lesssim0.7$\,\cite{Chen2007,Li2018EffectiveMass}.
For the incompressibility and skewness coefficients, we adopt
$220\,\mathrm{MeV}\lesssim K_0\lesssim260\,\mathrm{MeV}$\,\cite{Blaizot1980,Colo2004,
GargColo2018,Xu2021K0,XieLi2021,WangDeng2025,
Khan2012,KhanMargueron2013}
and $-600\,\mathrm{MeV}\lesssim J_0\lesssim0\,\mathrm{MeV}$\,\cite{Cai2017}, respectively.
In the isovector sector, the symmetry energy and its slope parameter are sampled within
$29\,\mathrm{MeV}\lesssim E_{\mathrm{sym}}(\rho_0)
\lesssim35\,\mathrm{MeV}$
and
$30\,\mathrm{MeV}\lesssim L\lesssim80\,\mathrm{MeV}$\,\cite{LiHan2013,LiCaiXieZhang2021}, respectively.

The average properties of finite nuclei provide relatively strong constraints on the symmetry energy and its density dependence at subsaturation densities, since the characteristic density probed by finite nuclei is generally lower than the saturation density $\rho_0$. We therefore impose additional constraints at the reference density $\rho_{\rm{r}}\approx0.11\,\mathrm{fm}^{-3}$, where the symmetry energy is sampled within
$26.15\,\mathrm{MeV}\lesssim E_{\mathrm{sym}}(\rho_{\rm{r}})
\lesssim27.15\,\mathrm{MeV}$
and its slope within
$35\,\mathrm{MeV}\lesssim L_{\rm{r}}\lesssim55\,\mathrm{MeV}$\,\cite{ZhangChen2013,ZhangChen2023},
where we define the slope as $L_{\rm{r}}\equiv
3\rho_{\rm{r}}
\mathrm{d}E_{\mathrm{sym}}(\rho)/\mathrm{d}\rho|
_{\rho_{\rm{r}}}$.
These intervals encompass the constraints inferred from isotope binding energy differences and neutron skin thicknesses.

For the numerical ensemble, we set $\Lambda_{\rm S}=0$. Although this coupling can appreciably affect the asymptotic high-density behavior of $E_{\rm{sym,4}}(\rho)$, its influence is expected to be relatively modest over the density range relevant to NS interiors. Consequently, this choice is not expected to qualitatively alter our main physical conclusions. Alternative coupling scenarios, such as allowing $\Lambda_{\rm S}$ to vary while fixing $\Lambda_{\rm V}$, may lead to physically comparable results within the observationally constrained density range. In the following analysis, we adopt this coupling scheme throughout.
We fix the nucleon mass at $M_{\rm N}\approx939\,\mathrm{MeV}$ and the meson masses at $m_\sigma\approx491.5\,\rm{MeV}$, $m_\omega\approx782.5\,\rm{MeV}$, $m_\rho\approx763.0\,\rm{MeV}$, and $m_\delta\approx980.0\,\rm{MeV}$. For each sampled set of nine nucleonic matter parameters, we solve the inverse problem to determine the nine RMF coupling parameters $g_\sigma$, $g_\omega$, $b_\sigma$, $c_\sigma$, $c_\omega$, $g_\rho$, $\Lambda_{\rm V}$, $g_\delta$, and $C_{\delta\sigma}$. Each successful solution defines a consistent RMF parameterization and its corresponding nucleonic matter EOS. Notably, all the resulting EOSs have positive values of both $\Lambda_{\rm V}$ and $c_\omega$, ensuring the presence of the $\omega$-$\rho$ cross coupling and the quartic $\omega$ self-interaction.

We then construct the EOS of cold NS matter. The core consists of neutrons, protons, electrons, and muons in $\beta$ equilibrium and under charge neutrality, $
    \mu_{\rm n}-\mu_{\rm p}=\mu_{\rm e}$ and $
    \rho_{\rm p}=\rho_{\rm e}+\rho_\mu$, 
with $\mu_\mu=\mu_{\rm e}$ when muons are present.
The leptons are treated as free relativistic Fermi gases.
For the outer crust, we adopt the Baym--Pethick--Sutherland and Feynman--Metropolis--Teller
EOSs\,\cite{Baym1971}.
The inner crust, between neutron drip and the core-crust transition, is described by a polytropic form $P=a+b\varepsilon^{4/3}$\,\cite{Iida1997,ZhangChen2016}, where the constants $a$ and $b$ are determined by matching continuously to the core and outer crust EOSs. The core-crust transition density is calculated self-consistently using the thermodynamic method\,\cite{Xu2009,Cai2012}. Consequently, $E_{\rm{sym,4}}(\rho)$ can affect the density range and thickness of the interpolated inner crust by modifying the location of the core-crust transition boundary.

\renewcommand*\figurename{\small FIG.}
\begin{figure}[h!]
    \centering
    \includegraphics[width=0.95\columnwidth]{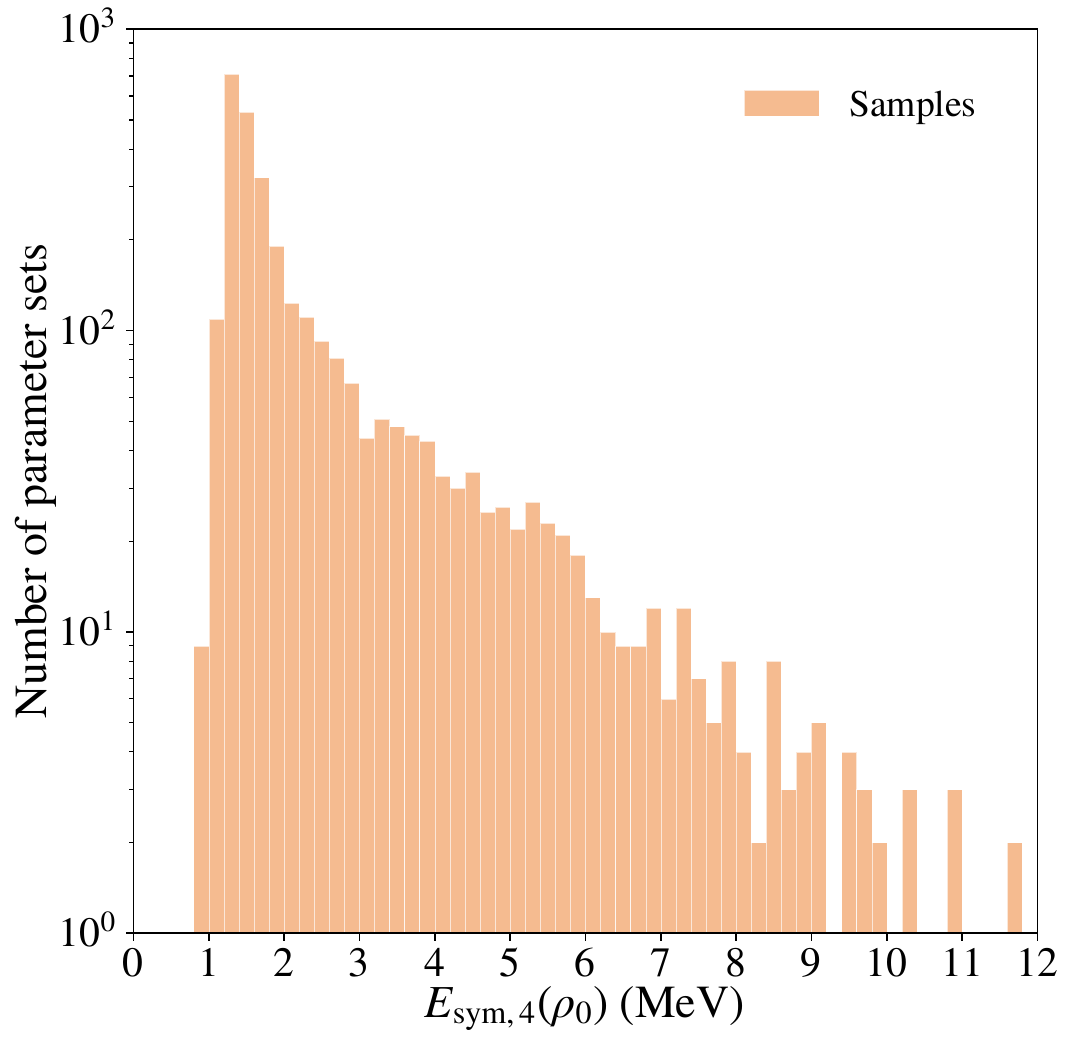}
    \caption{(Color online). Distribution of the fourth-order nuclear symmetry energy $E_{\rm{sym,4}}(\rho_0)$ at saturation density for the 3000 RMF parameter sets satisfying the adopted terrestrial and astrophysical constraints. The vertical axis shows the number of parameter sets on a logarithmic scale. Approximately $95\%$ of the sampled parameter sets have $E_{\rm{sym,4}}(\rho_0)\lesssim6\,\mathrm{MeV}$.}
    \label{fig:E4_rho0_distribution}
\end{figure}

For each complete EOS, we solve the TOV equations to obtain the NS M-R relation and the dimensionless tidal deformability $\Lambda$. We retain only models satisfying
$\Lambda_{1.4}\lesssim580$\,\cite{Abbott2017,Abbott2018} and
$M_{\rm{TOV}}\gtrsim2.01\,M_\odot$\,\cite{Fonseca2021},
where $\Lambda_{1.4}$ is the dimensionless tidal deformability of a canonical NS.
The resulting ensemble contains 3000 accepted RMF parameter sets and their corresponding EOSs.
Importantly, the accepted EOS ensemble satisfies the adopted terrestrial and astrophysical constraints, including those on NS masses and tidal deformabilities, which also restrict the allowed stellar radii. Such consistency is essential for obtaining physically meaningful constraints and predictions for $E_{\rm{sym,4}}(\rho)$. Our subsequent analysis therefore focuses on the $E_{\rm{sym,4}}(\rho)$ within this observationally constrained ensemble.

As shown in FIG.\,\ref{fig:E4_rho0_distribution}, the distribution of $E_{\rm{sym,4}}(\rho_0)$ is strongly right skewed, with most parameter sets concentrated between approximately $1$ and $4\,\mathrm{MeV}$ and a long tail extending to nearly about $12\,\mathrm{MeV}$. 
Parameter sets with $E_{\rm{sym,4}}(\rho_0)\lesssim1\,\mathrm{MeV}$ are particularly rare in the sampled ensemble.
Notably, $E_{\rm{sym,4}}(\rho_0)$ can reach about $6\,\mathrm{MeV}$ or even higher, substantially exceeding the empirical expectation of $E_{\rm{sym,4}}(\rho_0)\lesssim2\,\mathrm{MeV}$. Although larger values occur less frequently, their presence indicates that the adopted nucleonic matter and astrophysical constraints do not necessarily restrict the fourth-order symmetry energy to the small values commonly assumed at saturation. In particular, the pronounced asymmetry of the distribution suggests considerable model dependence in $E_{\rm{sym,4}}(\rho_0)$, even after imposing these constraints. Thus, a relatively large fourth-order symmetry energy at saturation remains compatible with current observations, motivating a closer examination of its density dependence and implications for neutron-rich matter.

\subsection{Density Dependence of $E_{\textmd{sym,4}}(\rho)$ and and Its Ratio to the Symmetry Energy}\label{subs_RMFE4}

\begin{figure}[h!]
    \centering
    \includegraphics[width=0.9\columnwidth]{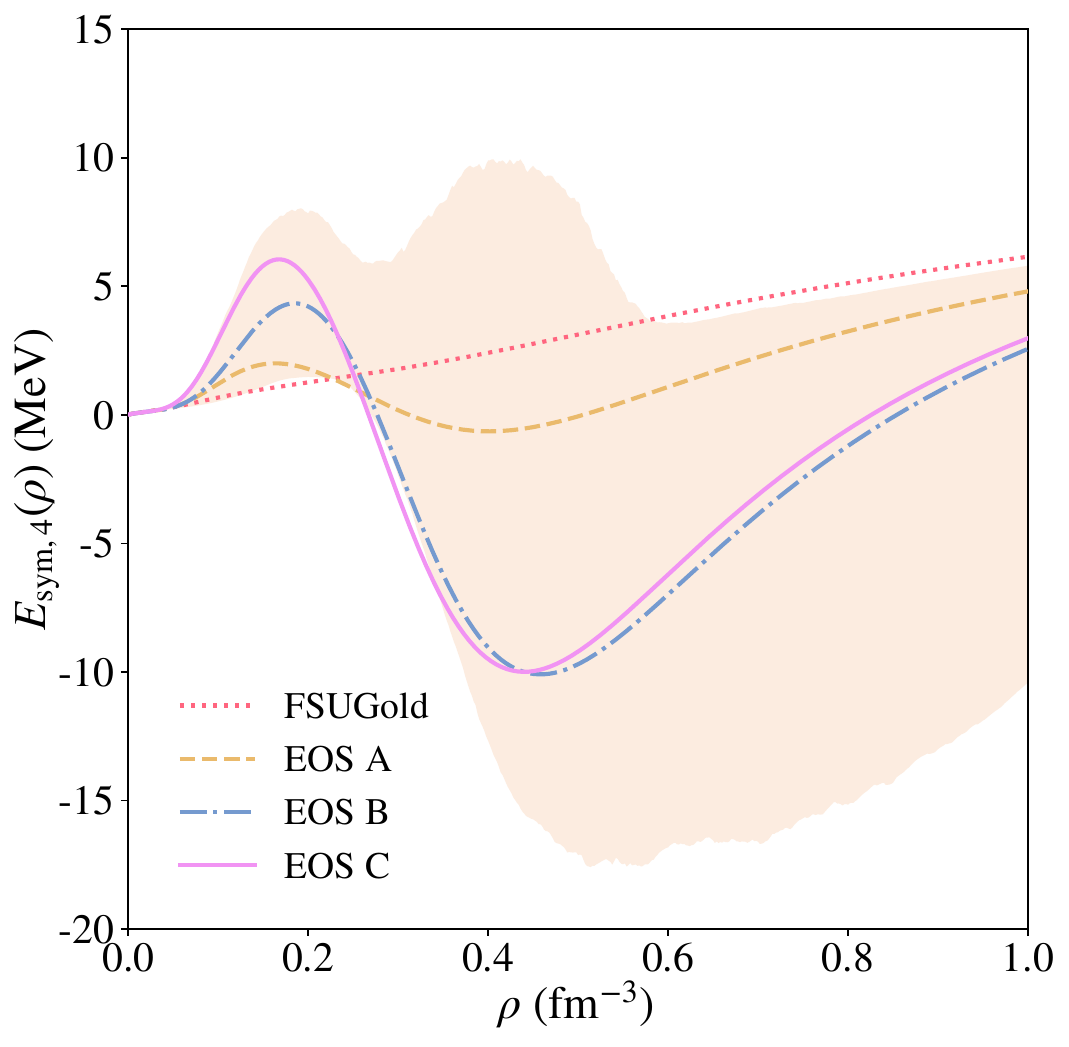}
    \caption{(Color online). Density dependence of the fourth-order nuclear symmetry energy $E_{\rm{sym,4}}(\rho)$. The shaded region represents the 95\% ensemble band of the accepted RMF models. Results for three representative parameterizations, EOS A, B, and C, are compared with those of FSUGold.}
    \label{fig:E4_band}
\end{figure}

In FIG.\,\ref{fig:E4_band}, we show the density dependence of the fourth-order symmetry energy $E_{\rm{sym,4}}(\rho)$ for the observationally constrained RMF ensemble, with the shaded region representing the 95\% ensemble band. For comparison, we also include the results from the well-known FSUGold parameterization\,\cite{ToddRutel2005} and three representative models, EOS A, B, and C. As summarized in TAB.\,\ref{tab:representative-eos}, these three EOSs are constructed to share identical conventional nucleonic matter properties at saturation and consistent symmetry energies at sub-saturation densities, while having distinct values of $E_{\rm{sym,4}}(\rho_0)=2$, $4$, and $6\,\mathrm{MeV}$, respectively. This deliberate construction allows us to explore the physical implications associated with different magnitudes of the fourth-order symmetry energy at saturation, particularly their manifestations in its density dependence and the resulting NS properties.

\renewcommand*\figurename{\small FIG.}

\begin{table}[h!]
    \centering
    \renewcommand{\arraystretch}{1.1}
    \setlength{\tabcolsep}{4pt}
    \begin{tabular}{cccc}
        \hline\hline
        Parameter & \multicolumn{3}{c}{Value} \\
        \hline
        $\rho_0$ ($\mathrm{fm}^{-3}$)
            & \multicolumn{3}{c}{0.16} \\
        $E_0(\rho_0)$ (MeV)
            & \multicolumn{3}{c}{$-16.5$} \\
        $K_0$ (MeV)
            & \multicolumn{3}{c}{240.0} \\
        $J_0$ (MeV)
            & \multicolumn{3}{c}{$-420.0$} \\
        $M_0^*(\rho_0)/M_\rm N$
            & \multicolumn{3}{c}{0.65} \\
        $E_{\mathrm{sym}}(\rho_0)$ (MeV)
            & \multicolumn{3}{c}{31.3} \\
        $L$ (MeV)
            & \multicolumn{3}{c}{46.0} \\
        \hline
        & EOS A & EOS B & EOS C \\
        $E_{\rm{sym,4}}(\rho_0)$ (MeV)
            & 2.0 & 4.0 & 6.0 \\
        \hline\hline
    \end{tabular}
\caption{
Saturation properties of the representative EOS A, B, and C.
The seven conventional saturation properties are identical,
while $E_{\rm{sym,4}}(\rho_0)$ differs among the three models.
All three EOSs also satisfy the adopted NS constraints,
$M_{\rm{TOV}}\gtrsim2.01M_\odot$ and $\Lambda_{1.4}\lesssim580$.
}
    \label{tab:representative-eos}
\end{table}

As shown in FIG.\,\ref{fig:E4_band}, the fourth-order symmetry energy exhibits considerable variation among the accepted models. For comparison, FSUGold is a conventional RMF parameterization incorporating nonlinear $\sigma$- and $\omega$-meson self-interactions and an $\omega$-$\rho$ cross coupling, but without an explicit $\delta$-meson field. Within this relatively simple Lagrangian framework, $E_{\rm{sym,4}}(\rho_0)$ is of order $1\,\mathrm{MeV}$ or smaller, and its density dependence is relatively weak and monotonically increasing. In contrast, the extended RMF model allows substantially larger values at saturation and strongly non-monotonic density dependences. In particular, EOS B and C exhibit pronounced peaks near saturation density, followed by a rapid decrease to negative values at higher densities and a subsequent recovery. These contrasting behaviors highlight the important role of additional meson interactions in shaping the density dependence of $E_{\rm{sym,4}}(\rho)$.

Despite these substantial variations, two important conclusions emerge from the observationally constrained EOS ensemble. First, the fourth-order symmetry energy at saturation can be considerably larger than the empirical expectation of about $2\,\mathrm{MeV}$, reaching
\begin{equation}
    E_{\rm{sym,4}}(\rho_0)\lesssim 6\,\mathrm{MeV},
    \text{ or even higher for selected EOSs}.
\end{equation}
Second, the overall magnitude of the fourth-order symmetry energy within the NS related densities remains relatively constrained. The 95\% ensemble band extends approximately from $-18$ to $10\,\mathrm{MeV}$ over the density range considered, suggesting
\begin{equation}
    |E_{\rm{sym,4}}(\rho)|\lesssim20\,\mathrm{MeV},
    \text{ within NS densities}.
\end{equation}
Thus, a relatively large fourth-order symmetry energy at saturation does not necessarily imply an excessively large contribution at higher densities. Potential inconsistencies associated with an extremely large quartic term are further examined in Appendix \ref{app:2}. The ensemble band also displays two broad maxima, reflecting the diverse non-monotonic behaviors allowed by the constrained RMF parameterizations.

\begin{figure}[h!]
    \centering
    \includegraphics[width=0.9\columnwidth]{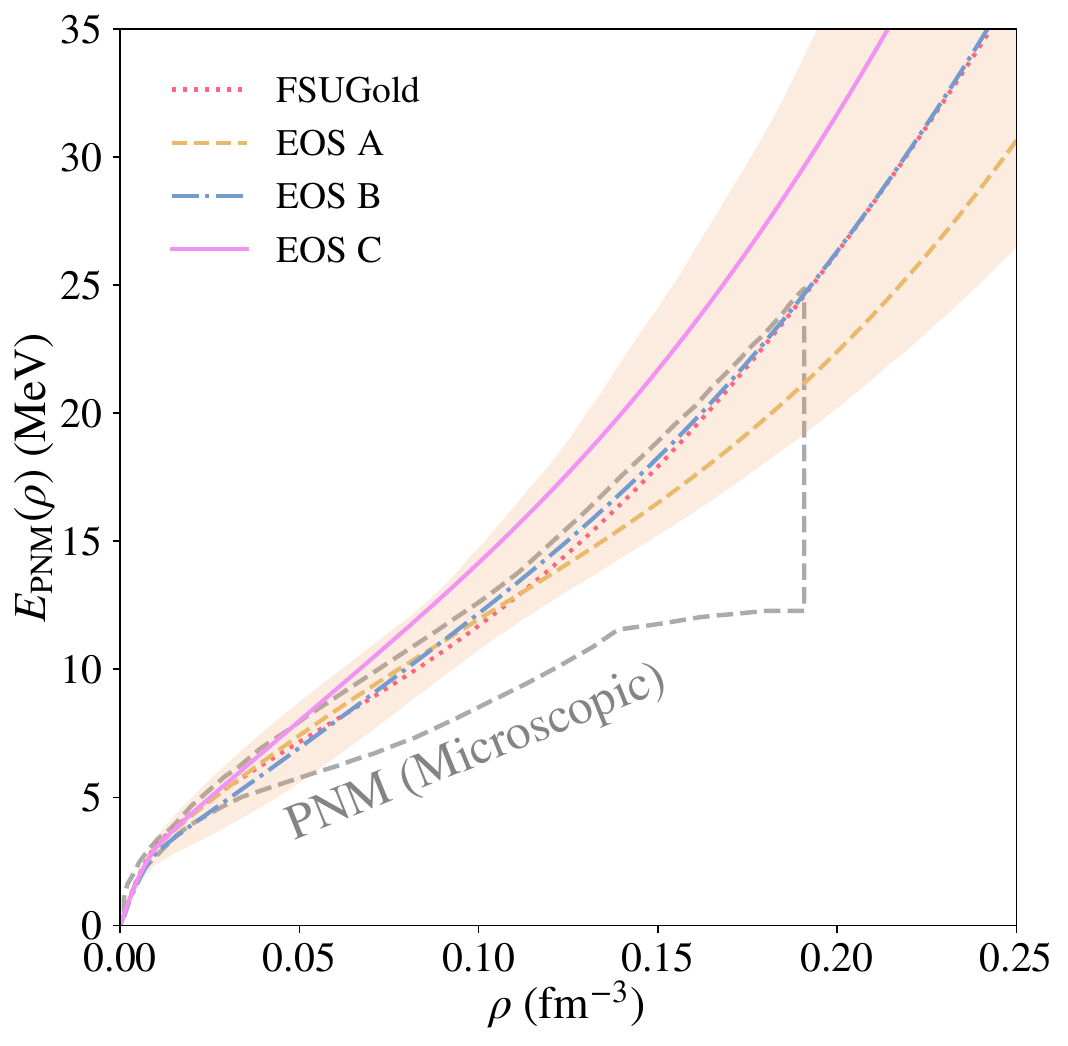}
    \caption{(Color online). Density dependence of the energy per nucleon of PNM, $E_{\rm{PNM}}(\rho)$, for the accepted RMF ensemble. The shaded region represents the ensemble band, while the curves show FSUGold and the three representative models, EOS A, B, and C. The gray dashed region indicates the microscopic PNM constraint\,\cite{WangChen2025,ZhangChen2023,Huth2021}.}
    \label{fig:E_PNM}
\end{figure}

The EOS of PNM provides an additional test of these predictions, since its energy per nucleon receives contributions from the quadratic, quartic, and higher-order symmetry energies. As shown in FIG.\,\ref{fig:E_PNM}, the predicted PNM EOSs are generally compatible with microscopic calculations at low densities, although some deviations occur, particularly for EOS C with $E_{\rm{sym,4}}(\rho_0)=6\,\mathrm{MeV}$. This suggests that microscopic PNM constraints\,\cite{WangChen2025,ZhangChen2023,Huth2021} may further restrict the allowed range of $E_{\rm{sym,4}}(\rho_0)$ beyond the terrestrial and astrophysical constraints adopted.

\renewcommand*\figurename{\small FIG.}

To illustrate this point, we approximate the PNM EOS by retaining only the first three terms in the isospin expansion, i.e., 
\begin{equation}
E_{\rm{PNM}}(\rho)\approx E_0(\rho)+E_{\rm{sym}}(\rho)+E_{\rm{sym,4}}(\rho).
\end{equation}
Combining the sampled ranges of $E_0(\rho_0)$ and $E_{\rm{sym}}(\rho_0)$ with the approximate microscopic constraint $10\,\mathrm{MeV}\lesssim E_{\rm{PNM}}(\rho_0)\lesssim20\,\mathrm{MeV}$ yields roughly
$
-10\,\mathrm{MeV}\lesssim E_{\rm{sym,4}}(\rho_0)\lesssim8\,\mathrm{MeV}$.
Importantly, this range represents only a rough estimate based on the truncated isospin expansion, rather than a rigorous constraint on the fourth-order symmetry energy. In particular, the rather broad range highlights the limited ability of the PNM EOS alone to isolate the quartic contribution. Contributions from sixth- and higher-order symmetry energies, which are neglected in this estimate, may further modify the inferred range. From this perspective, values of $E_{\rm{sym,4}}(\rho_0)$ around $6\,\mathrm{MeV}$ or even higher, as found in our constrained EOS ensemble, are not intrinsically excluded by microscopic PNM constraints.

\renewcommand*\figurename{\small FIG.}
\begin{figure}[h!]
    \centering
    \includegraphics[width=0.9\columnwidth]{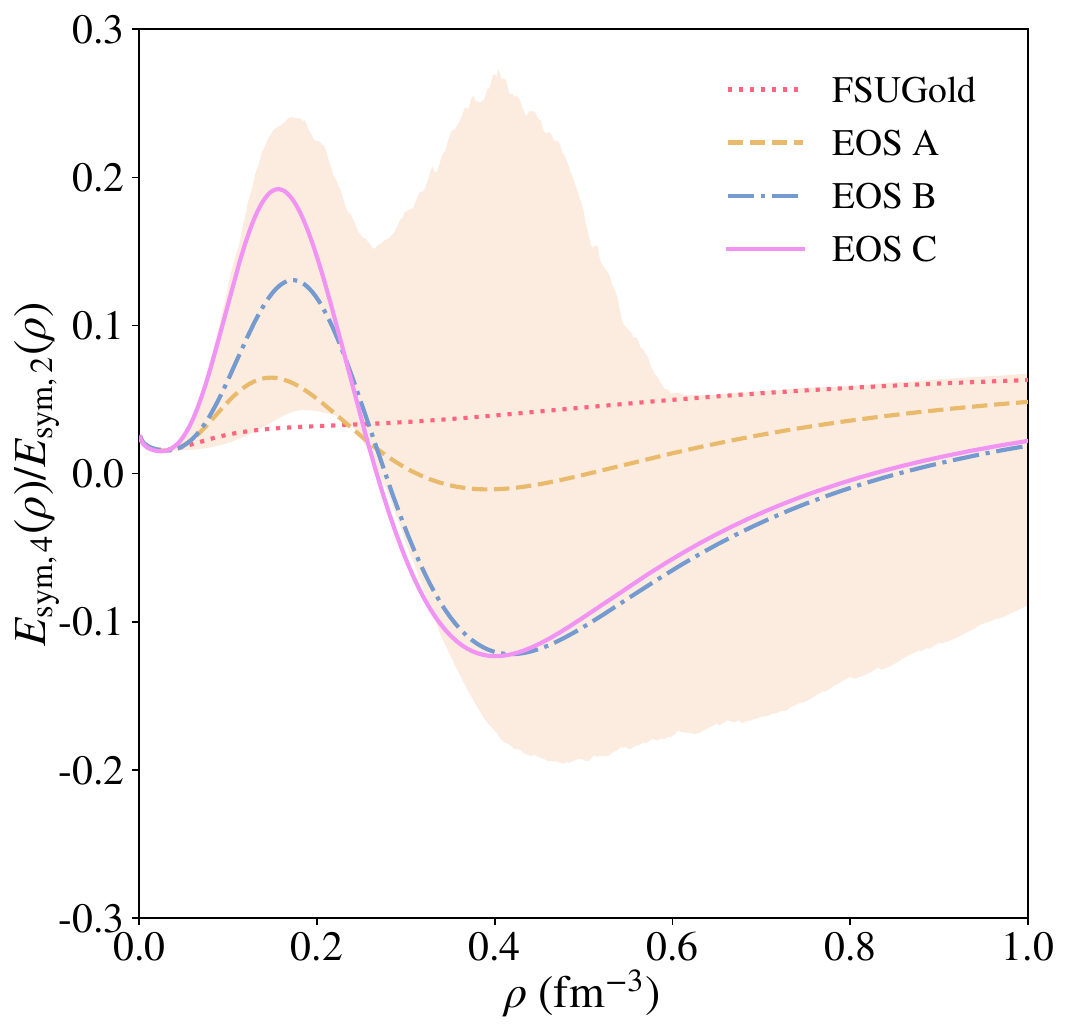}
    \caption{(Color online). Same as FIG.\,\ref{fig:E4_band}, but for the ratio of the quartic and quadratic terms $E_{\rm{sym,4}}(\rho)/E_{\rm{sym}}(\rho)$.}
    \label{fig:E4_E2}
\end{figure}

The ratio $E_{\rm{sym,4}}(\rho)/E_{\rm{sym}}(\rho)$, shown in FIG.\,\ref{fig:E4_E2}, exhibits a similarly non-monotonic density dependence, with a pronounced two-peak structure in the upper envelope of the ensemble band. More importantly, a third major finding is that the fourth-order symmetry energy can constitute a substantial fraction of the quadratic symmetry energy. Specifically, the 95\% ensemble band yields
\begin{equation}
-20\%\lesssim{E_{\rm{sym,4}}(\rho)}/{E_{\rm{sym}}(\rho)}
\lesssim30\%,\text{ within NS densities}.
\end{equation}
This result is particularly relevant for strongly neutron-rich matter, where the ratio of the quartic to quadratic contributions to the energy per nucleon is $E_{\rm{sym,4}}(\rho)\delta^2/E_{\rm{sym}}(\rho)$. Neglecting sixth- and higher-order symmetry energies, the corresponding relative correction to the neutron-proton chemical potential difference can be estimated as
\begin{equation}
\frac{\Delta(\mu_{\rm n}-\mu_{\rm p})}
{(\mu_{\rm n}-\mu_{\rm p})_{\rm{quadratic}}}
\approx 2\delta^2\frac{E_{\rm{sym,4}}(\rho)}{E_{\rm{sym}}(\rho)}
\sim (-40\%\text{ to }60\%)\delta^2,
\end{equation}
which can become appreciable as $\delta\to1$. For ultra-relativistic electrons at fixed baryon density, $\mu_{\rm e}\sim x_{\rm p}^{1/3}$, implying 
\begin{equation}
    \Delta x_{\rm p}/x_{\rm p}\approx3\Delta\mu_{\rm e}/\mu_{\rm e},
\end{equation} for small variations. These relations suggest that the fourth-order term may induce appreciable relative changes in $x_{\rm p}$, although a quantitative assessment requires a self-consistent treatment of the changing isospin asymmetry. The effects on $x_{\rm p}$, the core-crust transition density $\rho_{\rm{t}}$, and the corresponding transition pressure $P_{\rm{t}}$ are examined in Subsection \ref{subs_xp}.

\renewcommand*\figurename{\small FIG.}
\begin{figure}[h!]
    \centering
    \includegraphics[width=0.9\columnwidth]{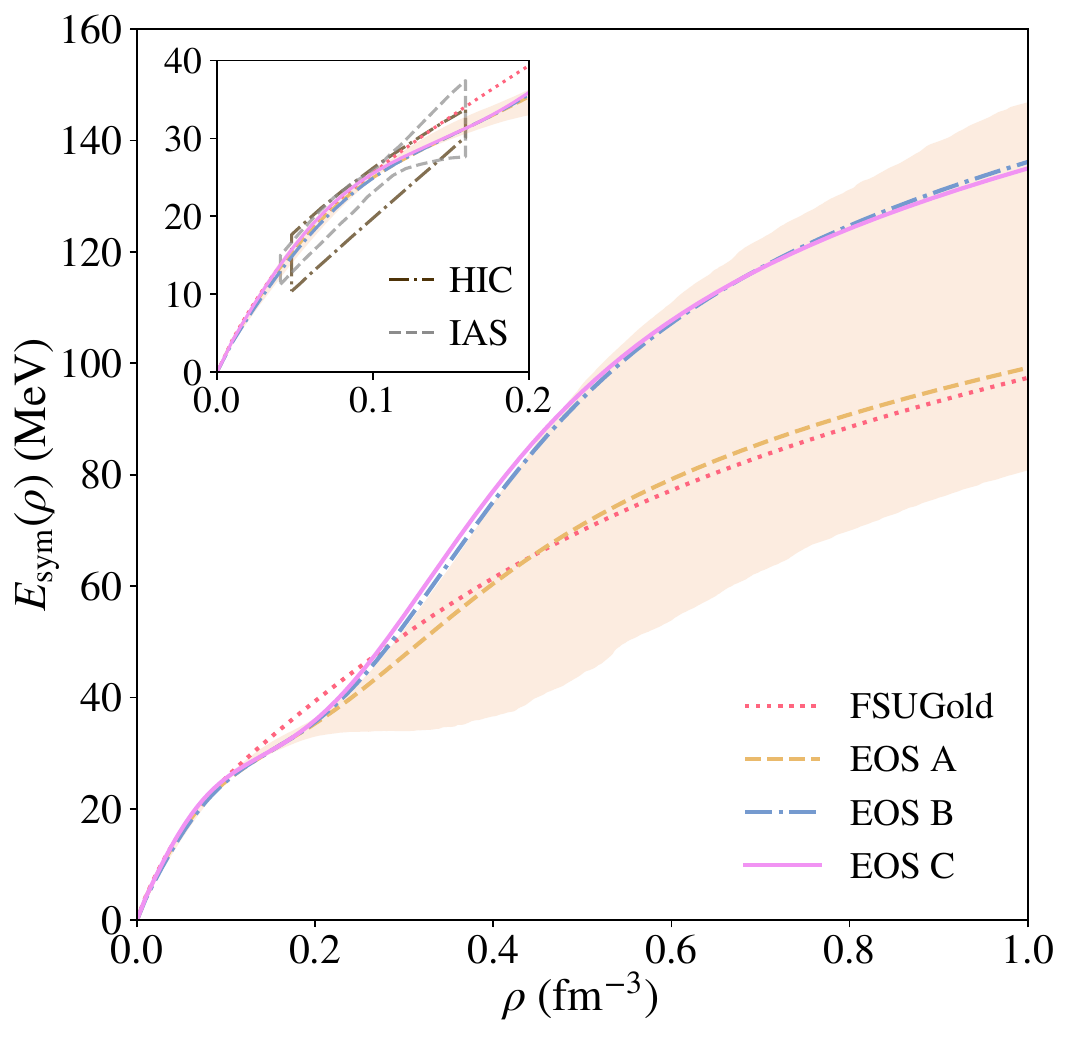}
    \caption{(Color online). Density dependence of the nuclear symmetry energy $E_{\rm{sym}}(\rho)$. The shaded region represents the 95\% ensemble band of the accepted RMF models consistent with astrophysical constraints on the maximum NS mass and tidal deformability. Results for three representative parameterizations, EOS A, B, and C, are compared with those of FSUGold. The inset shows the symmetry energy around nuclear saturation density, together with constraints from isobaric analog states (IAS)\,\cite{Danielewicz2014} and heavy ion collisions (HIC)\,\cite{Tsang2009}.}
    \label{fig:Esym}
\end{figure}

\begin{table}[h!]
    \centering
    \renewcommand{\arraystretch}{1.1}
    \setlength{\tabcolsep}{8pt}
    \begin{tabular}{ccccc}
        \hline\hline
        Parameter & FSUGold & EOS A & EOS B & EOS C \\
        \hline
        $K_{\mathrm{sym}}\,\mathrm{(MeV)}$   & $-51.4$  & $12.5$   & $57.1$   & $148.1$  \\
        $J_{\mathrm{sym}}\,\mathrm{(MeV)}$   & $426.6$  & $1532.6$ & $2919.8$ & $2751.2$ \\
        $L_{\mathrm{sym},4}\,\mathrm{(MeV)}$ & $2.54$   & $0.72$   & $12.35$  & $6.01$   \\
        $K_{\mathrm{sym},4}\,\mathrm{(MeV)}$ & $-4.8$   & $-90.8$  & $-202.9$ & $-370.3$ \\
        $I_0\,\mathrm{(MeV)}$               & $2817.1$ & $4074.6$ & $4074.6$ & $4074.6$ \\
        \hline\hline
    \end{tabular}
    \caption{Curvature and skewness parameters of the symmetry energy, slope and curvature parameters of the fourth-order symmetry energy, and the kurtosis coefficient $I_0$ of SNM for the four models at saturation density. Here, $K_{\rm{sym}}=9\rho_0^2\d^2E_{\rm{sym}}(\rho)/\d\rho^2|_{\rho_0}$, $J_{\rm{sym}}=27\rho_0^3\d^3E_{\rm{sym}}(\rho)/\d\rho^3|_{\rho_0}$, and $I_0=81\rho_0^4\d^4E_0(\rho)/\d\rho^4|_{\rho_0}$.}
    \label{tab:Esym4_KL}
\end{table}

Another interesting finding concerns the density dependence of the conventional symmetry energy $E_{\rm{sym}}(\rho)$, as shown in FIG.\,\ref{fig:Esym}. Although the four representative EOSs exhibit relatively similar behavior around nuclear saturation density, consistent with existing constraints from isobaric analog states (IAS)\,\cite{Danielewicz2014} and heavy ion collisions (HIC)\,\cite{Tsang2009}, their predictions gradually diverge at higher densities. Their similar symmetry energies around saturation density, together with their consistency with empirical constraints, suggest that these models can also provide reasonable descriptions of related finite-nucleus observables. In particular, EOSs with larger fourth-order symmetry energies at $\rho_0$ tend to exhibit stiffer symmetry energies at supra-saturation densities. As shown in TAB.\,\ref{tab:Esym4_KL}, the corresponding curvature parameters are $K_{\rm{sym}}\equiv9\rho_0^2\d^2E_{\rm{sym}}(\rho)/\d\rho^2|_{\rho_0}\approx-51.4$\,MeV for FSUGold and $K_{\rm{sym}}\approx12.5$\,MeV, $57.1$\,MeV, and $148.1$\,MeV for EOSs A, B, and C, respectively. Thus, all three representative EOSs exhibit positive $K_{\rm{sym}}$, despite satisfying the adopted astrophysical constraints on the maximum NS mass and tidal deformability of GW170817. The shaded 95\% ensemble band in FIG.\,\ref{fig:Esym} further illustrates the range of symmetry energy density dependences permitted by these constraints.

\renewcommand*\figurename{\small FIG.}
\begin{figure}[h!]
    \centering
    \includegraphics[width=\columnwidth]{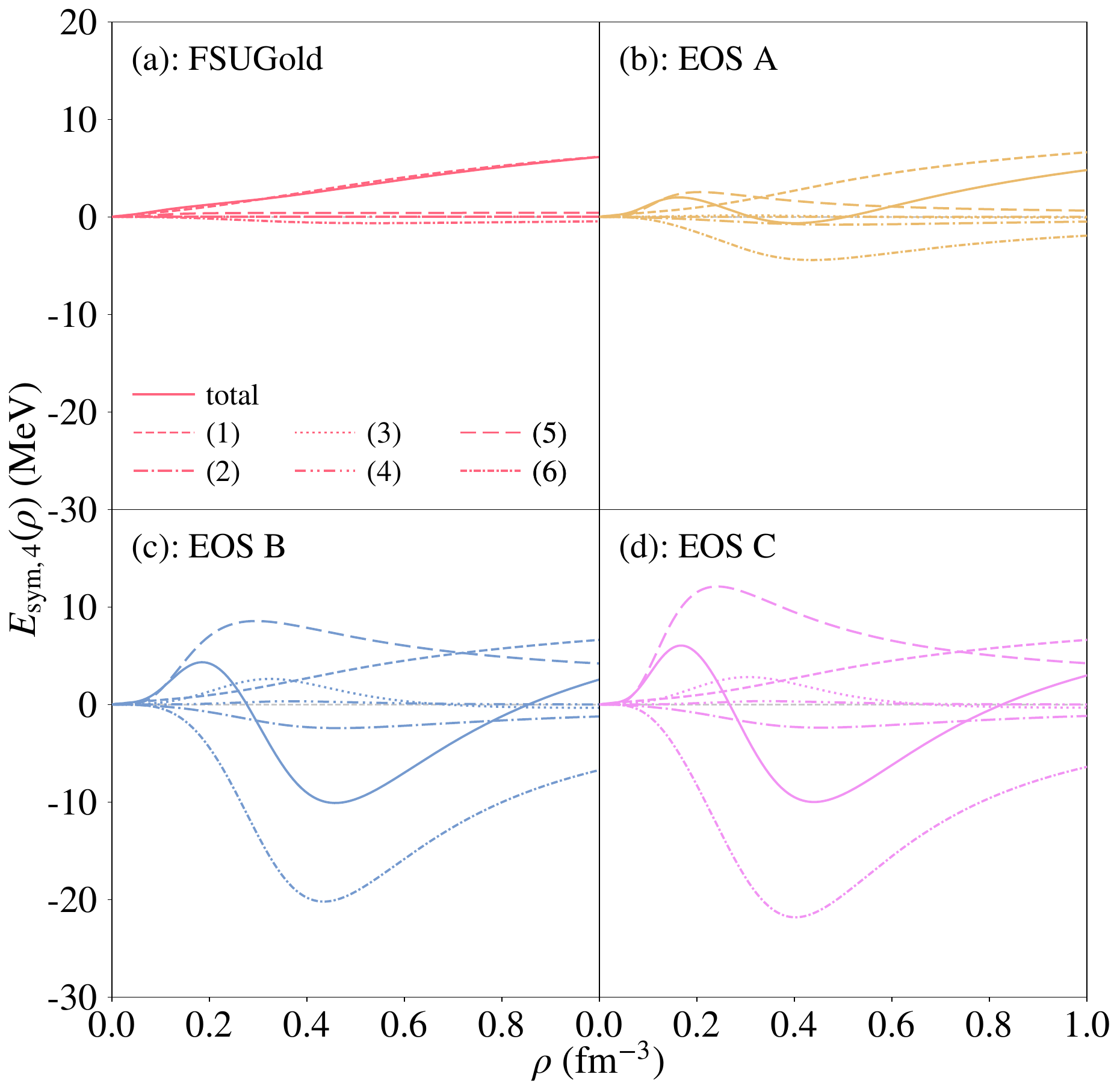}
    \caption{(Color online). Density dependence of the fourth-order nuclear symmetry energy $E_{\rm{sym,4}}(\rho)$ (solid lines) and its six decomposition terms (1)-(6) in Eq.\,(\ref{eq:Esym4-full}) for representative RMF parameterizations: (a) FSUGold, (b) EOS A, (c) EOS B, and (d) EOS C of TAB.\,\ref{tab:representative-eos}. Term (1) represents the kinetic contribution, terms (2)-(4) arise from the $\delta$-meson response, and terms (5) and (6) correspond to the induced $\omega$- and $\sigma$-meson responses, respectively. The comparison illustrates the different density dependences and competing contributions that determine the overall behavior of $E_{\rm{sym,4}}(\rho)$ across these RMF parameterizations.}
    \label{fig:E4_components}
\end{figure}

This result is particularly noteworthy because previous analyses of tidal deformability constraints, particularly those associated with GW170817, have often favored relatively small or negative values of $K_{\rm{sym}}$\,\cite{Zhou2019}, reflecting a preference for a softer density dependence of the symmetry energy. Our results suggest that this preference may be relaxed when a sizable fourth-order symmetry energy is taken into account. Specifically, the existence of EOSs with positive $K_{\rm{sym}}$ that remain compatible with the GW170817 tidal deformability constraint indicates that a relatively stiff symmetry energy need not necessarily imply an excessively large tidal deformability. The higher-order isospin contributions provide additional flexibility in the EOS of neutron-rich matter, potentially modifying the connection between $K_{\rm{sym}}$ and NS observables. This finding highlights the importance of treating the quadratic and fourth-order symmetry energies consistently when extracting information on the density dependence of nuclear symmetry energy from astrophysical observations.
For reference, our EOS ensemble yields $E_{\rm{sym}}(2\rho_0)\approx44.3\pm10.4$\,MeV from the 95\% ensemble band. The relatively broad range indicates that considerable uncertainties in the symmetry energy persist\,\cite{LiCaiXieZhang2021,Cai2021PRC} even after imposing astrophysical constraints on NS masses and tidal deformabilities. In particular, the presence of a sizable fourth-order symmetry energy allows greater flexibility in the density dependence of the symmetry energy, suggesting that astrophysical observations alone may not tightly constrain $E_{\rm{sym}}(\rho)$ without a more detailed understanding of higher-order isospin contributions.

Finally we shown in FIG.\,\ref{fig:E4_components} the density dependence of the six contributions to the fourth-order symmetry energy, 
\begin{equation}\label{eq:decompE4}
E_{\rm{sym,4}}(\rho)=\sum_{i=1}^{6}E_{\rm{sym,4}}^{(i)}(\rho),
\end{equation}
as given in Eq.\,(\ref{eq:Esym4-full}). A common feature is that the $\omega$- and $\sigma$-meson contributions, $E_{\rm{sym,4}}^{(5)}(\rho)$ and $E_{\rm{sym,4}}^{(6)}(\rho)$, have comparable magnitudes but opposite signs, resulting in a partial cancellation. For FSUGold, the meson contributions are generally smaller than the kinetic contribution $E_{\rm{sym,4}}^{(1)}(\rho)$, which largely determines the magnitude and density dependence of the total fourth-order symmetry energy $E_{\rm{sym,4}}(\rho)$.

For EOSs A, B, and C, the meson contributions become more significant, leading to a stronger competition among the six  terms of Eq.\,(\ref{eq:decompE4}). In particular, the negative $\sigma$-meson contribution $E_{\rm{sym,4}}^{(6)}(\rho)$ can substantially reduce the total fourth-order symmetry energy at intermediate densities, producing a non-monotonic density dependence that is especially pronounced for EOSs B and C. At higher densities, $E_{\rm{sym,4}}(\rho)$ increases again. Despite the relatively large magnitudes of some individual contributions, their partial cancellations can keep the total $E_{\rm{sym,4}}(\rho)$ much smaller. This illustrates why the fourth-order symmetry energy can exhibit markedly different density dependences among RMF parameterizations.

\subsection{Effects of the Fourth-order Symmetry Energy on NS Proton Fraction $x_{\textmd{p}}$, Core-crust Transition, M-R Relations, Tidal Deformabilities and Sound Speed Profiles}\label{subs_xp}

The fourth-order symmetry energy can affect NS properties by modifying the composition and pressure of highly ANM. Although $E_{\rm{sym,4}}(\rho)$ is generally smaller than the symmetry energy, its contribution may become important in neutron-rich matter, where the isospin asymmetry is large. In this subsection, we examine its effects on the proton fraction $x_{\rm{p}}$, core-crust transition, NS M-R relations, and tidal deformabilities, using FSUGold and the three parameterizations EOSs A, B, and C. Particular attention is paid to the differences between the conventional and fourth-order truncations of the nucleonic matter EOS.

The proton fraction $x_{\rm{p}}$ in cold, neutrino-free,
$\beta$-equilibrated NS matter is determined
by the chemical equilibrium and charge neutrality conditions.
For the weak processes
$\rm{n}\leftrightarrow\rm{p}+\rm{e}^{-}+\overline{\nu}_{\rm{e}}$,
the equilibrium condition reads\,\cite{Xu2009}
\begin{equation}\label{elechem1}
\mu_{\rm{n}}-\mu_{\rm{p}}
=\mu_{\rm{e}}
=2\frac{\partial E(\rho,\delta)}{\partial\delta}
\approx4\delta E_{\rm{sym}}(\rho)
+8\delta^3E_{\rm{sym,4}}(\rho),
\end{equation}
where $\delta=1-2x_{\rm{p}}$ and $\mu_J$ denotes the
chemical potential of particle species $J$.
Treating leptons as relativistic free Fermi gases,
their chemical potentials are given by
\begin{equation}\label{elechem2}
\mu_\ell=
\sqrt{m_\ell^2+(3\pi^2\rho x_\ell)^{2/3}},
~~ \ell=\rm{e},\mu.
\end{equation}
Once the electron chemical potential exceeds the muon
rest mass, muons appear in NS matter. The equilibrium
and charge neutrality conditions then become
\begin{equation}\label{elemuochem}
\mu_{\rm{n}}-\mu_{\rm{p}}
=\mu_{\rm{e}}=\mu_\mu,
~~
x_{\rm{p}}=x_{\rm{e}}+x_\mu.
\end{equation}
These equations determine the equilibrium proton
fraction and lepton abundances at a given baryon density.
In particular, the fourth-order symmetry energy enters
the neutron-proton chemical potential difference through
$8\delta^3E_{\rm{sym,4}}(\rho)$, which may become
important in highly neutron-rich matter.

The total energy density and pressure of the
npe$\mu$ system are expressed as
\begin{equation}
\varepsilon(\rho,\delta)
=\rho[M_{\rm N}+E(\rho,\delta)]
+\varepsilon_{\rm{e}}+\varepsilon_\mu,
\end{equation}
and
\begin{equation}
P=P_{\rm{b}}+P_{\rm{e}}+P_\mu,
\end{equation}
respectively, where the baryonic and leptonic
pressures satisfy $
P_{\rm{b}}=\rho_{\rm{n}}\mu_{\rm{n}}
+\rho_{\rm{p}}\mu_{\rm{p}}-\varepsilon_{\rm{b}}$ and $
P_\ell=\rho_\ell\mu_\ell-\varepsilon_\ell$.
The lepton energy densities are calculated using
the relativistic free Fermi gas model. Along the
$\beta$-equilibrium sequence, the total pressure
also satisfies the thermodynamic relation
$P=\rho^2\d(\varepsilon/\rho)/\d\rho$.

The core-crust transition density $\rho_{\rm{t}}$
marks the boundary between the uniform liquid core
and the inhomogeneous inner crust of NSs.
Here, we employ the thermodynamical method,
which requires the intrinsic stability conditions\,\cite{Xu2009}
\begin{equation}\label{ther2}
-\left(\frac{\partial P}{\partial v}\right)_{\mu_{\rm{np}}}>0,
~~
-\left(\frac{\partial\mu_{\rm{np}}}{\partial q_{\rm{c}}}\right)_v>0,
\end{equation}
where $v$ and $q_{\rm{c}}$ denote the volume and
electric charge per baryon, respectively, and
$\mu_{\rm{np}}=\mu_{\rm{n}}-\mu_{\rm{p}}$.
For $\beta$-equilibrated npe$\mu$ matter,
the lepton pressures depend only on the common
lepton chemical potential
$\mu_{\rm{e}}=\mu_\mu=\mu_{\rm{np}}$.
Consequently, at fixed $\mu_{\rm{np}}$, the
mechanical stability condition can be expressed
in terms of the nucleonic EOS as
\begin{align}\label{Vther}
V_{\rm{th}}
=&2\rho\frac{\partial E(\rho,x_{\rm{p}})}{\partial\rho}
+\rho^2\frac{\partial^2E(\rho,x_{\rm{p}})}{\partial\rho^2}
-\frac{(
\rho{\partial^2E(\rho,x_{\rm{p}})}/
{\partial\rho\partial x_{\rm{p}}}
)^2}
{{\partial^2E(\rho,x_{\rm{p}})}/{\partial x_{\rm{p}}^2}}\notag\\
=&\rho^2
\left(
\frac{\partial^2 E(\rho,x_{\rm p})}{\partial x_{\rm p}^2}
\right)^{-1}
\left[
\frac{\partial\mu_{\rm n}}{\partial\rho_{\rm n}}
\frac{\partial\mu_{\rm p}}{\partial\rho_{\rm p}}
-\left(
\frac{\partial\mu_{\rm n}}{\partial\rho_{\rm p}}
\right)^2
\right]>0.
\end{align}
The equilibrium composition entering this expression
is determined self-consistently from
Eq.\,(\ref{elemuochem}), including both electrons
and muons. The transition density is identified
by $V_{\rm{th}}(\rho_{\rm{t}})=0$
along the $\beta$-equilibrium trajectory.

Including the fourth-order symmetry energy,
the stability condition becomes\,\cite{Xu2009,Cai2012}
\begin{align}\label{Vther4th}
V_{\rm{th}}^{\rm{4th}}
=\frac{1}{9}\Bigg[&
K_0(\rho)+6L_0(\rho)
+\left(K_{\rm{sym}}(\rho)+6L(\rho)\right)\delta^2
\notag\\
&+\left(K_{\rm{sym,4}}(\rho)+6L_{\rm{sym,4}}(\rho)\right)\delta^4\notag\\
&-\frac{2\delta^2
[L(\rho)+2\delta^2L_{\rm{sym,4}}(\rho)]^2}
{E_{\rm{sym}}(\rho)+6\delta^2E_{\rm{sym,4}}(\rho)}
\Bigg]>0.
\end{align}
The second-order result $V_{\rm{th}}^{\rm{2nd}}$
follows by neglecting $E_{\rm{sym,4}}(\rho)$
and its density derivatives, i.e.,
\begin{align}\label{Vther2nd}
V_{\rm{th}}^{\rm{2nd}}
=\frac{1}{9}\Bigg[&
K_0(\rho)+6L_0(\rho)\notag\\
&+\delta^2\left(
K_{\rm{sym}}(\rho)+6L(\rho)
-\frac{2L^2(\rho)}{E_{\rm{sym}}(\rho)}
\right)
\Bigg]>0.
\end{align}

The corresponding core-crust transition pressure
is calculated from the total pressure of
$\beta$-equilibrated npe$\mu$ matter,
\begin{align}
P_{\rm{t}}^{\rm{4th}}
=&\left[
\rho^2\left(
\frac{\d E_0(\rho)}{\d\rho}
+\delta^2\frac{\d E_{\rm{sym}}(\rho)}{\d\rho}
+\delta^4\frac{\d E_{\rm{sym,4}}(\rho)}{\d\rho}
\right)
\right]_{\rho_{\rm{t}}^{\rm{4th}},\delta_{\rm{t}}}
\notag\\
&+\left[P_{\rm{e}}+P_\mu\right]
_{\rho_{\rm{t}}^{\rm{4th}},\delta_{\rm{t}}},
\end{align}
where the lepton pressures are evaluated at the corresponding equilibrium composition. The transition density $\rho_{\rm{t}}^{\rm{2nd}}$ and pressure $P_{\rm{t}}^{\rm{2nd}}$ in the parabolic approximation are obtained by consistently neglecting $E_{\rm{sym,4}}(\rho)$ in the equilibrium, stability, and pressure calculations. As indicated by Eq.\,(\ref{Vther4th}), the fourth-order symmetry energy affects the thermodynamical stability not only through its magnitude but also through its density dependence, characterized by $L_{\rm{sym,4}}(\rho)$ and $K_{\rm{sym,4}}(\rho)$. In particular, $K_{\rm{sym,4}}(\rho)$ enters the stability condition linearly, whereas $L_{\rm{sym,4}}(\rho)$ contributes both directly and through its nonlinear coupling with $L(\rho)$. In the present analysis, variations in $E_{\rm{sym,4}}(\rho_0)$ are accompanied by changes in the curvature parameter $K_{\rm{sym,4}}\equiv K_{\rm{sym,4}}(\rho_0)$, which is closely related to the behavior of $K_{\rm{sym,4}}(\rho)$ at subsaturation densities, where the core-crust transition occurs. Consequently, even a relatively small $E_{\rm{sym,4}}(\rho)$ may induce appreciable changes in the core-crust transition density and pressure through its density derivatives, particularly the curvature contribution. 
As summarized in TAB.\,\ref{tab:Esym4_KL}, the slope and curvature parameters of the fourth-order symmetry energy exhibit substantial model dependence. While $L_{\rm{sym,4}}$ varies from $0.72$ to $12.35$\,MeV among EOSs A, B, and C, the corresponding curvature parameters are $K_{\rm{sym,4}}\approx-90.8$\,MeV, $-202.9$\,MeV, and $-370.3$\,MeV, respectively, substantially more negative than the FSUGold value of $-4.8$\,MeV. The increasingly negative curvature from EOS A to EOS C reflects a progressively stronger downward bending of $E_{\rm{sym,4}}(\rho)$ near saturation, consistent with the pronounced nonmonotonic behavior shown in FIG.\,\ref{fig:E4_band}. Although these parameters are evaluated at $\rho_0$ rather than at the core-crust transition density, their substantial variations highlight the potentially important role of the density derivatives of $E_{\rm{sym,4}}(\rho)$ in determining the transition properties. This is consistent with the considerably larger changes in the transition pressure predicted by EOSs B and C compared with FSUGold. Comparisons with the corresponding full-EOS results further allow us to assess the contributions from higher-order symmetry energies beyond the fourth-order term.

\renewcommand*\figurename{\small FIG.}
\begin{figure}[h!]
    \centering
    \includegraphics[width=\columnwidth]{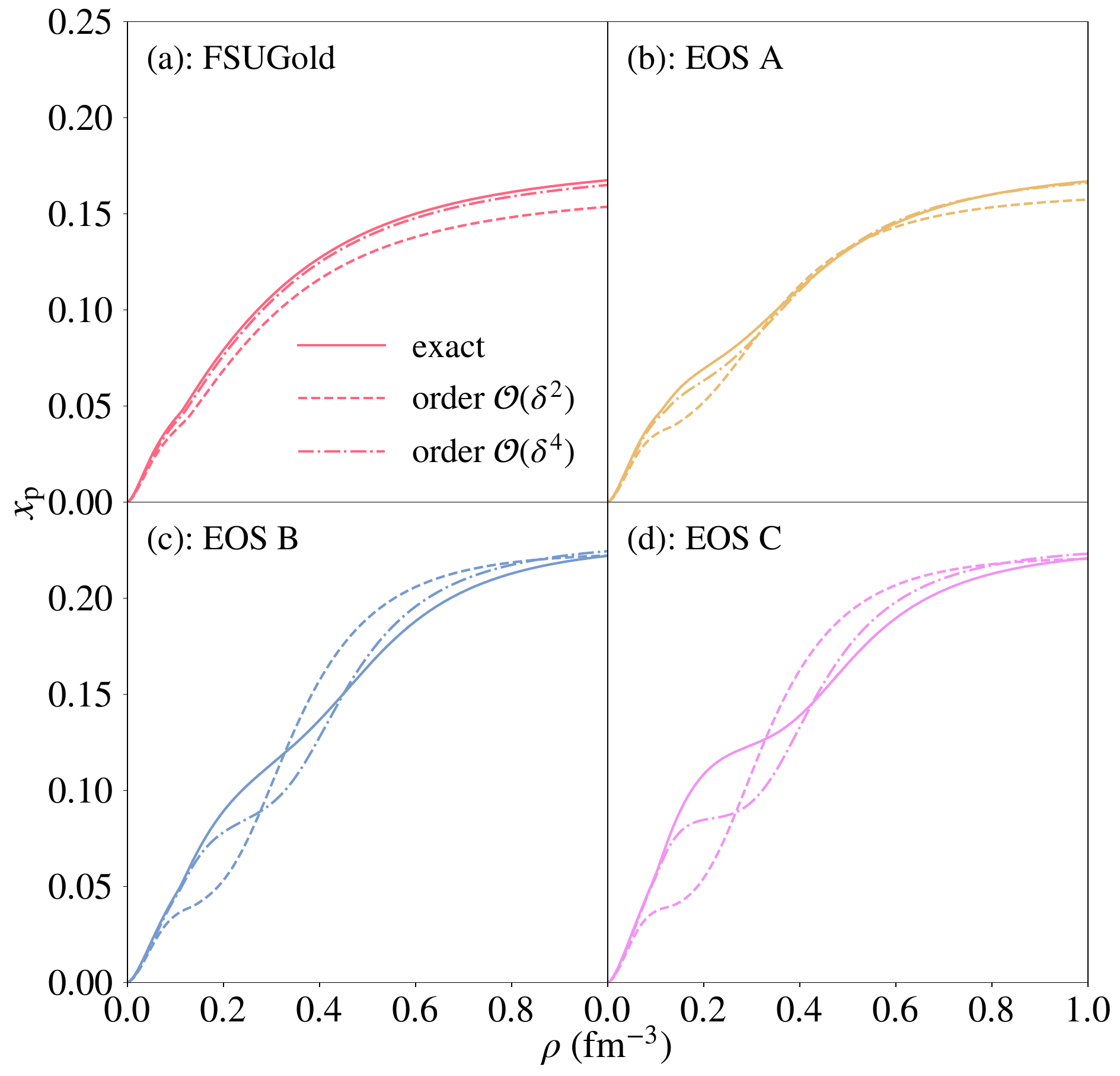}
    \caption{(Color online). Proton fraction $x_{\rm p}=\rho_{\rm p}/\rho$ as a function of baryon density $\rho$ in $\beta$-equilibrated NS matter for (a) FSUGold, (b) EOS A, (c) EOS B, and (d) EOS C. The dashed and dash-dotted curves represent the results obtained by truncating the isospin-asymmetry expansion of the nuclear EOS at order $\mathcal{O}(\delta^2)$ and order $\mathcal{O}(\delta^4)$, respectively, while the solid curves correspond to the full EOS.}
    \label{fig:xp_bands_E4}
\end{figure}

FIG.\,\ref{fig:xp_bands_E4} shows the density dependence of the proton fraction $x_{\rm p}$ in $\beta$-equilibrated NS matter for FSUGold and EOSs A, B, and C, comparing the predictions from the full EOS with those obtained using the second- and fourth-order expansions in $\delta$. For FSUGold, the fourth-order approximation reproduces the full result remarkably well over the entire density range considered, whereas the parabolic approximation systematically underestimates $x_{\rm p}$. A similar improvement is observed for EOS A, although some deviations remain at intermediate densities. In contrast, EOSs B and C exhibit more pronounced differences among the three calculations, particularly at intermediate densities, where the parabolic approximation substantially overestimates the proton fraction and fails to reproduce the detailed density dependence predicted by the full EOS.

The fourth-order symmetry energy also contributes to the pressure of neutron-rich matter. At saturation density, this contribution is given by $P_{\rm{sym,4}}(\rho_0,\delta)=\rho_0L_{\rm{sym,4}}\delta^4/3$, see Eq.\,(\ref{def:P4}). Taking $\rho_0\approx0.16,\rm{fm}^{-3}$ and using $\delta=1-2x_{\rm p}$ with the proton fractions estimated from the full-EOS curves in FIG.\,\ref{fig:xp_bands_E4}, we obtain $P_{\rm{sym,4}}(\rho_0,\delta)\approx0.025$-$0.30$\,MeV/fm$^3$ for the four models, with EOS B yielding the largest contribution. Relative to the quadratic symmetry pressure, the fourth-order contribution is characterized by $P_{\rm{sym,4}}(\rho_0,\delta)\delta^2/P_{\rm{sym}}(\rho_0,\delta)=\delta^2L_{\rm{sym,4}}/L$, amounting to approximately $1\%$-$18\%$ for EOSs A, B, and C. This illustrates the substantial model dependence of the fourth-order pressure contribution even near saturation density.

Including $E_{\rm{sym,4}}(\rho)$ generally improves the description of $x_{\rm p}$, but the degree of improvement depends strongly on the underlying EOS. For EOSs B and C, the fourth-order approximation captures part of the nontrivial density dependence of the full result, although noticeable discrepancies persist, indicating that symmetry energy contributions beyond fourth order may remain important. At higher densities, the predictions from the different truncations tend to converge. These results demonstrate that the convergence of the isospin-asymmetry expansion is model dependent and that the fourth-order symmetry energy can play an important role in determining the composition of neutron-rich NS matter.

\begin{table}[h!]
    \centering
    \setlength{\tabcolsep}{7pt}
    \renewcommand{\arraystretch}{1.1}
    \begin{tabular}{ccccc}
        \hline\hline
        $\rho$ or $P$ & FSUGold & EOS A & EOS B & EOS C \\
        \hline
        $\rho_{\rm t}\,(\mathrm{fm}^{-3})$
            & 0.0820 & 0.106 & 0.108 & 0.113 \\
        $\rho_{\rm t}^{\mathrm{2nd}}\,(\mathrm{fm}^{-3})$
            & 0.0866 & 0.105 & 0.110 & 0.108 \\
        $\rho_{\rm t}^{\mathrm{4th}}\,(\mathrm{fm}^{-3})$
            & 0.0858 & 0.108 & 0.113 & 0.118 \\
        $\Delta({\rho_{\rm t}})$
            & $-0.9\%$ & 2.9\% & 2.7\% & 9.3\% \\
        \hline
        $P_{\rm t}\,(\mathrm{MeV}/\mathrm{fm}^{3})$
            & 0.494 & 0.748 & 0.984 & 1.140 \\        
        $P_{\rm t}^{\mathrm{2nd}}\,(\mathrm{MeV}/\mathrm{fm}^{3})$
            & 0.528 & 0.553 & 0.663 & 0.484 \\
        $P_{\rm t}^{\mathrm{4th}}\,(\mathrm{MeV}/\mathrm{fm}^{3})$
            & 0.550 & 0.756 & 1.06 & 1.11 \\
        $\Delta({P_{\rm t}})$
            & 4.2\% & 36.7\% & 59.9\% & 129.3\% \\
        \hline\hline
    \end{tabular}
    \caption{Core-crust transition densities and pressures for
    FSUGold and EOSs A, B, and C. The quantities
    without superscripts are obtained from the full EOS,
    while the superscripts ``2nd'' and ``4th'' denote
    the corresponding truncations of the
     expansion. The relative changes for  $y=\rho_{\rm t}$ or $P_{\rm t}$
    are defined by
    $\Delta(y)=(y^{\rm {4th}}-y^{\rm {2nd}})/
    y^{\rm{2nd}}$.
    }
    \label{tab:transition}
\end{table}

We next examine the effects of the fourth-order symmetry energy on the core-crust transition density $\rho_{\rm t}$ and the corresponding pressure $P_{\rm t}$. TAB.\,\ref{tab:transition} summarizes the results for FSUGold and EOSs A, B, and C, together with the relative changes induced by including the fourth-order symmetry energy. Compared with the parabolic approximation, the fourth-order truncation changes the transition density by approximately $-0.9\%$, $2.9\%$, $2.7\%$, and $9.3\%$, respectively. In contrast, the corresponding transition pressures increase by approximately $4.2\%$, $36.7\%$, $59.9\%$, and $129.3\%$. Thus, the fourth-order symmetry energy has a considerably stronger impact on the transition pressure than on the transition density, particularly for EOSs B and C. This pronounced model dependence highlights the sensitivity of the thermodynamical stability condition to the density dependence of $E_{\rm{sym,4}}(\rho)$, especially its slope $L_{\rm{sym,4}}(\rho)$ and curvature $K_{\rm{sym,4}}(\rho)$.
Comparing the transition pressures obtained from the truncated expansions with those from the full EOS, we find that including the fourth-order term improves the agreement for EOSs A, B, and C, but not for FSUGold. In contrast, including the fourth-order term does not necessarily improve the agreement of the transition density with the exact result. For FSUGold, the fourth-order approximation slightly reduces the deviation, whereas for EOSs A, B, and C, it yields transition densities that are higher than the corresponding full-EOS predictions. This behavior indicates that higher-order terms beyond $E_{\rm{sym,4}}(\rho)$ may remain relevant near the core-crust transition, where the matter is highly neutron rich. Consequently, the convergence of the isospin-asymmetry expansion should be examined carefully when extracting core-crust transition properties from truncated nuclear EOSs.

The sizable effects of the fourth-order symmetry energy on the proton fraction and core-crust transition can be understood from derivatives of the EOS with respect to $\delta$. In determining $x_{\rm{p}}$, the neutron-proton chemical potential difference involves the first derivative of the EOS, giving $\partial E(\rho,\delta)/\partial x_{\rm{p}}\approx-4\delta E_{\rm{sym}}(\rho)-8\delta^3E_{\rm{sym,4}}(\rho)$. Thus, the fourth-order contribution is enhanced by an additional factor of two relative to its contribution to the EOS itself. The core-crust transition is even more sensitive to higher-order terms, since the thermodynamical stability condition also involves the second derivative, $\partial^2E(\rho,\delta)/\partial x_{\rm{p}}^2\approx8E_{\rm{sym}}(\rho)+48\delta^2E_{\rm{sym,4}}(\rho)$. Consequently, the fourth-order contribution is enhanced by a factor of six relative to its contribution to the EOS itself, apart from the magnitude of $E_{\rm{sym,4}}(\rho)$ and the additional factor $\delta^2$. These enhancements help explain why a relatively small fourth-order symmetry energy can nevertheless produce appreciable changes in the proton fraction and core-crust transition properties.

More generally, the relative importance of higher-order symmetry energies depends on the physical quantity under consideration. Physically, the EOS measures the energy cost of isospin asymmetry, while its first and second derivatives with respect to $\delta$ characterize the neutron-proton chemical potential difference and the response to changes in the isospin composition, respectively. This motivates the introduction of three effective symmetry energies associated with the EOS itself, the proton fraction, and the isospin curvature relevant to the core-crust transition:
\begin{align}
\widehat{E}_{\rm{sym}}^{\rm{(EOS)}}(\rho,\delta)
\equiv&\frac{E(\rho,\delta)-E_0(\rho)}{\delta^2}
\notag\\
=&E_{\rm{sym}}(\rho)
+\sum_{n=2}^{\infty}
E_{\rm{sym},2n}(\rho)\delta^{2n-2},
\label{Esymeff1}\\
\widehat{E}_{\rm{sym}}^{(x_{\rm{p}})}(\rho,\delta)
\equiv&\frac{1}{2\delta}
\frac{\partial E(\rho,\delta)}{\partial\delta}
\notag\\
=&E_{\rm{sym}}(\rho)
+\sum_{n=2}^{\infty}
nE_{\rm{sym},2n}(\rho)\delta^{2n-2},
\label{Esymeff2}\\
\widehat{E}_{\rm{sym}}^{(\rm{tran})}(\rho,\delta)
\equiv&\frac{1}{2}
\frac{\partial^2 E(\rho,\delta)}{\partial\delta^2}
\notag\\
=&E_{\rm{sym}}(\rho)
+\sum_{n=2}^{\infty}
(2n^2-n)E_{\rm{sym},2n}(\rho)\delta^{2n-2}.
\label{Esymeff3}
\end{align}
Here, the derivatives with respect to $\delta$ are taken at fixed baryon density $\rho$. These definitions reveal a systematic enhancement of higher-order symmetry energies as the order of differentiation increases. In particular, the fourth-order symmetry energy enters the three effective quantities with relative coefficients of $1$, $2$, and $6$, respectively. This enhancement has a simple physical interpretation: A higher-order term may contribute relatively little to the total energy but have a much stronger influence on how the energy changes when neutrons are converted into protons or when the isospin composition is perturbed. For example, taking $E_{\rm{sym,4}}(\rho)/E_{\rm{sym}}(\rho)\approx0.1$ near the core-crust transition (as shown in FIG.\,\ref{fig:E4_E2}) and a representative proton fraction $x_{\rm p}\approx0.04$, corresponding to $\delta\approx0.92$, the fourth-order contribution relative to the quadratic term in the energy is only $\delta^2E_{\rm{sym,4}}(\rho)/E_{\rm{sym}}(\rho)\approx8.5\%$. However, its relative contribution to the isospin curvature reaches $6\delta^2E_{\rm{sym,4}}(\rho)/E_{\rm{sym}}(\rho)\approx51\%$, indicating a potentially substantial effect on the transition.
This enhanced curvature contribution need not produce a comparably large shift in the transition density, which depends on the full stability condition, whereas the transition pressure can remain sensitive to changes in the EOS density derivatives and equilibrium composition. For quantities involving the third derivative of the EOS with respect to $\delta$, the quadratic symmetry energy no longer contributes, and the fourth-order term provides the leading non-vanishing contribution. Thus, higher-order symmetry energies, despite their relatively small magnitudes, can play important roles in the composition and stability of highly ANM.

Beyond their role in determining the core-crust transition, the slope and curvature parameters of the fourth-order symmetry energy also influence the incompressibility of ANM. The isospin dependent saturation density, defined through the relation ${\partial E(\rho,\delta)}/{\partial
\rho}|_{\rho=\rho_{\rm{sat}}(\delta)}=0$, can be expanded as\,\cite{Chen2009}
\begin{align}\label{fs_1}
\rho_{\rm{sat}}(\delta)\approx &\rho_0-\frac{3L\rho_0}{K_0}\delta^2\notag\\
&+\left(\frac{3K_{\rm{sym}}L}{K_0^2}-\frac{3L_{\rm{sym,4}}}{K_0}
-\frac{3J_0L^2}{2K_0^3}\right)\rho_0\delta^4+\mathcal{O}(\delta^6),
\end{align}
from which the incompressibility at saturation follows as
\begin{align}
K_{\rm{sat}}(\delta)\equiv&\left.9\rho_{\rm{sat}}^2(\delta)\frac{\partial^2E(\rho,\delta)}{\partial
\rho^2}\right|_{\rho=\rho_{\rm{sat}}(\delta)}\notag\\
\approx &K_0+K_{\rm{sat,2}}\delta^2
+K_{\rm{sat,4}}\delta^4+\mathcal{O}(\delta^6),
\end{align}
where the fourth-order coefficient $K_{\rm{sat,4}}$ is given by\,\cite{Chen2009,Cai2021PNM}
\begin{align}
K_{\rm{sat,4}}
=&K_{\rm{sym,4}}
-\left(6+\frac{J_0}{K_0}\right)L_{\rm{sym,4}}\notag\\
&+\left(9L-J_{\rm{sym}}
+\frac{J_0K_{\rm{sym}}}{K_0}\right)\left(\frac{L}{K_0}\right)
\notag\\
&+\left[\frac{I_0}{2}
+\left(3-\frac{J_0}{2K_0}\right)J_0\right]
\left(\frac{L}{K_0}\right)^2,
\end{align}
besides the second-order incompressibility coefficient\,\cite{Chen2009},
\begin{equation}
K_{\rm{sat,2}}=K_{\rm{sym}}-6L-{J_0L}/{K_0}.
\end{equation}
Here, $I_0=81\rho_0^4\d^4E_0(\rho)/\d\rho^4|_{\rho_0}$ is the kurtosis coefficient of the EOS of SNM at $\rho_0$, and $J_{\rm{sym}}=27\rho_0^3\d^3E_{\rm{sym}}(\rho)/\d\rho^3|_{\rho_0}$ is the skewness coefficient of the symmetry energy. Notably, $K_{\rm{sym,4}}$ enters $K_{\rm{sat,4}}$ with a unit coefficient, implying that its uncertainty propagates directly into $K_{\rm{sat,4}}$ with the same absolute magnitude, while the contribution from $L_{\rm{sym,4}}$ is weighted by the factor $-(6+J_0/K_0)$. As shown in TAB.\,\ref{tab:Esym4_KL}, $K_{\rm{sym,4}}$ exhibits substantial model dependence, spanning approximately $365$\,MeV between FSUGold and EOS C. This considerable variation highlights the importance of constraining $K_{\rm{sym,4}}$ for a reliable determination of the higher-order incompressibility coefficient $K_{\rm{sat,4}}$. Nevertheless, the latter also depends on higher-order density derivatives of the lower-order EOS terms, particularly $J_{\rm{sym}}$ and $I_0$, whose uncertainties may further affect its determination. Using the saturation parameters of the four representative EOSs, we obtain $K_{\rm{sat,4}}\approx-5.7$\,MeV, $-297.4$\,MeV, $-739.8$\,MeV, and $-878.4$\,MeV for FSUGold and EOSs A, B, and C, respectively. The corresponding second-order coefficients are $K_{\rm{sat,2}}\approx-274.3$\,MeV, $-183.0$\,MeV, $-138.4$\,MeV, and $-47.4$\,MeV. Remarkably, while the fourth-order coefficient is negligible compared with its second-order counterpart for FSUGold, it becomes substantially larger in magnitude for EOSs A, B, and C, revealing a pronounced model dependence. Interestingly, the increasing $K_{\rm{sat,2}}$ from EOS A to C tends to enhance the incompressibility, whereas the increasingly negative $K_{\rm{sat,4}}$ acts in the opposite direction. Their competition suggests that higher-order isospin contributions may substantially reduce the incompressibility of highly ANM at its isospin dependent saturation density. A more systematic investigation, including correlations among the relevant EOS parameters, is left for future studies.

\begin{figure}[h!]
    \centering
    \includegraphics[width=0.9\columnwidth]{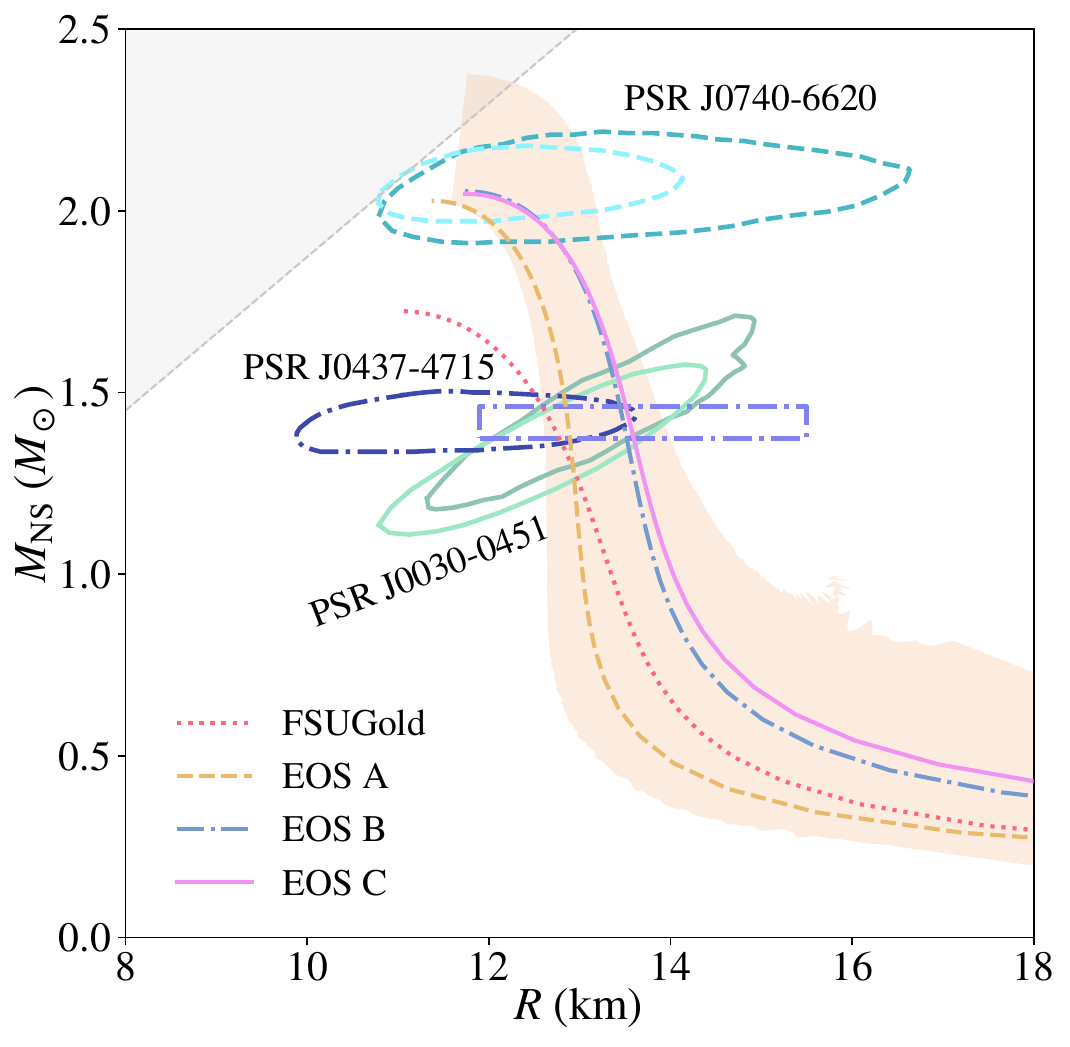}
    \caption{(Color online). M-R relations of NSs predicted by FSUGold and EOSs A, B, and C, together with observational constraints from PSR J0740-6620, PSR J0437-4715, and PSR J0030-0451. The shaded region represents the 95\% ensemble band obtained by incorporating all mutually consistent observational constraints. The gray dashed line in the upper left corner denotes the effective causality boundary $R/\rm{km}\gtrsim4.73\,M_{\rm{NS}}/M_\odot+1.14$ derived within the IPAD-TOV framework\,\cite{CaiLiZhang2023ApJ,CaiLiZhang2023PRD}, with the gray shaded region above the line disfavored by this constraint.}
    \label{fig:MR_bands}
\end{figure}

We next investigate the M-R relations of NSs by solving the Tolman-Oppenheimer-Volkoff (TOV) equations\,\cite{Tolman1939,Oppenheimer1939,Misner1973},
\begin{align}
\frac{\d P}{\d r}
=&-\frac{\varepsilon(r)M(r)}{r^2}
\left(1+\frac{P(r)}{\varepsilon(r)}\right)\notag\\
&\times\left(1+\frac{4\pi r^3P(r)}{M(r)}\right)
\left(1-\frac{2M(r)}{r}\right)^{-1},\\
\frac{\d M}{\d r}
=&4\pi r^2\varepsilon(r),
\end{align}
where $c=G=1$, and $P(r)$, $\varepsilon(r)$, and $M(r)$ denote the pressure, energy density, and enclosed gravitational mass at radius $r$, respectively. For a given EOS $P(\varepsilon)$, these equations are integrated from the stellar center, with $M(0)=0$ and $P(0)=P_{\rm c}$, to the surface defined by $P(R)=0$, yielding the NS $M_{\rm{NS}}\equiv M(R)$ and radius $R$. 
FIG.\,\ref{fig:MR_bands} shows the resulting M-R relations for FSUGold and EOSs A, B, and C, together with several observational constraints. The predicted relations exhibit appreciable model dependence, particularly in the maximum mass and the radii of NSs with intermediate masses. Among the four models, FSUGold and EOS A predict relatively low maximum masses, whereas EOSs B and C support more massive NSs, with maximum masses slightly above $2M_\odot$. These differences reflect variations in the underlying EOSs and their density dependence.

\renewcommand*\figurename{\small FIG.}

It is important to emphasize that the sizable effects of the fourth-order symmetry energy $E_{\rm{sym,4}}(\rho)$ on the proton fraction and core-crust transition do not necessarily imply comparably large effects on the global structure of NSs. The proton fraction and core-crust transition are particularly sensitive to derivatives of the EOS with respect to the isospin asymmetry, which can enhance the contributions from higher-order symmetry energies. In contrast, the M-R relation is governed by the pressure and energy density over a broad density range, with the behavior of the EOS at supra-saturation densities playing a particularly important role. Specifically, the stiffness of SNM, characterized by $E_0(\rho)$, and the density dependence of $E_{\rm{sym}}(\rho)$ can strongly influence the resulting stellar masses and radii. Although $E_{\rm{sym,4}}(\rho)$ also contributes to the EOS of neutron-rich matter, its effects on the M-R relation cannot be readily disentangled from those of the lower-order contributions. Consequently, the differences among the M-R curves in FIG.\,\ref{fig:MR_bands} should be attributed to the combined effects of the underlying EOSs, rather than solely to the fourth-order symmetry energy.

\renewcommand*\figurename{\small FIG.}
\begin{figure}[h!]
    \centering
    \includegraphics[width=0.9\columnwidth]{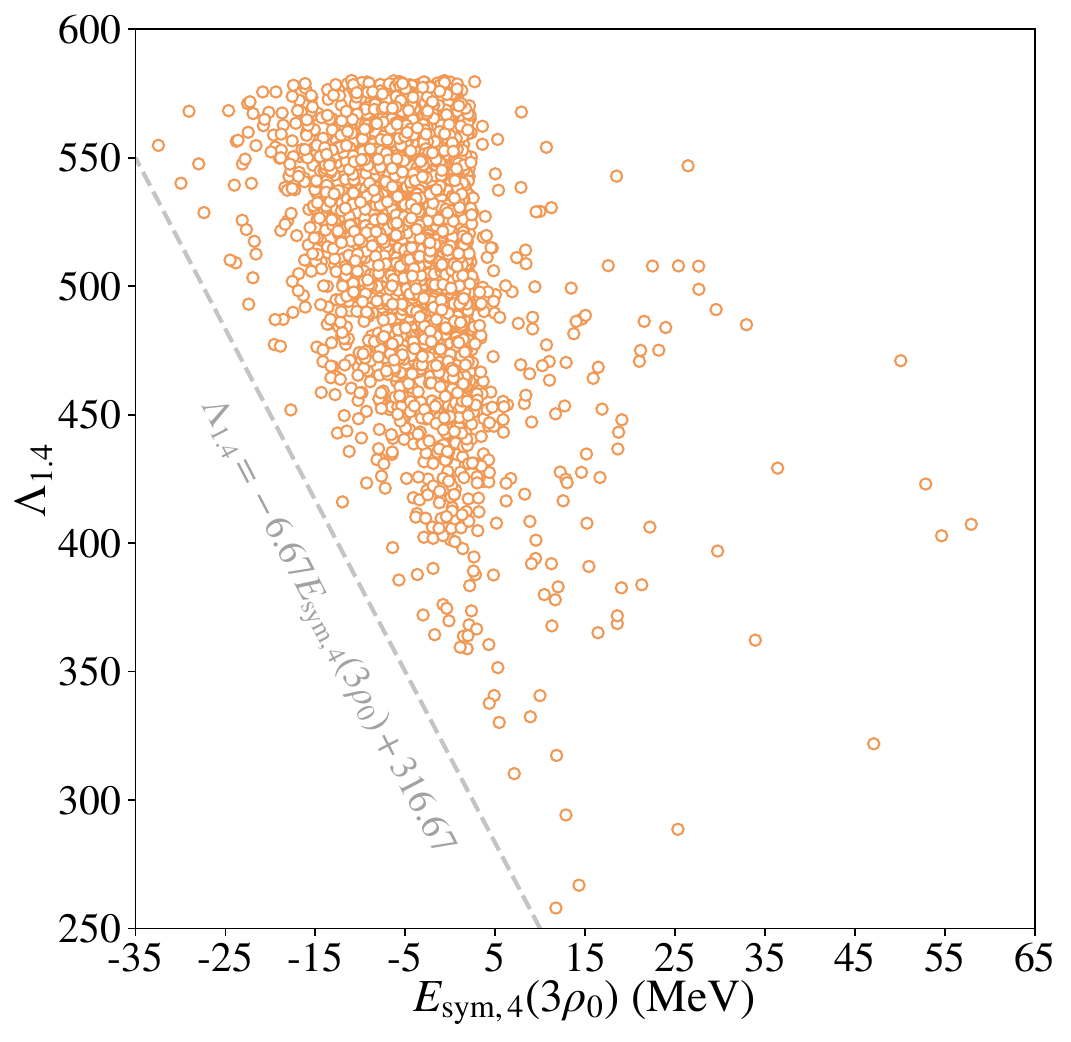}
    \caption{(Color online). Scatter plot of the dimensionless tidal deformability $\Lambda_{1.4}$ of a canonical $1.4M_\odot$ NS and the fourth-order symmetry energy $E_{\rm{sym,4}}(3\rho_0)$ at three times the nuclear saturation density. The open circles represent the sampled EOSs, while the dashed line indicates the approximate lower envelope, $\Lambda_{1.4}\approx-6.67E_{\rm{sym,4}}(3\rho_0)+316.67$, with $E_{\rm{sym,4}}(3\rho_0)$ expressed in MeV.}
    \label{fig:Lambda}
\end{figure}

We can also explore the possible connection between the fourth-order symmetry energy and the tidal deformability of NSs. The dimensionless tidal deformability is defined as\,\cite{Hinderer2008}
\begin{equation}
\Lambda=\frac{2}{3}k_2\left(\frac{R}{M_{\rm{NS}}}\right)^5
=\frac{2k_2}{3\xi^5},
\end{equation}
where $k_2$ is the quadrupolar tidal Love number and $\xi=M_{\rm{NS}}/R$ is the NS compactness\,\cite{CaiLi2025Trace}. The Love number $k_2$ is determined by the stellar structure and the underlying EOS through the linear tidal perturbation equations. Thus, $\Lambda$ depends sensitively on the stellar radius and the internal distribution of matter\,\cite{Shi2026}.

FIG.\,\ref{fig:Lambda} shows $\Lambda_{1.4}$ as a function of $E_{\rm{sym,4}}(3\rho_0)$ for the sampled EOSs, where $\Lambda_{1.4}$ denotes the tidal deformability of a canonical NS. Here, $3\rho_0$ is chosen as a representative supra-saturation density, without any particular physical significance. Similar weak correlations are also observed when $E_{\rm{sym,4}}(\rho)$ is evaluated at other densities, indicating that the weak correlation is not specific to the choice of $3\rho_0$. As a higher-order quantity in the expansion of the EOS with respect to the isospin asymmetry, $E_{\rm{sym,4}}(\rho)$ is generally less directly constrained by the tidal deformability than the leading contributions from SNM and the symmetry energy. Indeed, the broad scatter in the figure indicates a weak correlation between $\Lambda_{1.4}$ and $E_{\rm{sym,4}}(3\rho_0)$. Nevertheless, the sampled EOSs exhibit an approximate lower envelope, $
\Lambda_{1.4}\gtrsim
6.67E_{\rm{sym,4}}(3\rho_0)+316.67$,
where $E_{\rm{sym,4}}(3\rho_0)$ is expressed in MeV. 
This empirical relation suggests that tidal deformability measurements may still provide indirect constraints on the fourth-order symmetry energy. For example, the approximate boundary corresponds to $\Lambda_{1.4}\gtrsim317$ for $E_{\rm{sym,4}}(3\rho_0)\approx0$\,MeV and $\Lambda_{1.4}\gtrsim250$ for $E_{\rm{sym,4}}(3\rho_0)\approx10$\,MeV. Conversely, an observational upper bound on $\Lambda_{1.4}$ could translate into a lower limit on $E_{\rm{sym,4}}(3\rho_0)$ within the sampled EOS ensemble. For example, adopting the constraint $\Lambda_{1.4}\lesssim580$ from GW170817, the empirical boundary yields $E_{\rm{sym,4}}(3\rho_0)\gtrsim-39.5$\,MeV, indicating that the fourth-order symmetry energy cannot become arbitrarily negative at this density. Although this lower bound is relatively weak, it illustrates the potential of tidal deformability measurements to constrain higher-order symmetry energies. Moreover, the constraint remains model dependent, since the approximate lower boundary is inferred from the sampled EOS ensemble rather than established as a universal relation. Its robustness should therefore be examined using broader EOS ensembles and updated multi-messenger observations. Future measurements of tidal deformabilities, combined with independent constraints on the density dependence of the symmetry energy, may help disentangle the contributions from higher-order symmetry energies and improve our understanding of dense neutron-rich matter.

\begin{figure}[h!]
    \centering
    \includegraphics[width=0.9\columnwidth]{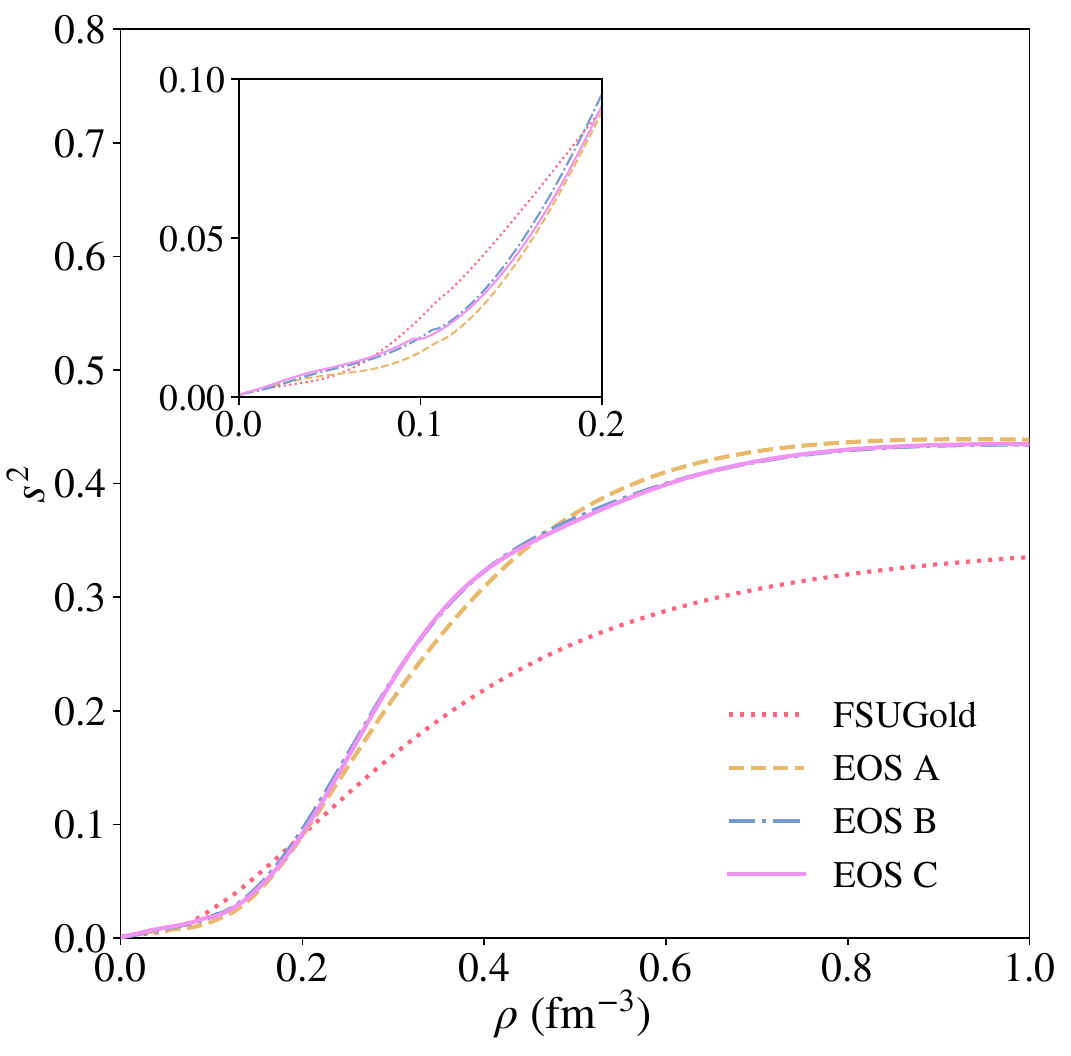}
    \caption{(Color online). Squared speed of sound $s^2$ as a function of baryon density $\rho$ for FSUGold and the three representative EOSs A, B, and C. The inset magnifies the low-density region, highlighting the differences among the models at subsaturation densities.}
    \label{fig:s2}
\end{figure}

Finally, we examine the squared speed of sound (SSS), $s^2=\d P/\d\varepsilon$, which characterizes the response of dense matter to compression and provides an important measure of the stiffness of the EOS. Its density dependence is particularly relevant for NS properties, while causality and thermodynamic stability require $0\leq s^2\leq1$\,\cite{CaiLi2025Review}. FIG.\,\ref{fig:s2} shows the density dependence of $s^2$ for FSUGold and the three representative EOSs A, B, and C. All four models satisfy these requirements over the density range considered. For EOSs A, B, and C, $s^2$ increases rapidly at intermediate densities, particularly around $\rho\approx0.2$-$0.5\,\rm{fm}^{-3}$, before gradually flattening to $s^2\approx0.45$ at higher densities. In contrast, FSUGold exhibits a more gradual increase, reaching only $s^2\approx0.34$ at $\rho\approx1\,\rm{fm}^{-3}$.

\renewcommand*\figurename{\small FIG.}

The inset of FIG.\,\ref{fig:s2} highlights the low-density region, where appreciable relative differences emerge despite the small absolute values of $s^2$. At $\rho\approx0.1\,\rm{fm}^{-3}$, FSUGold predicts $s^2\approx0.0251$, compared with approximately $0.0143$, $0.0188$, and $0.0183$ for EOSs A, B, and C, respectively, corresponding to reductions of about $43\%$, $25\%$, and $27\%$ relative to FSUGold. More importantly, EOSs A, B, and C share the same conventional saturation properties but differ substantially in $E_{\rm{sym,4}}(\rho_0)$, taking values of $2$, $4$, and $6$\,MeV, respectively. Their distinct sound speeds at subsaturation densities suggest that $E_{\rm{sym,4}}(\rho_0)$ can play an important role in determining the low-density response of NS matter to compression, beyond the constraints provided by conventional nucleonic matter properties. It should be noted, however, that differences in other higher-order EOS parameters, such as the symmetry energy curvature $K_{\rm{sym}}$, may also contribute to these variations in the sound speed. In contrast, the sound speeds predicted by these three EOSs become remarkably similar at higher densities, despite their substantial differences in the fourth-order symmetry energy and its density derivatives at $\rho_0$. This suggests that the sensitivity of $s^2$ to variations associated with $E_{\rm{sym,4}}(\rho_0)$ is more pronounced at low densities, whereas the combined contributions from different components of the EOS may reduce this sensitivity at higher densities. Nevertheless, the convergence of the sound speeds does not necessarily imply that the fourth-order symmetry energy becomes unimportant in NS cores. Rather, it raises an interesting question concerning how its density dependence contributes to the compressional response of highly ANM.

To explore this connection, we consider the incompressibility of ANM at arbitrary densities, defined as\,\cite{Perego2022}
\begin{equation}
K_{\rm{NM}}(\rho,\delta)\equiv\left.9\rho^2\frac{\partial^2E(\rho,\delta)}{\partial\rho^2}\right|_{\delta},
\end{equation}
where the derivative is taken at fixed $\delta$. It differs from the previously defined $K_{\rm{sat}}(\delta)$ in that the latter is evaluated specifically at the isospin-dependent saturation density, namely $K_{\rm{sat}}(\delta)=K_{\rm{NM}}(\rho_{\rm{sat}}(\delta),\delta)$. Unlike the SSS, which characterizes the response along the $\beta$-equilibrium trajectory, $K_{\rm{NM}}$ directly measures the density curvature of the nuclear EOS at fixed composition. Under $\beta$ equilibrium and charge neutrality, the thermodynamic relations imply $\d\varepsilon/\d\rho=\mu_{\rm{n}}=(\varepsilon+P)/\rho$, where $\mu_{\rm{n}}$ is the neutron chemical potential. Combining this relation with $s^2=\d P/\d\varepsilon$, we obtain
\begin{align}
K_{\rm{NM}}=&9\mu_{\rm{n}}\left(s^2-\frac{2\phi}{1+\phi}\right)
+\frac{18P_\ell}{\rho}-9\frac{\d P_\ell}{\d\rho}\notag\\
&-9\rho^2\frac{\partial^2E(\rho,\delta)}{\partial\rho\partial\delta}
\frac{\d\delta}{\d\rho},
\end{align}
where $\phi\equiv P/\varepsilon$ is EOS-parameter of the whole system. The last three terms account for the lepton contributions and the variation of the isospin asymmetry along the $\beta$-equilibrium trajectory. In NS cores, the density dependence of $\delta$ may become relatively weak (i.e., it flattens), suppressing the contribution proportional to $\d\delta/\d\rho$. Moreover, the lepton pressure is generally expected to be smaller than the nucleonic contribution, with its density derivative also expected to provide a subleading correction. Neglecting these terms as a first approximation, we arrive at the simple relation\,\cite{CaiLi2025Review}
\begin{equation}
K_{\rm{NM}}\approx9\mu_{\rm{n}}\left(s^2-\frac{2\phi}{1+\phi}\right).
\end{equation}

This relation provides an intuitive connection between the density curvature of the nuclear EOS and the SSS of NS matter. Expanding the incompressibility in powers of the isospin asymmetry yields $
K_{\rm{NM}}(\rho,\delta)\approx K_0(\rho)+K_{\rm{sym}}(\rho)\delta^2
+K_{\rm{sym,4}}(\rho)\delta^4+\mathcal{O}(\delta^6)$,
where $K_{\rm{sym,4}}(\rho)$ denotes the density dependent curvature of the fourth-order symmetry energy. Combining these expressions, we obtain
\begin{equation}
s^2-\frac{2\phi}{1+\phi}\approx\frac{1}{9\mu_{\rm n}}\left(
K_0(\rho)+K_{\rm{sym}}(\rho)\delta^2
+K_{\rm{sym,4}}(\rho)\delta^4+\cdots\right).
\end{equation}
This expression reveals that the fourth-order symmetry energy can influence the SSS through its density curvature, rather than merely through its magnitude. Since NS cores are highly neutron-rich, the $\delta^4$ contribution need not be negligible. In particular, a sufficiently negative $K_{\rm{sym,4}}(\rho)$ at core densities can reduce $K_{\rm{NM}}$, thereby lowering $s^2-2\phi/(1+\phi)$. If the combined contributions drive $K_{\rm{NM}}<0$, the approximate relation yields $s^2<2\phi/(1+\phi)$. A negative fourth-order symmetry energy curvature may therefore provide a mechanism for suppressing the SSS in dense ANM.
It is worth emphasizing that $K_{\rm{NM}}<0$ does not necessarily indicate mechanical instability, since $K_{\rm{NM}}$ characterizes the energy curvature at fixed composition rather than the total pressure response along the $\beta$-equilibrium trajectory.

This connection becomes particularly interesting when the EOS-parameter $\phi$ approaches the conformal value of $1/3$\,\cite{CollinsPerry1975,Bjorken1983}, as predicted for an ultra-relativistic free Fermi gas. Writing $\phi=1/3+\delta\phi$, with $|\delta\phi|\ll1$, and assuming that $\phi$ approaches $1/3$ from below, as may occur in NS cores, we have $\delta\phi<0$. Consequently, if $K_{\rm{NM}}<0$, the approximate thermodynamic relation yields
\begin{equation}
s^2\lesssim
\frac{1}{2}\left(1+\frac{9}{4}\delta\phi\right)<\frac12.
\end{equation}
This offers a simple physical interpretation of how a sufficiently negative nuclear incompressibility can be associated with an SSS below $1/2$\,\cite{Saes22,Saes24,Marc24}, consistent with the high density behavior of four EOSs shown in FIG.\,\ref{fig:s2}. Importantly, $\phi\approx1/3$ does not equivalently imply $s^2\approx1/3$, since the former characterizes the EOS-parameter, whereas the latter describes the differential response of pressure to energy density.

The same thermodynamic relation also provides insight into $K_{\rm{sym,4}}$ from the low-density limit. At sufficiently low densities, nucleonic matter approaches the non-relativistic regime, for which the kinetic contribution gives $P\sim\rho^{5/3}$ and $\varepsilon\approx M_{\rm N}\rho$, leading to $s^2\approx5\phi/3$ with $\phi\ll1$. It then follows that $s^2<2\phi/(1+\phi)$ and thus $K_{\rm{NM}}<0$. For highly asymmetric matter with $\delta\approx1$, expanding the isoscalar, quadratic, and fourth-order symmetry energies around saturation density and retaining terms through their skewness parameters gives the approximate relation
\begin{equation}
K_{\rm{sym,4}}
\lesssim
-K_0-K_{\rm{sym}}
+3^{-1}({J_0+J_{\rm{sym}}+J_{\rm{sym,4}}}),
\end{equation}
here $J_{\rm{sym,4}}$ is the skewness coefficient of $E_{\rm{sym},4}(\rho)$ at $\rho_0$.
This relation is satisfied by all three representative EOSs considered here. Interestingly, the corresponding values of $K_{\rm{sym,4}}$ can be sizable rather than negligibly small, indicating that the fourth-order symmetry energy can contribute significantly to the incompressibility of highly asymmetric matter.
Similarly, if $s^2>2\phi/(1+\phi)$ at nuclear saturation density, one obtains the approximate constraint $K_{\rm{sym,4}}\gtrsim-K_0-K_{\rm{sym}}$ for $\delta^2\approx\delta^4\approx1$.

The above analysis is intended to provide physical intuition rather than a precise constraint on the SSS. Although EOS C exhibits the most negative fourth-order symmetry energy curvature at saturation density among the three representative EOSs, the magnitude and even the sign of $K_{\rm{sym,4}}(\rho)$ at NS core densities remain uncertain. Moreover, the conventional symmetry energy curvature $K_{\rm{sym}}(\rho)$ contributes through the $\delta^2$ term and may compete with or reinforce the fourth-order contribution. The isoscalar curvature, higher-order isospin terms, lepton effects, and variations in the equilibrium composition can also affect the SSS. These contributions may partially compensate one another, helping to explain why different EOSs can yield remarkably similar sound speeds at high densities. Nevertheless, the approximate thermodynamic connection derived above highlights a potentially important role of the fourth-order symmetry energy curvature in shaping the compressional response of NS matter and motivates further investigations of its density dependence at supra-saturation densities.

\section{Summary and Outlook}\label{SEC5}

In this work, we have derived the analytical expression for the fourth-order nuclear symmetry energy $E_{\rm{sym,4}}(\rho)$ within the nonlinear Walecka model, incorporating the relevant meson interactions and cross couplings. By constructing 3000 EOSs constrained by terrestrial nuclear experimental data and astrophysical observations, we find that $E_{\rm{sym,4}}(\rho_0)$ can reach approximately $6$\,MeV or even higher, considerably exceeding the conventionally expected upper value of about $2$\,MeV. Nevertheless, its magnitude remains generally below about $20$\,MeV over the density range relevant to NSs, while its ratio to the conventional symmetry energy can reach approximately $25\%$ in magnitude. These findings suggest that the fourth-order symmetry energy can be appreciable, although its magnitude is substantially constrained by existing nuclear and astrophysical observations. Interestingly, EOSs with sizable $E_{\rm{sym,4}}(\rho_0)$ can exhibit positive curvature parameters $K_{\rm{sym}}$ of the  symmetry energy while remaining consistent with the tidal deformability constraint from GW170817. This suggests that accounting for higher-order isospin contributions may relax the preference for a negative $K_{\rm{sym}}$ found in previous analyses.

The analytical expressions further reveal that the scalar $\sigma$ and $\delta$ mesons, together with the $\sigma$-$\rho$ cross coupling governed by the $\Lambda_{\rm S}$ parameter, can generate negative contributions to $E_{\rm{sym,4}}(\rho)$. In particular, the $\Lambda_{\rm S}$ coupling can drive $E_{\rm{sym,4}}(\rho)$ toward substantially more negative values at high densities. Although the magnitude of $E_{\rm{sym,4}}(\rho)$ itself is relatively limited, its density derivatives, especially the curvature parameter $K_{\rm{sym,4}}$, can be sizable and significantly affect the fourth-order isospin incompressibility coefficient $K_{\rm{sat,4}}$.
We have also examined the implications of $E_{\rm{sym,4}}(\rho)$ for NS properties. Its appreciable effects on the proton fraction and core-crust transition are clarified through the different effective symmetry energies governing the EOS, beta equilibrium, and thermodynamic stability. Moreover, variations in $E_{\rm{sym,4}}(\rho_0)$ are associated with changes in the low-density squared speed of sound reaching approximately $40\%$ or more, whereas their effects on the global NS M-R relations and tidal deformabilities are comparatively modest.

The potential importance of these contributions can also be understood from the Taylor expansion of $E(\rho,\delta)$ around SNM at saturation density. Treating $\delta=(\rho_{\rm n}-\rho_{\rm p})/(\rho_{\rm n}+\rho_{\rm p})$ and $\chi=(\rho-\rho_0)/3\rho_0$ as two small expansion parameters of the same order, one can group terms proportional to $\delta^i\chi^j$ according to their total order $i+j$\,\cite{Cai2017}. This provides a useful way to organize the expansion near $\rho_0$ and for sufficiently small isospin asymmetries. For example, the terms governed by $K_0$ and $J_0$ are of second and third order, respectively, while those governed by $K_{\rm{sym}}$ and $E_{\rm{sym,4}}(\rho_0)$ are both of fourth order: the former multiplies $\delta^2\chi^2$, whereas the latter multiplies $\delta^4$. Thus, retaining the symmetry-energy curvature while neglecting the fourth-order symmetry energy at saturation omits a contribution of the same total order. This comparison suggests that $E_{\rm{sym,4}}(\rho_0)$ deserves attention alongside the density dependence of the conventional symmetry energy. Similarly, $K_{\rm{sym,4}}$ multiplies $\delta^4\chi^2$ and enters at sixth order, but its contribution can become more prominent in quantities involving density derivatives, such as the incompressibility. This ordering provides qualitative guidance near $\rho_0$, rather than a quantitative hierarchy of physical effects. The actual importance of each term depends on its coefficient, the density and isospin asymmetry, and the observable considered.

Several important questions remain open. While the relatively small fourth-order symmetry energy at low densities can be understood from the dominance of its kinetic contribution, the microscopic mechanisms capable of generating a sizable $E_{\rm{sym,4}}(\rho)$ at higher densities deserve further investigation. In particular, the possible roles of nucleon short-range correlations\,\cite{Cai2016,Hen15PRC,Cai15PRC,Hen17RMP,Cai2026EPJST} and other many-body effects\,\cite{Benhar93,Polls1995PPNP,Pandharipande97,Soma2006PRC} beyond the mean field approximation, as well as the connection between higher-order symmetry energies and the EOS of PNM, remain to be clarified. It would also be valuable to investigate $E_{\rm{sym,4}}(\rho)$ within other theoretical EOS frameworks to assess the model dependence and robustness of the present findings. Moreover, our results indicate that the density derivatives of $E_{\rm{sym,4}}(\rho)$ can be substantial even when its magnitude is relatively small, raising the question of whether observables sensitive to these derivatives could provide more effective constraints. Promising directions include the core-crust transition, the low-density sound speed, and the onset of direct Urca processes and associated NS cooling behavior. Identifying observables that can disentangle the fourth-order symmetry energy from lower-order EOS contributions will be essential for establishing more direct constraints on its magnitude and density dependence.

\section*{Acknowledgment} 
This work was supported in part by the National Natural Science Foundation of China under contract Nos. 12688101 and 12547102, and the STCSM under Grant Nos.~23590780100 and 23JC1400200. 

\appendix
\renewcommand{\theequation}{\Alph{section}.\arabic{equation}}
\setcounter{equation}{0}

\section{Detailed Derivation on the Fourth-order Symmetry Energy}
\label{app:E4}

At fixed baryon density $\rho$, we introduce the scaled mean fields
\begin{align}
    S_\sigma&=g_\sigma\overline{\sigma},~~
V_\omega=g_\omega\overline{\omega}_0,
~~
    S_\delta=g_\delta\overline{\delta}^{(3)},~~
V_\rho=g_\rho\overline{\rho}_0^{(3)},
\end{align}
and 
\begin{equation}
a_\phi=m_\phi^2/g_\phi^2,
\text{ with }\phi=\sigma,\omega,\delta,\rho.
\end{equation}
Under neutron-proton exchange, the isoscalar fields are even
functions of $\delta$, whereas the isovector fields are odd.
Their expansions around SNM therefore read
\begin{align}
    S_\sigma
    &\approx S_\sigma^{[0]}+S_\sigma^{[2]}\delta^2
    +S_\sigma^{[4]}\delta^4
      +\mathcal{O}(\delta^6),
    \notag\\
    V_\omega
    &\approx V_\omega^{[0]}+V_\omega^{[2]}\delta^2
    +V_\omega^{[4]}\delta^4
      +\mathcal{O}(\delta^6),
    \notag\\
    S_\delta
    &\approx S_\delta^{[1]}\delta+S_\delta^{[3]}\delta^3
    +\mathcal{O}(\delta^5),
    \notag\\
    V_\rho
    &\approx V_\rho^{[1]}\delta+V_\rho^{[3]}\delta^3
    +\mathcal{O}(\delta^5).
    \label{eq:app_field_expansions}
\end{align}
Here, the superscript (or sometimes subscript) $[\ell]$ denotes the expansion coefficient at $\delta=0$,
including the factor $1/\ell!$.
The SNM effective mass and Fermi energy are
\begin{equation}
    M_0^*=M_{\rm N}-S_\sigma^{[0]},
~~
    \varepsilon_{\rm F}^*
    =\sqrt{k_{\rm F}^2+M_0^{*,2}}.
\end{equation}
The subscript ``0'' on $M_0^*$ denotes SNM at the density $\rho$.
Although the mean fields are expanded to fourth order in the isospin asymmetry, the final expression for $E_{\mathrm{sym},4}(\rho)$ requires only the first-order isovector and second-order isoscalar field coefficients. As demonstrated below, contributions involving higher-order field coefficients vanish as a consequence of the mean-field equations.

Starting from the kinetic energy density,
\begin{equation}\label{def-epskin}
 \varepsilon_J^{\rm{kin}}= \frac{2}{(2\pi)^3} \int_0^{k_{\rm{F}}^J} \d\v{k} \sqrt{k^2+M_J^{\ast,2}},~~k_{\rm{F}}^J=(3\pi^2 \rho_J)^{1/3},
\end{equation}
the total scalar density in SNM is
\begin{align}\label{def:rhos0}
    \rho_{\rm s}^{[0]}
    \notag&= \left[\frac{\partial \varepsilon_{\rm{n}}^{\rm{kin}}}{\partial M_{\rm{n}}^\ast} +\frac{\partial \varepsilon_{\rm{p}}^{\rm{kin}}}{\partial M_{\rm{p}}^\ast}\right]_{\delta=0}
    =2\frac{\partial\varepsilon_{[0]}^{\rm{kin}}}{\partial M_0^\ast}
    \\
    &=
    \frac{M_0^*}{\pi^2}
    \left[
    k_{\rm F}\varepsilon_{\rm F}^*
    -M_0^{*,2}
    \ln\left(
    \frac{k_{\rm F}+\varepsilon_{\rm F}^*}{M_0^*}
    \right)
    \right],
\end{align}
where $\varepsilon_{[0]}^{\rm{kin}}$ is the kinetic energy density (either for neutrons or protons) in SNM.
The derivative of the total scalar density with respect to the effective mass $M_0^\ast$ at fixed Fermi momentum $k_{\rm{F}}$ defines the function $A$: 
\begin{align}
    A \equiv 2\frac{\partial^2 \varepsilon_{[0]}^{\rm{kin}}}{\partial M_0^{\ast,2}}
    \equiv
    \left.
    \frac{\partial\rho_{\rm s}^{[0]}}
    {\partial M_0^*}
    \right|_{k_{\rm F}}
    = 3 \left( \frac{\rho^{[0]}_\rm s}{M_0^*} 
    -\frac{\rho}{\varepsilon^*_\rm F} \right).
    \label{eq:app_A}
\end{align}
The scalar density and baryon density are given by
\begin{align}
\rho_{\rm s}^{[0]}=&\frac{4}{(2\pi)^3}
\int_{|\v{k}|\leq k_{\rm F}}\d\v{k}
\frac{M_0^\ast}{\sqrt{k^2+M_0^{\ast,2}}},\\
\rho=&\frac{4}{(2\pi)^3}
\int_{|\v{k}|\leq k_{\rm F}}\d\v{k}
=\frac{2k_{\rm F}^3}{3\pi^2},
\end{align}
respectively.
Differentiating the scalar density at fixed $k_{\rm F}$ and using $\d\v{k}=4\pi k^2\d k$ for the angular integration, we obtain
\begin{align}
\left.
\frac{\partial\rho_{\rm s}^{[0]}}{\partial M_0^\ast}
\right|_{k_{\rm F}}
=&\frac{4}{(2\pi)^3}
\int_{|\v{k}|\leq k_{\rm F}}\d\v{k}
\frac{k^2}{(k^2+M_0^{\ast,2})^{3/2}}\notag\\
=&\frac{4}{(2\pi)^3}4\pi
\int_0^{k_{\rm F}}\d k
\frac{k^4}{(k^2+M_0^{\ast,2})^{3/2}}.
\end{align}
Using integration by parts, we have
\begin{align}
\int_0^{k_{\rm F}}\d k
\frac{k^4}{(k^2+M_0^{\ast,2})^{3/2}}
=&3\int_0^{k_{\rm F}}\d k
\frac{k^2}{\sqrt{k^2+M_0^{\ast,2}}}
-\frac{k_{\rm F}^3}{\varepsilon_{\rm F}^\ast}.
\end{align}
Consequently,
\begin{align}
\left.
\frac{\partial\rho_{\rm s}^{[0]}}{\partial M_0^\ast}
\right|_{k_{\rm F}}
=&\frac{4}{(2\pi)^3}4\pi
\left[
3\int_0^{k_{\rm F}}\d k
\frac{k^2}{\sqrt{k^2+M_0^{\ast,2}}}
-\frac{k_{\rm F}^3}{\varepsilon_{\rm F}^\ast}
\right]\notag\\
=&3\left(
\frac{\rho_{\rm s}^{[0]}}{M_0^\ast}
-\frac{\rho}{\varepsilon_{\rm F}^\ast}
\right),
\end{align}
where the last equality follows from the definitions of $\rho_{\rm s}^{[0]}$ and $\rho$, thus recovering Eq.\,(\ref{eq:app_A}); see also Eq.\,(\ref{def:pp}).

The first-order and second-order derivatives of $A$ with respect to the effective mass $M_0^*$ at fixed $k_{\rm{F}}$ are similarily given as:
\begin{align}
    \left.\frac{\partial A}{\partial M_0^*}\right|_{k_\rm F}
    &=
    -\frac{6M_0^*}{\pi^2}
    \left[
    \ln\left(
    \frac{k_{\rm F}+\varepsilon_{\rm F}^*}{M_0^*}
    \right)
    -\frac{k_{\rm F}}{\varepsilon_{\rm F}^*}
    -\frac{k_{\rm F}^3}{3\varepsilon_{\rm F}^{*,3}}
    \right],
    \label{eq:app_AM}\\
    \left.\frac{\partial^2 A}{\partial M_0^{*,2}}\right|_{k_\rm F}
    &=
    \frac{6}{\pi^2}
    \left[
    -\ln\left(
    \frac{k_{\rm F}+\varepsilon_{\rm F}^*}{M_0^*}
    \right)
    +\frac{k_{\rm F}}{\varepsilon_{\rm F}^*}
    +\frac{k_{\rm F}^3}{3\varepsilon_{\rm F}^{*,3}}
    +\frac{k_{\rm F}^5}{\varepsilon_{\rm F}^{*,5}}
    \right].
    \label{eq:app_AMM}
\end{align}
The non-relativistic expansions of $A$ and its first two derivatives at fixed $k_{\rm{F}}$ are given by
\begin{align}
A\approx&\frac{2k_{\rm F}^5}{5\pi^2M_0^{*,3}}
\left[
1-\frac{15k_{\rm F}^2}{14M_0^{*,2}}
+\mathcal{O}\left(\left({k_{\rm F}}/{M_0^{*}}\right)^4\right)
\right],\label{apxA}\\
\left.\frac{\partial A}{\partial M_0^*}\right|_{k_{\rm F}}
\approx&-\frac{6k_{\rm F}^5}{5\pi^2M_0^{*,4}}
\left[
1-\frac{25k_{\rm F}^2}{14M_0^{*,2}}
+\mathcal{O}\left(\left({k_{\rm F}}/{M_0^{*}}\right)^4\right)
\right],\label{apxA1}\\
\left.\frac{\partial^2 A}{\partial M_0^{*,2}}\right|_{k_{\rm F}}
\approx&\frac{24k_{\rm F}^5}{5\pi^2M_0^{*,5}}
\left[
1-\frac{75k_{\rm F}^2}{28M_0^{*,2}}+
\mathcal{O}\left(\left({k_{\rm F}}/{M_0^{*}}\right)^4\right)
\right].\label{apxA2}
\end{align}
The leading-order non-relativistic expressions for these functions are obtained by further setting $M_0^{\ast}\approx M_{\rm{N}}$.

To derive the fourth-order expansion in $\delta$, we first
collect the partial derivatives of $\varepsilon_J^{\rm{kin}}(\rho_J,M_J^\ast)$ up to fourth order.
All derivatives in the following expressions are finally evaluated
at $\delta=0$ (SNM), where the density of each nucleon
species is $\rho/2$ and its effective mass is $M_0^*$.

The derivatives with respect to the effective mass at fixed
$\rho_J$ (or equivalently, fixed $k_{\rm F}^J$) are already
determined by the quantities given above: The first-order
derivative is $\rho_{\rm s}^{[0]}/2$, the second-order derivative
is $A/2$, and the third- and fourth-order derivatives are
$(\partial A/\partial M_0^*)_{k_\rm F}/2$
and
$(\partial^2 A/\partial M_0^{*,2})_{k_\rm F}/2$,
respectively.
We next calculate the derivatives with respect to $\rho_J$
at fixed effective mass $M_J^*$.
Partially differentiating the kinetic energy density $\varepsilon_J^{\rm{kin}}(\rho_J,M_J^\ast)$ given in Eq.\,(\ref{def-epskin}) with respect to $\rho_J$ at fixed effective mass $M_J^\ast$, we obtain the Fermi energy,
\begin{align}
\left.
\frac{\partial\varepsilon_J^{\rm{kin}}}{\partial\rho_J}
\right|_{M_J^\ast}
=&
\frac{\partial\varepsilon_J^{\rm{kin}}/\partial k_{\rm F}^J}
{\partial\rho_J/\partial k_{\rm F}^J}
=\sqrt{k_{\rm F}^{J,2}+M_J^{\ast,2}}
\equiv\varepsilon_{\rm F}^{J,\ast}.
\end{align}
Therefore,
\begin{equation}
\left.{\partial\varepsilon_J^{\rm{kin}}}/{\partial\rho_J}
\right|_{M_J^\ast;\,\delta=0}
=\varepsilon_{\rm F}^\ast.
\end{equation}
Similarly, we may obtain:
\begin{align}
    \left.
\frac{\partial^2\varepsilon_J^{\rm{kin}}}{\partial\rho_J^2}
\right|_{M_J^\ast;\,\delta=0}
=&\frac{2k_{\rm{F}}^2}{3\rho\varepsilon_{\rm{F}}^{\ast}},\\
\left.
\frac{\partial^3\varepsilon_J^{\rm{kin}}}{\partial\rho_J^3}
\right|_{M_J^\ast;\,\delta=0}
=&-\frac{
    4k_{\rm F}^2
    (M_0^{*,2}+2k_{\rm F}^2)}
    {9\rho^2\varepsilon_{\rm F}^{*,3}}, \\
 \left.
\frac{\partial^4\varepsilon_J^{\rm{kin}}}{\partial\rho_J^4}
\right|_{M_J^\ast;\,\delta=0}
   =&
    \frac{
    8k_{\rm F}^2
    (
    4M_0^{*,4}
    +11M_0^{*,2}k_{\rm F}^2
    +10k_{\rm F}^4
    )}
    {27\rho^3\varepsilon_{\rm F}^{*,5}}.
    \label{eq:app_f_density_derivatives}
\end{align}

The mixed partial derivatives could also be obtained:
\begin{align}
     \left.
\frac{\partial^2\varepsilon_J^{\rm{kin}}}{\partial\rho_J\partial M_J^\ast}
\right|_{\delta=0}=&\frac{M_0^\ast}{\varepsilon_{\rm{F}}^\ast},\\
 \left.
\frac{\partial^3\varepsilon_J^{\rm{kin}}}{\partial\rho_J\partial M_J^{\ast,2}}
\right|_{\delta=0}=&\frac{k_{\rm F}^2}{\varepsilon_{\rm F}^{*,3}},\\
 \left.
\frac{\partial^4\varepsilon_J^{\rm{kin}}}{\partial\rho_J\partial M_J^{\ast,3}}
\right|_{\delta=0}=&-\frac{3M_0^*k_{\rm F}^2}
    {\varepsilon_{\rm F}^{*,5}},
    \label{eq:app_f_mixed_derivatives_1}
\\
    \left.
\frac{\partial^3\varepsilon_J^{\rm{kin}}}{\partial\rho_J^2\partial M_J^\ast}
\right|_{\delta=0}=&-\frac{2M_0^*k_{\rm F}^2}
    {3\rho\varepsilon_{\rm F}^{*,3}},
    \end{align}
as well as
\begin{align}
    \left.
\frac{\partial^4\varepsilon_J^{\rm{kin}}}{\partial\rho_J^2\partial M_J^{\ast,2}}
\right|_{\delta=0}=&\frac{
    2k_{\rm F}^2
    (2M_0^{*,2}-k_{\rm F}^2)}
    {3\rho\varepsilon_{\rm F}^{*,5}},\\
    \left.
\frac{\partial^4\varepsilon_J^{\rm{kin}}}{\partial\rho_J^3\partial M_J^\ast}
\right|_{\delta=0}=&\frac{
    4M_0^*k_{\rm F}^2
    (M_0^{*,2}+4k_{\rm F}^2)}
    {9\rho^2\varepsilon_{\rm F}^{*,5}}.
    \label{eq:app_f_mixed_derivatives_2}
\end{align}

For convenience, we introduce the following factors:
\begin{align}
    H_\sigma
    &=
    A+a_\sigma
    +2b_\sigma M_{\rm N}S_\sigma^{[0]}
    +3c_\sigma S_\sigma^{[0],2},
     \label{eq:app_Hsigma}\\
    H_\omega
    &=
    a_\omega+3c_\omega V_\omega^{[0],2},
     \label{eq:app_Homega}\\
    H_\delta
    &=
    a_\delta
    -C_{\delta\sigma}S_\sigma^{[0],2}+A,
     \label{eq:app_Hdelta}\\
    H_\rho
    &=
    a_\rho
    +\Lambda_{\rm S} S_\sigma^{[0],2}
    +\Lambda_{\rm V} V_\omega^{[0],2}.
    \label{eq:app_Hrho}
\end{align}
We now expand the full nucleonic kinetic energy density
to fourth order in $\delta$.
The increments in the neutron and proton densities and effective
masses relative to SNM can be expressed as $
    \Delta\rho_{\rm n}
    =\rho\delta/2$ and $
    \Delta\rho_{\rm p}
    =-\rho\delta/2$,
and
\begin{align}
    \Delta M_{\rm n}
    &\approx
    S_\delta^{[1]}\delta
    -S_\sigma^{[2]}\delta^2
    +S_\delta^{[3]}\delta^3
    -S_\sigma^{[4]}\delta^4
    +\mathcal{O}(\delta^5),
    \notag\\
    \Delta M_{\rm p}
    &\approx
    -S_\delta^{[1]}\delta
    -S_\sigma^{[2]}\delta^2
    -S_\delta^{[3]}\delta^3
    -S_\sigma^{[4]}\delta^4
    +\mathcal{O}(\delta^5),
    \label{eq:app_kinetic_increments}
\end{align}
where $\Delta M_J^\ast=M_J^*-M_0^*$ and
$\Delta\rho_J=\rho_J-\rho/2$, with $J=\mathrm{n,p}$.
Since both the nucleon densities and effective masses depend on the isospin asymmetry, the fourth-order kinetic contribution must account for their simultaneous variations. We therefore expand the kinetic energy density in both variables and retain all terms contributing to order $\delta^4$.
For nucleon $J$, the Taylor
expansion around SNM to order $\delta^4$ is 
\begin{align}\label{eq:app_kinetic_taylor}
\varepsilon_J^{\rm{kin}}
\approx{}&
\varepsilon_{[0]}^{\rm{kin}}
+\left.
\frac{\partial\varepsilon_J^{\rm{kin}}}{\partial\rho_J}
\right|_{\delta=0}\cdot\Delta\rho_J
+\left.
\frac{\partial\varepsilon_J^{\rm{kin}}}{\partial M_J^\ast}
\right|_{\delta=0}\cdot\Delta M_J^\ast
\notag\\
&+\left.
\frac{1}{2}\frac{\partial^2\varepsilon_J^{\rm{kin}}}{\partial\rho_J^2}
\right|_{\delta=0}\cdot\left(\Delta\rho_J\right)^2
+\left.
\frac{\partial^2\varepsilon_J^{\rm{kin}}}
{\partial\rho_J\partial M_J^\ast}
\right|_{\delta=0}\cdot\Delta\rho_J\Delta M_J^\ast
\notag\\
&+\left.
\frac{1}{2}\frac{\partial^2\varepsilon_J^{\rm{kin}}}{\partial M_J^{\ast,2}}
\right|_{\delta=0}\cdot\left(\Delta M_{J}^\ast\right)^{2}
\notag\\
&+\left.
\frac{1}{6}\frac{\partial^3\varepsilon_J^{\rm{kin}}}{\partial\rho_J^{3}}
\right|_{\delta=0}\cdot\left(\Delta\rho_J\right)^3
+\left.
\frac{1}{2}\frac{\partial^3\varepsilon_J^{\rm{kin}}}
{\partial\rho_J^2\partial M_J^\ast}
\right|_{\delta=0}\cdot\left(\Delta\rho_J\right)^2\Delta M_J^\ast
\notag\\
&+\left.
\frac{1}{2}\frac{\partial^3\varepsilon_J^{\rm{kin}}}
{\partial\rho_J\partial M_J^{\ast,2}}
\right|_{\delta=0}\cdot\Delta\rho_J\left(\Delta M_J^\ast\right)^2
+\left.
\frac{1}{6}\frac{\partial^3\varepsilon_J^{\rm{kin}}}{\partial M_J^{\ast,3}}
\right|_{\delta=0}\cdot\left(\Delta M_J^\ast\right)^3
\notag\\
&+\left.
\frac{1}{24}\frac{\partial^4\varepsilon_J^{\rm{kin}}}{\partial\rho_J^4}
\right|_{\delta=0}\cdot\left(\Delta\rho_J\right)^4
+\left.
\frac{1}{6}\frac{\partial^4\varepsilon_J^{\rm{kin}}}
{\partial\rho_J^3\partial M_J^\ast}
\right|_{\delta=0}\cdot\left(\Delta\rho_J\right)^3\Delta M_J^\ast
\notag\\
&+\left.
\frac{1}{4}\frac{\partial^4\varepsilon_J^{\rm{kin}}}
{\partial\rho_J^2\partial M_J^{\ast,2}}
\right|_{\delta=0}\cdot\left(\Delta\rho_J\right)^2\left(\Delta M_J^\ast\right)^2
\notag\\
&+\left.
\frac{1}{6}\frac{\partial^4\varepsilon_J^{\rm{kin}}}
{\partial\rho_J\partial M_J^{\ast,3}}
\right|_{\delta=0}\cdot\Delta\rho_J\left(\Delta M_J^\ast\right)^3
\notag\\
&+\left.
\frac{1}{24}\frac{\partial^4\varepsilon_J^{\rm{kin}}}{\partial M_J^{\ast,4}}
\right|_{\delta=0}\cdot\left(\Delta M_J^\ast\right)^4
+\mathcal{O}(\delta^5).
\end{align}

We now collect the contributions to the coefficient of $\delta^4$
from each order of the Taylor expansion, summing over the neutron
and proton species.
The linear terms contribute
\begin{equation}
    -\left.
    2\frac{\partial\varepsilon_J^{\rm{kin}}}{\partial M_J^\ast}
    \right|_{\delta=0}\cdot S_\sigma^{[4]}.
\end{equation}
The quadratic terms give
\begin{align}
    &\left.
    \rho\frac{\partial^2\varepsilon_J^{\rm{kin}}}
    {\partial\rho_J\partial M_J^\ast}
    \right|_{\delta=0}\cdot S_\delta^{[3]}
    +\left.
    \frac{\partial^2\varepsilon_J^{\rm{kin}}}{\partial M_J^{\ast,2}}
    \right|_{\delta=0}\cdot
    S_\sigma^{[2],2}
    \notag\\
    &+\left.
    2\frac{\partial^2\varepsilon_J^{\rm{kin}}}{\partial M_J^{\ast,2}}
    \right|_{\delta=0}\cdot
    S_\delta^{[1]}S_\delta^{[3]}.
\end{align}
The cubic terms can be combined into
$B_\mathrm{kin}S_\sigma^{[2]}$, where
\begin{align}
    B_\mathrm{kin}
    \equiv{}&
    -\left.
    \frac{\rho^2}{4}
    \frac{\partial^3\varepsilon_J^{\rm{kin}}}
    {\partial\rho_J^2\partial M_J^\ast}
    \right|_{\delta=0}
    -\left.
    \rho\frac{\partial^3\varepsilon_J^{\rm{kin}}}
    {\partial\rho_J\partial M_J^{\ast,2}}
    \right|_{\delta=0}\cdot S_\delta^{[1]}
    \notag\\
    &-\left.
    \frac{\partial^3\varepsilon_J^{\rm{kin}}}{\partial M_J^{\ast,3}}
    \right|_{\delta=0}\cdot
    S_\delta^{[1],2}.
    \label{eq:app_Bkin_derivatives}
\end{align}
The quartic terms contribute
\begin{align}
    &\left.
    \frac{\rho^4}{192}
    \frac{\partial^4\varepsilon_J^{\rm{kin}}}
    {\partial\rho_J^4}
    \right|_{\delta=0}
    +\left.
    \frac{\rho^3}{24}
    \frac{\partial^4\varepsilon_J^{\rm{kin}}}
    {\partial\rho_J^3\partial M_J^\ast}
    \right|_{\delta=0}\cdot S_\delta^{[1]}
    \notag\\
    &+\left.
    \frac{\rho^2}{8}
    \frac{\partial^4\varepsilon_J^{\rm{kin}}}
    {\partial\rho_J^2\partial M_J^{\ast,2}}
    \right|_{\delta=0}\cdot
    S_\delta^{[1],2}
    +\left.
    \frac{\rho}{6}
    \frac{\partial^4\varepsilon_J^{\rm{kin}}}
    {\partial\rho_J\partial M_J^{\ast,3}}
    \right|_{\delta=0}\cdot
    S_\delta^{[1],3}
    \notag\\
    &+\left.
    \frac{1}{12}
    \frac{\partial^4\varepsilon_J^{\rm{kin}}}
    {\partial M_J^{\ast,4}}
    \right|_{\delta=0}\cdot
    S_\delta^{[1],4}
    \label{eq:app_kinetic_quartic}
\end{align}

Combining these contributions and substituting the SNM derivatives
obtained above, we arrive at
\begin{align}
    \varepsilon^{\mathrm{kin}}\left[\text{at order }\delta^4\right]
    ={}&
    -\rho_{\rm s}^{[0]}S_\sigma^{[4]}
    +\frac{A}{2}S_\sigma^{[2],2}
    +B_\mathrm{kin}S_\sigma^{[2]}
    \notag\\
    &+
    \left(
    \frac{\rho M_0^*}{\varepsilon_{\rm F}^*}
    +A S_\delta^{[1]}
    \right)S_\delta^{[3]}
    \notag\\
    &+
    \frac{
    \rho k_{\rm F}^2
    (
    4M_0^{*,4}
    +11M_0^{*,2}k_{\rm F}^2
    +10k_{\rm F}^4
    )}
    {648\varepsilon_{\rm F}^{*,5}}
    \notag\\
    &+
    \frac{
    \rho M_0^*k_{\rm F}^2
    (M_0^{*,2}+4k_{\rm F}^2)}
    {54\varepsilon_{\rm F}^{*,5}}
    S_\delta^{[1]}
    \notag\\
    &+
    \frac{
    \rho k_{\rm F}^2
    (2M_0^{*,2}-k_{\rm F}^2)}
    {12\varepsilon_{\rm F}^{*,5}}
    S_\delta^{[1],2}
    \notag\\
    &-
    \frac{\rho M_0^*k_{\rm F}^2}
    {2\varepsilon_{\rm F}^{*,5}}
    S_\delta^{[1],3}
    +\left.\frac{1}{24}\frac{\partial^2 A}{\partial M_0^{\ast,2}}\right|_{k_\rm F}\cdot
    S_\delta^{[1],4},
    \label{eq:app_kinetic_fourth}
\end{align}
where
\begin{equation}
    B_\mathrm{kin}
    =
    \frac{\rho M_0^*k_{\rm F}^2}
    {6\varepsilon_{\rm F}^{*,3}}
    -
    \frac{\rho k_{\rm F}^2}
    {\varepsilon_{\rm F}^{*,3}}S_\delta^{[1]}
    -
   \left. \frac{1}{2}\frac{\partial A}{\partial M_0^*}\right|_{k_\rm F}\cdot S_\delta^{[1],2}.
    \label{eq:app_Bkin}
\end{equation}

We next expand the mesonic contributions.
For this purpose, it is convenient to use an equivalent
form of Eq.\,\eqref{eq:eps},
\begin{equation}
    \varepsilon
    =
    \varepsilon^{\mathrm{kin}}
    +\varepsilon_\delta+\varepsilon_\sigma
    +\varepsilon_\omega+\varepsilon_\rho,
\end{equation}
where $\varepsilon^{\rm{kin}}=\varepsilon_{\rm n}^{\rm{kin}}+\varepsilon_{\rm p}^{\rm{kin}}$, and 
\begin{align}
    \varepsilon_\delta
    &=
    \frac{1}{2}
    \left(a_\delta-C_{\delta\sigma}S_\sigma^2\right)
    S_\delta^2,
    \notag\\
    \varepsilon_\sigma
    &=
    \frac{1}{2}a_\sigma S_\sigma^2
    +\frac{1}{3}b_\sigma M_{\rm N}S_\sigma^3
    +\frac{1}{4}c_\sigma S_\sigma^4,
    \notag\\
    \varepsilon_\omega
    &=
    \rho V_\omega
    -\frac{1}{2}a_\omega V_\omega^2
    -\frac{1}{4}c_\omega V_\omega^4,
    \notag\\
    \varepsilon_\rho
    &=
    -\rho\delta V_\rho
    -\frac{1}{2}
    \left(
    a_\rho+\Lambda_{\rm S}S_\sigma^2
    +\Lambda_{\rm V}V_\omega^2
    \right)V_\rho^2.
    \label{eq:app_mesonic_energy}
\end{align}
Using the $\omega$ and $\rho$ meson field equations
in Eq.\,\eqref{eq:app_mesonic_energy} reproduces
Eq.\,\eqref{eq:eps}.

The $\delta$ meson contribution, including the
$\sigma$-$\delta$ mixing term, gives
\begin{align}
    \varepsilon_\delta\left[\text{at order }\delta^4\right]
    ={}&
    \left(
    a_\delta
    -C_{\delta\sigma}S_\sigma^{[0],2}
    \right)S_\delta^{[1]}S_\delta^{[3]}\notag\\
&    -
    C_{\delta\sigma}S_\sigma^{[0]}
    S_\delta^{[1],2}S_\sigma^{[2]}.
    \label{eq:app_delta_fourth}
\end{align}
Combining this with the kinetic contribution,
the coefficient of $S_\delta^{[3]}$ becomes
$\rho M_0^*/\varepsilon_{\rm F}^*+H_\delta S_\delta^{[1]}$ which vanishes by virtue of the $\delta$ meson field equation Eq.\,\eqref{eq:deqm}.
For the $\sigma$ meson contribution, the expansion gives $\rho_{\rm s}^{[0]}S_\sigma^{[4]}
+(H_\sigma-A)S_\sigma^{[2],2}/2$.
Thus, the term proportional to $S_\sigma^{[4]}$
cancels its kinetic counterpart, while the combined
coefficient of $S_\sigma^{[2],2}$
is $H_\sigma/2$.
The $\omega$ meson contribution yields $-H_\omega V_\omega^{[2],2}/2$. Here, the SNM $\sigma$ and $\omega$ meson field equations
have been used, with the latter eliminating the term
proportional to $V_\omega^{[4]}$.
Finally, the $\rho$ meson contribution, including
the $\sigma$-$\rho$ and $\omega$-$\rho$ mixing terms,
gives
\begin{align}
    \varepsilon_\rho\left[\text{at order }\delta^4\right]
    =&{}
    -\left(\rho+H_\rho V_\rho^{[1]}\right)V_\rho^{[3]}
    \notag\\
    &-
    \Lambda_{\rm S}S_\sigma^{[0]}
    V_\rho^{[1],2}S_\sigma^{[2]}
    -
    \Lambda_{\rm V}V_\omega^{[0]}
    V_\rho^{[1],2}V_\omega^{[2]}.
    \label{eq:app_rho_fourth}
\end{align}
The term proportional to $V_\rho^{[3]}$ vanishes
because of the $\rho$ meson field equation Eq.\,\eqref{eq:reqm}.
An important simplification follows from these cancellations: All contributions involving the fourth-order isoscalar and third-order isovector field coefficients vanish upon applying the mean field equations. Consequently, the fourth-order symmetry energy depends only on the first-order isovector and second-order isoscalar field responses, which we determine below.

Combining the mesonic and kinetic contributions
and using $E_{\mathrm{sym},4}(\rho)
=\varepsilon[\text{at order }\delta^4]/\rho$, we obtain the expression for $E_{\mathrm{sym},4}(\rho)$. 
The only remaining unknown quantities in this expression are $S_\sigma^{[2]}$, $V_\omega^{[2]}$, $S_\delta^{[1]}$, and $V_\rho^{[1]}$. We next derive these four coefficients from the meson field equations.

We first determine the first-order isovector field coefficients.
The neutron and proton effective masses are
\begin{equation}
    M_{\rm n}^*=M_{\rm N}-S_\sigma+S_\delta,
    ~~
    M_{\rm p}^*=M_{\rm N}-S_\sigma-S_\delta.
\end{equation}
Since the isoscalar field has no term linear in $\delta$,
their expansions to first order are
\begin{equation}
    M_{\rm n}^*
    \approx M_0^*+S_\delta^{[1]}\delta
    +\mathcal{O}(\delta^2),
    ~~
    M_{\rm p}^*
    \approx M_0^*-S_\delta^{[1]}\delta
    +\mathcal{O}(\delta^2).
\end{equation}
Expanding the scalar density of each nucleon species around
SNM to first order in $\delta$, we obtain
\begin{align}
    \rho_{\rm n}^{\rm s}
    \approx{}&
    \left.
    \frac{\partial\varepsilon_J^{\rm{kin}}}{\partial M_J^\ast}
    \right|_{\delta=0}
   +\left[
    \frac{\rho}{2}
    \frac{\partial^2\varepsilon_J^{\rm{kin}}}
    {\partial\rho_J\partial M_J^\ast}
    +
    \frac{\partial^2\varepsilon_J^{\rm{kin}}}
    {\partial M_J^{\ast,2}}
    \cdot S_\delta^{[1]}
    \right]_{\delta=0}\cdot\delta
    +\mathcal{O}(\delta^2)
    \notag\\
    ={}&
    \frac{\rho_{\rm s}^{[0]}}{2}
    +\frac{1}{2}
    \left(
    \frac{\rho M_0^*}{\varepsilon_{\rm F}^*}
    +A S_\delta^{[1]}
    \right)\delta
    +\mathcal{O}(\delta^2),
    \notag\\
    \rho_{\rm p}^{\rm s}
    ={}&
    \frac{\rho_{\rm s}^{[0]}}{2}
    -\frac{1}{2}
    \left(
    \frac{\rho M_0^*}{\varepsilon_{\rm F}^*}
    +A S_\delta^{[1]}
    \right)\delta
    +\mathcal{O}(\delta^2).
    \label{eq:app_scalar_density_expansions}
\end{align}
The scalar density difference is
\begin{equation}
    \rho_{\rm p}^{\rm s}-\rho_{\rm n}^{\rm s}
    \approx
    -\left(
    \frac{\rho M_0^*}{\varepsilon_{\rm F}^*}
    +A S_\delta^{[1]}
    \right)\delta
    +\mathcal{O}(\delta^3).
    \label{eq:app_scalar_density_difference}
\end{equation}

Substituting the field and scalar density expansions into Eq.\,\eqref{eq:deqm} and collecting terms linear in $\delta$, we obtain
\begin{equation}
    \left(
    a_\delta
    -C_{\delta\sigma}S_\sigma^{[0],2}
    \right)S_\delta^{[1]}
    =
    -\frac{\rho M_0^*}{\varepsilon_{\rm F}^*}
    -A S_\delta^{[1]}.
\end{equation}
Rearranging gives $H_\delta S_\delta^{[1]}=-\rho M_0^*/\varepsilon_{\rm F}^*$. Similarly, expanding the $\rho$ meson field equation, Eq.\,\eqref{eq:reqm}, to first order in $\delta$ yields $H_\rho V_\rho^{[1]}=-\rho$. The first-order isovector field coefficients are therefore
\begin{equation}
    S_\delta^{[1]}
    =
    -\frac{\rho M_0^*}
    {\varepsilon_{\rm F}^*H_\delta},
   ~~
    V_\rho^{[1]}
    =
    -\frac{\rho}{H_\rho}.
    \label{eq:app_first_order_fields}
\end{equation}

We next expand the isoscalar field equations to second order
in $\delta$. Following the same procedure, we obtain
\begin{equation}
    H_\sigma S_\sigma^{[2]}=-B_\sigma,
    ~~
    H_\omega V_\omega^{[2]}=B_\omega,
    \label{eq:app_second_order_fields}
\end{equation}
where
\begin{align}
    B_\sigma
    &=
    B_{\mathrm{kin}}
    -C_{\delta\sigma}S_\sigma^{[0]}
    S_\delta^{[1],2}
    -\Lambda_{\rm S}S_\sigma^{[0]}
   V_\rho^{[1],2},
    \label{eq:app_Bsigma}\\
    B_\omega
    &=
    -\Lambda_{\rm V}V_\omega^{[0]}
    V_\rho^{[1],2}.
    \label{eq:app_Bomega}
\end{align}
Substituting these coefficients into the corresponding
energy density contributions yields
\begin{align}
    B_\sigma S_\sigma^{[2]}
    +\frac{H_\sigma}{2}S_\sigma^{[2],2}
    &=
    -\frac{B_\sigma^2}{2H_\sigma},
    \notag\\
    B_\omega V_\omega^{[2]}
    -\frac{H_\omega}{2}V_\omega^{[2],2}
    &=
    \frac{B_\omega^2}{2H_\omega}.
\end{align}

These relations eliminate the second-order isoscalar field coefficients in favor of the corresponding source terms and response functions.
With the field coefficients determined above, we obtain
the complete expression for the fourth-order symmetry energy,
\begin{align}
    E_{\mathrm{sym},4}(\rho)
    ={}&
    \frac{
    k_{\rm F}^2
    (
    4M_0^{*,4}+11M_0^{*,2}k_{\rm F}^2+10k_{\rm F}^4
    )}
    {648\varepsilon_{\rm F}^{*,5}}
    \notag\\
    &+
    \frac{
    M_0^*k_{\rm F}^2
    (M_0^{*,2}+4k_{\rm F}^2)}
    {54\varepsilon_{\rm F}^{*,5}}
    S_\delta^{[1]}
    +
    \frac{
    k_{\rm F}^2(2M_0^{*,2}-k_{\rm F}^2)}
    {12\varepsilon_{\rm F}^{*,5}}
    S_\delta^{[1], 2}
    \notag\\
    &
    -
    \frac{M_0^*k_{\rm F}^2}
    {2\varepsilon_{\rm F}^{*,5}}
    S_\delta^{[1], 3}
   +
    \left.
    \frac{1}{24\rho}
    \frac{\partial^2 A}{\partial M_0^{\ast,2}}
    \right|_{k_{\rm F}}
    \cdot S_\delta^{[1], 4}\notag\\
    &
    +\frac{1}{2\rho}\left(
\frac{B_\omega^2}{H_\omega}
-\frac{B_\sigma^2}{H_\sigma}
\right).
    \label{eq:app_Esym4}
\end{align}
Writing out the detailed expressions here gives the formula (\ref{eq:Esym4-full}) in the main text.

\section{Fine-tuning and Potential Inconsistencies Associated with Extremely Large Higher-order Symmetry Energies}\label{app:2}

In this appendix, we examine the extent to which fine-tuned cancellations among anomalously large higher-order symmetry energies can preserve a finite PNM EOS, and investigate whether such cancellations can remain effective for other physical quantities. Although different higher-order terms may cancel precisely in the PNM energy, the same cancellation generally does not persist in the neutron-proton chemical potential difference, since differentiation with respect to the isospin asymmetry introduces different weights for different orders. Consequently, maintaining a finite chemical potential difference and beta equilibrium may require an increasingly restrictive composition and may lead to anomalous behavior at intermediate isospin asymmetries. For instance, in the sixth-order model, the nonzero beta-equilibrium branch is forced to approach $\delta_\beta\approx\sqrt{2/3}$ within an increasingly narrow interval as the quartic symmetry energy grows. We first illustrate these effects using two analytical models involving terms up to sixth and eighth order, respectively, and then generalize the analysis to arbitrarily high orders.

The EOS of ANM can be expanded in powers of the isospin asymmetry $\delta$ as
\begin{align}
E(\rho,\delta)\approx& E_0(\rho)
+E_{\rm{sym}}(\rho)\delta^2
+E_{\rm{sym,4}}(\rho)\delta^4\notag\\
&+E_{\rm{sym,6}}(\rho)\delta^6
+E_{\rm{sym,8}}(\rho)\delta^8+\cdots.
\end{align}
For PNM, $\delta=1$, and the higher-order contributions enter only through their sum. Consequently, a smooth and apparently reasonable PNM EOS does not necessarily exclude anomalously large individual higher-order symmetry energies, provided that strong cancellations occur among them. However, such cancellations at $\delta=1$ do not generally persist at intermediate isospin asymmetries. In particular, the neutron-proton chemical potential difference and the isospin curvature,
\begin{equation}
\mu_{\rm{n}}-\mu_{\rm{p}}
\equiv 2\dot{E}(\rho,\delta)
\equiv 2\frac{\partial E(\rho,\delta)}{\partial\delta},
~~
\ddot{E}(\rho,\delta)
\equiv\frac{\partial^2 E(\rho,\delta)}{\partial\delta^2},
\end{equation}
involve different combinations of the higher-order coefficients. Since $\delta=1-2x_{\rm p}$, the latter is related to the composition curvature in the thermodynamic stability condition, Eq.\,(\ref{Vther}), through $\partial^2 E/\partial x_{\rm p}^2=4\ddot E$. The requirement $\ddot E>0$ ensures local convexity of the nuclear EOS with respect to the isospin composition, whereas $\ddot E<0$ violates this stability requirement and invalidates the usual interpretation of Eq.\,(\ref{Vther}). Although the lepton contribution must also be included when assessing the stability of charge-neutral beta-equilibrated matter, it cannot compensate a negatively divergent nuclear curvature at a fixed interior composition. We illustrate how anomalously large higher-order symmetry energies can violate this requirement using two simple analytical models and then generalize the argument to arbitrary higher-order terms. Throughout the following analysis, $\rho_{\rm{f}}$ denotes a fixed reference density at which the anomalously large higher-order coefficients are considered. For the two polynomial models below, the finite contributions and their first two isospin derivatives are assumed to remain bounded as the quartic coefficient diverges.

We first consider $E_{\rm{sym,4}}(\rho_{\rm f})\to+\infty$. For the opposite limit, the leading higher-order energies and nuclear isospin curvatures derived below reverse sign, while the conclusion that the compensated EOS develops singular behavior remains unchanged.

{\it Sixth-order model.}
Consider the limiting behavior
\begin{align}
E_{\rm{sym,4}}(\rho_{\rm{f}})&\to+\infty\text{ (or extremely large)},\\
E_{\rm{sym,6}}(\rho_{\rm{f}})&
\approx-E_{\rm{sym,4}}(\rho_{\rm{f}})+\epsilon,
\end{align}
where $\epsilon$ (which is small) and the lower order coefficients remain finite. The PNM contribution from the fourth- and sixth-order terms is then finite,
\begin{align}
E_{\rm{sym,4}}(\rho_{\rm{f}})
+E_{\rm{sym,6}}(\rho_{\rm{f}})\approx\epsilon.
\end{align}
The corresponding higher-order contribution at arbitrary $\delta$ is then approximately given by
\begin{equation}
E_{\rm{high}}^{(\rm{6th})}(\rho_{\rm{f}},\delta)
\approx E_{\rm{sym,4}}(\rho_{\rm{f}})
\delta^4(1-\delta^2)+\epsilon\delta^6.
\end{equation}
Although the leading divergent contribution vanishes at PNM with $\delta=1$, leaving a finite residual contribution of approximately $\epsilon$, the higher-order energy develops an increasingly large positive barrier at intermediate isospin asymmetries as $E_{\rm{sym,4}}(\rho_{\rm{f}})$ increases.

The neutron-proton chemical potential difference is
\begin{equation}
\mu_{\rm{n}}-\mu_{\rm{p}}
\approx4E_{\rm{sym}}(\rho_{\rm{f}})\delta
+4E_{\rm{sym,4}}(\rho_{\rm{f}})
\delta^3(2-3\delta^2)
+12\epsilon\delta^5.
\end{equation}
For $\beta$-equilibrated matter, $\mu_{\rm{n}}-\mu_{\rm{p}}=\mu_{\rm e}$, where $\mu_{\rm e}$ is the electron chemical potential. For simplicity, we consider electrons only, since the inclusion of muons does not qualitatively alter our conclusions. 
As the quartic symmetry energy becomes increasingly large, maintaining a finite neutron-proton chemical potential difference requires the leading divergent term to be strongly suppressed. This can occur only when its asymmetry dependent factor approaches zero, giving rise to two possible asymptotic stationary branches: $\delta_\beta\to0$ and $\delta_\beta\to\sqrt{2/3}$. The former corresponds to $x_{\rm{p}}\to50\%$, an anomalously proton-rich composition for conventional NS matter. The latter corresponds to
\begin{equation}
\delta_\beta\to\delta_*
\equiv\sqrt{\frac{2}{3}},
~~
x_{\rm{p}}\to x_{\rm{p}}^\ast
\equiv\frac{1-\sqrt{2/3}}{2}
\approx9.2\%.
\end{equation}
For this nonzero branch,
\begin{equation}
\delta_\beta\approx \delta_*
+\mathcal{O}\left(
E_{\rm{sym,4}}^{-1}(\rho_{\rm{f}})
\right).
\end{equation}
Although the limiting proton fraction appears reasonable, the leading higher-order energy at the corresponding beta-equilibrium stationary point satisfies
\begin{equation}
E_{\rm{high}}^{(\rm{6th})}(\rho_{\rm{f}},\delta_\beta)
\approx\frac{4}{27}E_{\rm{sym,4}}(\rho_{\rm{f}})
+\mathcal{O}(1
)
\to+\infty.
\end{equation}
Moreover, the isospin curvature is
\begin{equation}
\ddot{E}(\rho_{\rm{f}},\delta_\beta)
\approx-\frac{16}{3}E_{\rm{sym,4}}(\rho_{\rm{f}})
+\mathcal{O}(1)
\to-\infty.
\end{equation}
Thus, for $E_{\rm{sym,4}}(\rho_{\rm f})\to+\infty$, the nonzero beta-equilibrium stationary branch approaches the maximum of a divergent positive energy barrier. The finite stabilizing lepton contribution cannot compensate the negatively divergent nuclear isospin curvature, rendering this branch locally unstable despite its apparently reasonable proton fraction (roughly about 9.2\%). For $E_{\rm{sym,4}}(\rho_{\rm f})\to-\infty$, the same branch instead approaches an increasingly deep energy minimum with positive isospin curvature, but its energy diverges to negative infinity, making the assumed EOS physically unacceptable in this limit.

The PNM endpoint $\delta=1$ requires separate consideration. Although its higher-order energy contribution remains finite, $E_{\rm{high}}^{(\rm{6th})}(\rho_{\rm f},1)\approx\epsilon$, the neutron-proton chemical potential difference behaves as $\mu_{\rm n}-\mu_{\rm p}\approx-4E_{\rm{sym,4}}(\rho_{\rm f})+\mathcal O(1)$ and diverges in magnitude. Thus, this endpoint cannot be approached by an interior beta-equilibrium branch ($0<x_{\rm p}<1/2$) with a finite electron chemical potential. Its boundary stability depends on the sign of the quartic symmetry energy. Since $\delta=1-2x_{\rm p}$, introducing a small proton fraction changes the leading higher-order energy by approximately $4E_{\rm{sym,4}}(\rho_{\rm f})x_{\rm p}$, while the lower-order and lepton contributions cannot compensate this divergent response. Consequently, the PNM endpoint is a locally stable boundary minimum for $E_{\rm{sym,4}}(\rho_{\rm f})\to+\infty$, but becomes unstable against the appearance of a small proton and electron fraction for $E_{\rm{sym,4}}(\rho_{\rm f})\to-\infty$. Thus, even at the PNM endpoint, the cancellation preserves a finite energy but leaves a divergent chemical potential difference, revealing a singular and physically problematic composition response in the assumed limit.

{\it Eighth-order model.}
We next consider
\begin{align}
E_{\rm{sym,4}}(\rho_{\rm{f}})&\to+\infty,~~
E_{\rm{sym,6}}(\rho_{\rm{f}})
\approx rE_{\rm{sym,4}}(\rho_{\rm{f}}),\\
E_{\rm{sym,8}}(\rho_{\rm{f}})&
\approx -(1+r)E_{\rm{sym,4}}(\rho_{\rm{f}})
+\epsilon,
\end{align}
where $r$ and $\epsilon$ remain finite ($\epsilon$ is assumed to be small).  Again, the sum
\begin{equation}
E_{\rm{sym,4}}(\rho_{\rm{f}})
+E_{\rm{sym,6}}(\rho_{\rm{f}})
+E_{\rm{sym,8}}(\rho_{\rm{f}})
\approx\epsilon
\end{equation}
is finite, whereas the higher-order contribution at arbitrary asymmetry is
\begin{align}
E_{\rm{high}}^{(\rm{8th})}(\rho_{\rm{f}},\delta)
\approx&E_{\rm{sym,4}}(\rho_{\rm{f}})
\left[\delta^4+r\delta^6
-(1+r)\delta^8\right]+\epsilon\delta^8\notag\\
=&E_{\rm{sym,4}}(\rho_{\rm{f}})
\vartheta^2(1-\vartheta)
\left[1+(1+r)\vartheta\right]+\epsilon\vartheta^4,
\end{align}
where $\vartheta\equiv\delta^2$. The leading divergent contribution vanishes at both $\vartheta=0$ and $\vartheta=1$, while the full higher-order contribution satisfies $E_{\rm{high}}^{(\rm{8th})}(\rho_{\rm{f}},0)\approx0$ and $E_{\rm{high}}^{(\rm{8th})}(\rho_{\rm{f}},1)\approx\epsilon$. Nevertheless, it can grow without bound at intermediate isospin asymmetries in the assumed limit $E_{\rm{sym,4}}(\rho_{\rm{f}})\to+\infty$.

The neutron-proton chemical potential difference is
\begin{equation}
\mu_{\rm{n}}-\mu_{\rm{p}}
\approx4E_{\rm{sym,4}}(\rho_{\rm{f}})
\delta^3\mathcal{P}(\vartheta)
+4E_{\rm{sym}}(\rho_{\rm{f}})\delta
+16\epsilon\delta^7,
\end{equation}
where
\begin{equation}
\mathcal{P}(\vartheta)=2+3r\vartheta
-4(1+r)\vartheta^2.
\end{equation}
For a branch approaching a fixed interior asymmetry, a finite beta-equilibrium chemical potential requires a nonzero limiting root $\vartheta_*$ satisfying
\begin{equation}
\mathcal{P}(\vartheta_*)\approx0,~~
0<\vartheta_*<1.
\end{equation}
This condition suppresses the leading divergent contribution to the chemical potential difference, but does not necessarily suppress the corresponding higher-order energy or its isospin curvature. Their limiting behavior therefore provides an additional test of the physical consistency of the stationary branch.
For a fixed interior root,
 $
\vartheta_\beta\equiv\delta_\beta^2
=\vartheta_*+\mathcal{O}(
E_{\rm{sym,4}}^{-1}(\rho_{\rm{f}})
)$.
Using the root condition, the leading higher-order energy evaluated at the beta-equilibrium stationary composition becomes
\begin{equation}
E_{\rm{high}}^{(\rm{8th})}(\rho_{\rm{f}},\delta_\beta)
\approx E_{\rm{sym,4}}(\rho_{\rm{f}})
\frac{\vartheta_*^2(1-\vartheta_*)^2}
{3-4\vartheta_*}
+\mathcal{O}(1).
\end{equation}
The corresponding isospin curvature is
\begin{align}
\ddot{E}(\rho_{\rm{f}},\delta_\beta)
\approx&-4E_{\rm{sym,4}}(\rho_{\rm{f}})
\vartheta_*(4+3r\vartheta_*)
+\mathcal{O}(1)\notag\\
=&-4E_{\rm{sym,4}}(\rho_{\rm{f}})
\vartheta_*
\frac{6-16\vartheta_*+12\vartheta_*^2}
{3-4\vartheta_*}
+\mathcal{O}(1).
\end{align}
Since
\begin{equation}
6-16\vartheta_*+12\vartheta_*^2
=12\left(\vartheta_*-\frac{2}{3}\right)^2
+\frac{2}{3}>0,
\end{equation}
the signs of the leading higher-order energy and the isospin curvature are opposite. Specifically, for a fixed interior root,
\begin{equation}
\begin{cases}
E_{\rm{high}}^{(\rm{8th})}(\rho_{\rm{f}},\delta_\beta)\to+\infty,
\ddot{E}(\rho_{\rm{f}},\delta_\beta)\to-\infty,
& 0<\vartheta_*<3/4,\\
E_{\rm{high}}^{(\rm{8th})}(\rho_{\rm{f}},\delta_\beta)\to-\infty,
\ddot{E}(\rho_{\rm{f}},\delta_\beta)\to+\infty,
& 3/4<\vartheta_*<1.
\end{cases}
\end{equation}

In fact, the ratio between the higher-order energy contribution and the isospin curvature approaches a finite, negative value,
\begin{equation}
\lim_{E_{\rm{sym,4}}(\rho_{\rm{f}})\to+\infty}
\frac{E_{\rm{high}}^{(\rm{8th})}(\rho_{\rm{f}},\delta_\beta)}
{\ddot{E}(\rho_{\rm{f}},\delta_\beta)}
\approx-\frac{1}{4}\frac{\vartheta_*(1-\vartheta_*)^2}
{6-16\vartheta_*+12\vartheta_*^2}
<0,
\end{equation}
for any fixed interior root $0<\vartheta_*<1$. This demonstrates that the divergent higher-order energy and isospin curvature have opposite signs, irrespective of the value of $r$ admitting such an interior root. The two asymptotic regimes are formally separated by $\vartheta_*=3/4$, corresponding to
\begin{equation}
x_{\rm{p}}^\ast=\frac{2-\sqrt{3}}{4}
\approx6.7\%.
\end{equation}
Specifically, $\vartheta_*<3/4$ corresponds to divergent positive energy and negative curvature, whereas $\vartheta_*>3/4$ corresponds to divergent negative energy and positive curvature. However, the dividing value itself cannot satisfy the stationary point condition for any finite $r$, since $\mathcal{P}(3/4)=-1/4\neq0$. Indeed, the root condition gives $r=(4\vartheta_*^2-2)/\vartheta_*(3-4\vartheta_*)$, implying $|r|\to\infty$ as $\vartheta_*\to3/4$. Thus, $\vartheta_*=3/4$ represents a formal boundary between the two asymptotic regimes rather than an attainable stationary point for finite $r$.

For $x_{\rm{p}}^\ast\gtrsim6.7\%$, the nonzero stationary branch exhibits a divergent positive energy contribution and negative composition curvature, indicating local instability. Conversely, for $0<x_{\rm{p}}^\ast\lesssim6.7\%$, the branch has positive curvature but develops an increasingly deep negative energy minimum. In this latter regime, local stability against composition fluctuations is restored, yet the energy per baryon at the stationary composition tends to negative infinity in the assumed limit, indicating an anomalous energy scale. 

The PNM endpoint corresponding to $\vartheta_\ast=1$ requires separate consideration. Its higher-order energy contribution remains finite, $E_{\rm{high}}^{(\rm{8th})}(\rho_{\rm f},1)\approx\epsilon$, whereas the neutron-proton chemical potential difference behaves as 
\begin{equation}
\mu_{\rm n}-\mu_{\rm p}\approx-4(2+r)E_{\rm{sym,4}}(\rho_{\rm f})+\mathcal O(1).\end{equation} For fixed $r\neq-2$, this difference diverges in magnitude, preventing an interior beta-equilibrium branch ($0<x_{\rm p}<1/2$) with a finite electron chemical potential from approaching PNM. Introducing a small proton fraction changes the leading higher-order energy by approximately $4(2+r)E_{\rm{sym,4}}(\rho_{\rm f})x_{\rm p}$. Thus, the endpoint is locally stable when $(2+r)E_{\rm{sym,4}}\to+\infty$ and unstable when this product tends to negative infinity.
The exceptional case $r\approx-2$ also cancels the leading divergent contribution to the chemical potential difference. The higher-order energy then becomes $E_{\rm{high}}^{(\rm{8th})}\approx E_{\rm{sym,4}}(\rho_{\rm f})\delta^4(1-\delta^2)^2+\epsilon\delta^8$, so both the endpoint energy and its first isospin derivative remain finite. Nevertheless, the curvature satisfies $\ddot E(\rho_{\rm f},1)\approx8E_{\rm{sym,4}}(\rho_{\rm f})+\mathcal O(1)$ and diverges in magnitude.
A negative divergence signals a loss of convexity of the nuclear energy, whereas a positive divergence implies an infinitely stiff nuclear isospin response; neither case restores a finite, regular curvature.
The boundary stability must therefore be assessed from the one-sided change in the full energy, including the lepton contribution. A finite endpoint chemical potential difference alone does not establish either beta equilibrium or boundary stability. Consequently, even this additional cancellation does not regularize the nuclear composition response: For any fixed finite $r$, either the neutron-proton chemical potential difference or the nuclear isospin curvature remains singular in the assumed limit.

These two analytical examples demonstrate that cancellations among anomalously large higher-order symmetry energies can conceal substantial irregularities in the isospin dependence of the EOS. A finite and smooth PNM energy does not necessarily guarantee a physically reasonable beta-equilibrated composition or regular behavior of the energy and its isospin derivatives, either at intermediate asymmetries or at the PNM endpoint. These conclusions apply specifically to the singular limit of strongly compensated higher-order coefficients and do not imply that a finite but sizable quartic symmetry energy is intrinsically problematic.

{\it Generalization to arbitrary higher-order terms.}
The preceding examples can be generalized to an arbitrary number of higher-order symmetry energies. Consider
\begin{equation}
E(\rho,\delta)=E_0(\rho)
+E_{\rm{sym}}(\rho)\delta^2
+\sum_{n=2}^{\infty}
E_{\rm{sym},2n}(\rho)\delta^{2n}.
\end{equation}
At the fixed density $\rho_{\rm{f}}$, suppose the higher-order coefficients share a common divergent scale,
$
E_{\rm{sym},2n}(\rho_{\rm{f}})
\approx r_nE_{\rm{sym,4}}(\rho_{\rm{f}})
+\mathcal{O}(1)$, where $r_2\approx 1$ as $E_{\rm{sym,4}}(\rho_{\rm{f}})\to+\infty$. Here, the normalized coefficients $r_n$ are fixed independently of the divergent scale, with $r_2\approx1$. We assume that the summed remainder remains uniformly bounded and that the normalized series
\begin{equation}
\mathcal{C}(\delta)\equiv
\sum_{n=2}^{\infty}r_n\delta^{2n}
\end{equation}
converges to a continuous function on $0\leq\delta\leq1$, with sufficient regularity in the interior and a convergent expansion near the SNM with $\delta=0$.

If the divergent higher-order contributions cancel in PNM, the normalized coefficients must satisfy
$
\mathcal{C}(1)=\sum_{n=2}^{\infty}r_n=0$.
Importantly, this condition constrains only the sum of the higher-order coefficients at $\delta=1$, rather than their contributions at arbitrary isospin asymmetry. In particular, since $r_2\approx1$, the behavior of $\mathcal{C}(\delta)$ near SNM is governed by
\begin{equation}
\mathcal{C}(\delta)\approx\delta^4+\mathcal{O}(\delta^6),
~~
\lim_{\delta\to0^+}
[{\mathcal{C}(\delta)}/{\delta^4}]\approx1.
\end{equation}
Consequently, for any fixed constant $d\in(0,1)$, there exists a sufficiently small but still finite $\delta_0\in(0,1)$, independent of $E_{\rm{sym,4}}(\rho_{\rm{f}})$, that
$
\mathcal{C}(\delta_0)\gtrsim d\delta_0^4>0$.
Moreover, since $\mathcal{C}(0)=\mathcal{C}(1)=0$ and $\mathcal{C}(\delta_0)>0$, the continuity of $\mathcal{C}(\delta)$ guarantees its maximum is attained at some intermediate asymmetry $\delta_*\in(0,1)$, where
\begin{equation}
\mathcal{C}'(\delta_*)\approx0,~~
\mathcal{C}(\delta_*)\gtrsim \mathcal{C}(\delta_0)
\gtrsim d\delta_0^4>0.
\end{equation}
Consequently, the higher-order contribution at this intermediate asymmetry satisfies
\begin{align}
E_{\rm{high}}(\rho_{\rm{f}},\delta_*)
&=E_{\rm{sym,4}}(\rho_{\rm{f}})
\mathcal{C}(\delta_*)+\mathcal{O}(1)\notag\\
&\gtrsim d\delta_0^4
E_{\rm{sym,4}}(\rho_{\rm{f}})
+\mathcal{O}(1)
\to+\infty.
\end{align}
Therefore, although the divergent higher-order contributions cancel in PNM, they necessarily generate an increasingly large positive energy barrier at some intermediate isospin asymmetry $\delta_*$. This conclusion is independent of the number of higher-order terms retained, provided that the normalized expansion converges with the assumed regularity and its remainder remains uniformly bounded. Thus, including arbitrarily high-order symmetry energies cannot eliminate this anomalous behavior merely through cancellations at the PNM limit.
The same argument applies when $E_{\rm{sym,4}}(\rho_{\rm{f}})\to-\infty$. In this case, the positive maximum of $\mathcal{C}(\delta)$ instead produces an increasingly deep negative energy minimum at an intermediate isospin asymmetry, so the higher-order contribution again becomes unbounded in magnitude.

More generally, under the stated assumptions, cancellations among arbitrarily many higher-order symmetry energies cannot simultaneously preserve a finite PNM contribution and prevent the higher-order energy from becoming unbounded at intermediate isospin asymmetries. Such cancellations may require substantial fine-tuning among the higher-order coefficients, particularly when their individual magnitudes become anomalously large. Nevertheless, the present analysis is not exhaustive, as it assumes a common divergent scale and a sufficiently regular higher-order expansion. More elaborate fine-tuning mechanisms, including nonuniform cancellations among infinitely many terms, may evade the specific argument presented here and deserve further investigation. Thus, although these results do not establish a universal upper bound on a finite $E_{\rm{sym,4}}(\rho)$, they reveal potential inconsistencies associated with anomalously large, strongly compensated higher-order symmetry energies. They also underscore the importance of examining the full isospin dependence of the EOS, particularly its beta-equilibrated composition and stability properties, rather than relying solely on the PNM limit.

\end{document}